\documentclass[aps, prc, reprint, superscriptaddress]{revtex4-1}
\usepackage{amsmath}
\usepackage{amsfonts}
\usepackage{amssymb}
\usepackage{dcolumn}
\usepackage{lineno}
\usepackage{multirow}
\usepackage{array}
\usepackage{graphicx}
\usepackage[caption=false]{subfig}
\usepackage[colorlinks]{hyperref}
\hypersetup{linkcolor=blue, citecolor=blue, filecolor=blue, urlcolor=blue}
\usepackage[all]{hypcap}
\usepackage{color}
\usepackage[utf8]{inputenc}

\newcommand{\snn}{\ensuremath{\sqrt{s_{{\rm NN}}}}}

\newcommand\la{\langle}
\newcommand\ra{\rangle}
\newcommand{\KV}{{\mbox{$\kappa \sigma^{2}$}}}

\newcommand{\sNN}{{{$\sqrt{s_{{\mathrm{NN}}}}$}}}

\newcommand{ \be }{\begin{equation}}
\newcommand{ \ee }{\end{equation}}

\begin{document}
\title{Cumulants and Correlation Functions of Net-proton, Proton and Antiproton Multiplicity Distributions in Au+Au Collisions at energies available at the BNL Relativistic Heavy Ion Collider}
\affiliation{Abilene Christian University, Abilene, Texas   79699}
\affiliation{AGH University of Science and Technology, FPACS, Cracow 30-059, Poland}
\affiliation{Alikhanov Institute for Theoretical and Experimental Physics NRC "Kurchatov Institute", Moscow 117218, Russia}
\affiliation{Argonne National Laboratory, Argonne, Illinois 60439}
\affiliation{American University of Cairo, New Cairo 11835, New Cairo, Egypt}
\affiliation{Brookhaven National Laboratory, Upton, New York 11973}
\affiliation{University of California, Berkeley, California 94720}
\affiliation{University of California, Davis, California 95616}
\affiliation{University of California, Los Angeles, California 90095}
\affiliation{University of California, Riverside, California 92521}
\affiliation{Central China Normal University, Wuhan, Hubei 430079 }
\affiliation{University of Illinois at Chicago, Chicago, Illinois 60607}
\affiliation{Creighton University, Omaha, Nebraska 68178}
\affiliation{Czech Technical University in Prague, FNSPE, Prague 115 19, Czech Republic}
\affiliation{Technische Universit\"at Darmstadt, Darmstadt 64289, Germany}
\affiliation{ELTE E\"otv\"os Lor\'and University, Budapest, Hungary H-1117}
\affiliation{Frankfurt Institute for Advanced Studies FIAS, Frankfurt 60438, Germany}
\affiliation{Fudan University, Shanghai, 200433 }
\affiliation{University of Heidelberg, Heidelberg 69120, Germany }
\affiliation{University of Houston, Houston, Texas 77204}
\affiliation{Huzhou University, Huzhou, Zhejiang  313000}
\affiliation{Indian Institute of Science Education and Research (IISER), Berhampur 760010 , India}
\affiliation{Indian Institute of Science Education and Research (IISER) Tirupati, Tirupati 517507, India}
\affiliation{Indian Institute Technology, Patna, Bihar 801106, India}
\affiliation{Indiana University, Bloomington, Indiana 47408}
\affiliation{Institute of Modern Physics, Chinese Academy of Sciences, Lanzhou, Gansu 730000 }
\affiliation{University of Jammu, Jammu 180001, India}
\affiliation{Joint Institute for Nuclear Research, Dubna 141 980, Russia}
\affiliation{Kent State University, Kent, Ohio 44242}
\affiliation{University of Kentucky, Lexington, Kentucky 40506-0055}
\affiliation{Lawrence Berkeley National Laboratory, Berkeley, California 94720}
\affiliation{Lehigh University, Bethlehem, Pennsylvania 18015}
\affiliation{Max-Planck-Institut f\"ur Physik, Munich 80805, Germany}
\affiliation{Michigan State University, East Lansing, Michigan 48824}
\affiliation{National Research Nuclear University MEPhI, Moscow 115409, Russia}
\affiliation{National Institute of Science Education and Research, HBNI, Jatni 752050, India}
\affiliation{National Cheng Kung University, Tainan 70101 }
\affiliation{Nuclear Physics Institute of the CAS, Rez 250 68, Czech Republic}
\affiliation{Ohio State University, Columbus, Ohio 43210}
\affiliation{Institute of Nuclear Physics PAN, Cracow 31-342, Poland}
\affiliation{Panjab University, Chandigarh 160014, India}
\affiliation{Pennsylvania State University, University Park, Pennsylvania 16802}
\affiliation{NRC "Kurchatov Institute", Institute of High Energy Physics, Protvino 142281, Russia}
\affiliation{Purdue University, West Lafayette, Indiana 47907}
\affiliation{Rice University, Houston, Texas 77251}
\affiliation{Rutgers University, Piscataway, New Jersey 08854}
\affiliation{Universidade de S\~ao Paulo, S\~ao Paulo, Brazil 05314-970}
\affiliation{University of Science and Technology of China, Hefei, Anhui 230026}
\affiliation{Shandong University, Qingdao, Shandong 266237}
\affiliation{Shanghai Institute of Applied Physics, Chinese Academy of Sciences, Shanghai 201800}
\affiliation{Southern Connecticut State University, New Haven, Connecticut 06515}
\affiliation{State University of New York, Stony Brook, New York 11794}
\affiliation{Instituto de Alta Investigaci\'on, Universidad de Tarapac\'a, Arica 1000000, Chile}
\affiliation{Temple University, Philadelphia, Pennsylvania 19122}
\affiliation{Texas A\&M University, College Station, Texas 77843}
\affiliation{University of Texas, Austin, Texas 78712}
\affiliation{Tsinghua University, Beijing 100084}
\affiliation{University of Tsukuba, Tsukuba, Ibaraki 305-8571, Japan}
\affiliation{United States Naval Academy, Annapolis, Maryland 21402}
\affiliation{Valparaiso University, Valparaiso, Indiana 46383}
\affiliation{Variable Energy Cyclotron Centre, Kolkata 700064, India}
\affiliation{Warsaw University of Technology, Warsaw 00-661, Poland}
\affiliation{Wayne State University, Detroit, Michigan 48201}
\affiliation{Yale University, New Haven, Connecticut 06520}

\author{M.~S.~Abdallah}\affiliation{American University of Cairo, New Cairo 11835, New Cairo, Egypt}
\author{J.~Adam}\affiliation{Brookhaven National Laboratory, Upton, New York 11973}
\author{L.~Adamczyk}\affiliation{AGH University of Science and Technology, FPACS, Cracow 30-059, Poland}
\author{J.~R.~Adams}\affiliation{Ohio State University, Columbus, Ohio 43210}
\author{J.~K.~Adkins}\affiliation{University of Kentucky, Lexington, Kentucky 40506-0055}
\author{G.~Agakishiev}\affiliation{Joint Institute for Nuclear Research, Dubna 141 980, Russia}
\author{I.~Aggarwal}\affiliation{Panjab University, Chandigarh 160014, India}
\author{M.~M.~Aggarwal}\affiliation{Panjab University, Chandigarh 160014, India}
\author{Z.~Ahammed}\affiliation{Variable Energy Cyclotron Centre, Kolkata 700064, India}
\author{I.~Alekseev}\affiliation{Alikhanov Institute for Theoretical and Experimental Physics NRC "Kurchatov Institute", Moscow 117218, Russia}\affiliation{National Research Nuclear University MEPhI, Moscow 115409, Russia}
% ...

\collaboration{STAR Collaboration}\noaffiliation

\medskip

\date{\today}
\begin{abstract}
We report a systematic measurement of cumulants, $C_{n}$, for net-proton, proton and antiproton multiplicity distributions, and correlation functions, $\kappa_n$, for proton and antiproton multiplicity distributions up to the fourth order in Au+Au collisions at $\sqrt{s_{\mathrm {NN}}}$ = 7.7, 11.5, 14.5, 19.6, 27, 39, 54.4, 62.4 and 200 GeV. The $C_{n}$ and $\kappa_n$ are presented as a function of collision energy, centrality and kinematic acceptance in rapidity, $y$, and transverse momentum, $p_{T}$. The data were taken during the first phase of the Beam Energy Scan (BES) program (2010 -- 2017) at the BNL Relativistic Heavy Ion Collider (RHIC) facility. The measurements are carried out at midrapidity ($|y| <$ 0.5) and transverse momentum 0.4 $<$ $p_{\rm T}$ $<$ 2.0 GeV/$c$, using the STAR detector at RHIC. We observe a non-monotonic energy dependence ({\snn} = 7.7 -- 62.4 GeV) of the net-proton $C_{4}$/$C_{2}$ with the significance of 3.1$\sigma$ for the 0-5\% central Au+Au collisions. This is consistent with the expectations of critical fluctuations in a QCD-inspired model. Thermal and transport model calculations show a monotonic variation with \snn. For the multiparticle correlation functions, we observe significant negative values for a two-particle correlation function, $\kappa_2$, of protons and antiprotons, which are mainly due to the effects of baryon number conservation. Furthermore, it is found that the four-particle correlation function, $\kappa_4$, of protons plays a role in determining the energy dependence of proton $C_4/C_1$ below 19.6 GeV, which cannot be understood by the effect of baryon number conservation. \end{abstract}
\maketitle

\section{Introduction}
The main goal of the Beam Energy Scan (BES) program at the BNL Relativistic Heavy Ion Collider (RHIC) is to study
the QCD phase structure~\cite{Aggarwal:2010cw,bes2}.  This is expected to lead to the
mapping of the phase diagram for strong interactions in the space of temperature ($T$) versus baryon chemical potential 
($\mu_{\rm B}$). Both theoretically and experimentally, several advancements have
been made towards this goal. Lattice QCD calculations have established that at high temperatures, there occurs a crossover transition from 
hadronic matter to a deconfined state of quarks and gluons at
$\mu_{\rm B}$ = 0 MeV~\cite{Aoki:2006we}.  Experimental data from RHIC
and the Large Hadron Collider (LHC) have provided evidence of this matter with quark and gluon degrees of freedom called the quark-gluon
plasma (QGP)~\cite{Arsene:2004fa,Back:2004je,Adcox:2004mh,Adams:2005dq}. The QGP has been found to
hadronize into a gas of hadrons, which undergoes
chemical freeze-out (inelastic collisions cease)~\cite{Adamczyk:2017iwn} at a temperature
close to the lattice QCD-estimated quark-hadron transition temperature at
$\mu_{\rm B}$ = 0 MeV~\cite{Borsanyi:2010bp,Bazavov:2018mes}. A suite of interesting results from the BES program indicate a change of equation of state of QCD matter, with collision energy from partonic-interaction-dominated matter at higher collision energies to a hadronic-interaction regime at lower energies. These include the
observations of breakdown in the number of constituent-quark scaling of the elliptic flow at lower \snn~\cite{Adamczyk:2013gv}, non-monotonic variation of the slope of the directed flow for protons and net-protons at
midrapidity as a function of \snn~\cite{Adamczyk:2014ipa}, nuclear modification factor changing values from smaller than unity to larger than unity at high $p_{\mathrm T}$ as we go to lower \snn~\cite{Adamczyk:2017nof},
and finite to vanishing values of the three-particle correlations with respect to the event plane~\cite{Adamczyk:2014mzf} as we go to lower \snn. 

The QCD phase structure at finite temperature and baryon chemical potential has been extensively studied by various QCD-based model calculations, such as the Dyson-Schwinger equation (DSE) method~\cite{Fischer:2014ata,Shi:2014zpa,Gao:2016qkh,Fischer:2018sdj,Gao:2020qsj}, functional renormalization group (FRG)~\cite{Fu:2019hdw}, Nambu-Jona-Lasinio (NJL)~\cite{Buballa:2003qv},  Polyakov Nambu-Jona-Lasinio (PNJL)~\cite{Fu:2007xc,Herbst:2010rf,Li:2018ygx} and other effective models~\cite{Fukushima:2010bq,Fukushima:2013rx}. One of the most important studies of the QCD phase structure relates to the first-order phase boundary and the expected existence of the critical point (CP)~\cite{Stephanov:1999zu,Stephanov:2004wx,Fodor:2004nz,Stephanov:2007fk,Gavai:2008zr,Gupta:2009mu}. This is the end point of
a first-order phase boundary between quark-gluon and hadronic
phases~\cite{Ejiri:2008xt,Bowman:2008kc}. Experimental confirmation of the CP would be a landmark 
of exploring the QCD phase structure. Previous studies of higher-order cumulants of net-proton multiplicity 
distributions suggest that the possible CP region is unlikely to be
below $\mu_{\rm B}$ = 200~MeV~\cite{Aggarwal:2010wy}, which is consistent with the theoretical findings~\cite{Fodor:2004nz,Gavai:2008zr,Bazavov:2017dus,Fu:2019hdw,Gao:2020qsj}.
The versatility of the RHIC machine has permitted the colliding energies of ions to be varied below the injection energy of $\sqrt{s_{\mathrm {NN}}}$ = 19.6 GeV~\cite{Abelev:2009bw}, and thereby the RHIC BES program provides the possibility to scan the QCD phase diagram up to $\mu_{\rm B}$ = 420 MeV with the collider mode, and $\mu_{\rm B}$ = 720 MeV with the fixed-target mode~\cite{bes2,STAR:2020dav}. This, in turn, opens the possibility to find the experimental signatures of a first-order phase transition and the CP~\cite{Luo:2017faz,Bzdak:2019pkr}.

Higher-order cumulants of the distributions of conserved charge, such as net-baryon ($B$), net-charge ($Q$), and net-strangeness ($S$) numbers, are sensitive to the QCD phase transition and CP~\cite{Asakawa:2000wh,Hatta:2002sj,Hatta:2003wn,Ejiri:2005wq,Koch:2005vg,Stephanov:2008qz,Asakawa:2009aj,Athanasiou:2010kw,Friman:2011pf,Gupta:2011wh,Ding:2015ona}. 
The signatures of conserved-charge fluctuations near CP have been studied by various model calculations~\cite{Stephanov:2008qz,Asakawa:2009aj,Schaefer:2011ex,Chen:2014ufa,Lu:2015naa,Chen:2015dra,Vovchenko:2015pya,Jiang:2015hri,Mukherjee:2016nhb,Herold:2016uvv,Fan:2017kym,Zhang:2017icm,Shao:2017yzv,Isserstedt:2019pgx,Mroczek:2020rpm,Fu:2021oaw}.  However, these model calculations are based on the assumption of thermal equilibrium with a static and infinite medium. 
In heavy-ion collisions, finite-size and time effects will put constraints on the significance of the signals~\cite{Palhares:2010zz,Pan:2016ecs}. A theoretical calculation suggests 
the non-equilibrium correlation length $\xi$ $\approx$ 2-3 fm for heavy-ion collisions~\cite{Berdnikov:1999ph}. Dynamical modeling of heavy-ion collisions with the physics of a critical point and non-equilibrium effects is in progress~\cite{Mukherjee:2016kyu,Stephanov:2017ghc,Wu:2018twy,Rajagopal:2019xwg,An:2019csj}. The signatures of a phase transition or a CP are detectable if they survive the evolution of the system~\cite{Stephanov:2009ra}. Due to a stronger dependence on the correlation length ($\xi$)~\cite{Stephanov:2008qz,Asakawa:2009aj,Athanasiou:2010kw}, it is proposed to study the higher moments -- skewness (${\it {S}}$ = $\left\langle (\delta N)^3 \right\rangle/\sigma^{3}$) 
and kurtosis ($\kappa$ = $\left\langle (\delta N)^4 \right\rangle/\sigma^{4}$ -- 3) 
with $\delta N$ = $N$ -- $\langle N \rangle$, or cumulants $C_{n}$ (defined
in Sec.~\ref{sub:def}) of distributions of conserved quantities. Both the magnitude
and the sign of the moments or $C_{n}$~\cite{Asakawa:2009aj,Stephanov:2011pb}, which quantify the shape of the multiplicity
distributions, are important for understanding the phase transition and CP
effects. The aim is to search for signatures of the CP 
over a broad range of $\mu_{B}$ in the QCD phase diagram~\cite{Aggarwal:2010wy}. 

Furthermore, the products of the moments or ratios of  $C_{n}$ can be related to susceptibilities associated 
with the conserved numbers. The product ($\kappa$$\sigma^2$), or
equivalently, the ratio ($C_{4}$/$C_{2}$) of the net-baryon number
distribution is related to the ratio of fourth-order ($\chi^{\mathrm B}_4$) 
to second-order ($\chi^{\mathrm B}_2$) baryon number susceptibilities~\cite{Ejiri:2005wq,Cheng:2008zh,Stokic:2008jh,Gupta:2011wh,Gavai:2010zn}. 
The ratio, $\chi^{\mathrm B}_{4}$/$\chi^{\mathrm B}_{2}$, is expected to deviate
from unity near the CP. It has different values for the hadronic and partonic phases~\cite{Gavai:2010zn}. 
Similarly, the products $\it{S}$$\sigma$ ($C_{3}$/$C_{2}$) and $\sigma^{2}$/$\langle N \rangle$ ($C_{2}$/$C_{1}$) are related to $\chi^{\mathrm B}_{3}$/$\chi^{\mathrm B}_{2}$ and $\chi^{\mathrm B}_{2}$/$\chi^{B}_{1}$, respectively. Experimentally, it is not
possible to measure the net-baryon distributions, however, theoretical calculations 
have shown that net-proton multiplicity ($N_{p} - N_{\bar{p}}$ = $\Delta N_{p}$) fluctuations reflect the singularity of 
the charge and baryon number susceptibility,  as expected at 
the CP~\cite{Hatta:2003wn}. 
References~\cite{Kitazawa:2012at,Bzdak:2012ab} discuss the effect of using net-proton as the approximation for the net-baryon distributions and the acceptance dependence for the moments of the protons and antiprotons.

In an early publication from the STAR experiment on the higher moments
of net-proton distributions, the selected kinematics of the (anti)proton are $|y| <$ 0.5 and 0.4 $<$ $p_{\rm T}$ $<$ 0.8 GeV/$c$, where only the Time Projection Chamber (TPC)~\cite{Ackermann:2002ad,Anderson:2003ur} was used for (anti)protons identification. Interesting hints of a non-monotonic variation of
$\kappa$$\sigma^2$ (or $C_{4}$/$C_{2}$) was observed~\cite{Adamczyk:2013dal}.   In this paper,
we report measurements of the energy dependence of $C_{n}$
up to fourth order of the net-proton multiplicity distributions from Au+Au collisions with a larger acceptance of 0.4 $<$ $p_{\rm T}$ $<$ 2.0 GeV/$c$~\cite{Adam:2020unf}. This is achieved by adding the information from STAR's
Time-of-Flight (TOF) detector~\cite{Llope:2012zz}.  We present results from Au+Au collisions at 9 different collision energies, \snn\ = 7.7, 11.5, 14.5, 19.6, 27, 39, 54.4, 62.4 and 200 GeV.

The paper is organized as follows. In the next section, we discuss the
data sets used, event selection criteria, centrality selection procedure, proton
identification method, measurement of raw cumulants of the net-proton
distributions,  corrections for the effects of centrality bin width (CBW) and efficiency,  and estimation of statistical and systematic uncertainties on the measurements. In Sec. III, we present the results of cumulants and
their ratios for net protons, protons  and antiprotons in Au+Au collisions as a function of
collision energy ($\sqrt{s_{\mathrm {NN}}}$), centrality, transverse momentum ($p_{T}$) acceptance and rapidity acceptance
($\Delta y$).  In addition, we present the extracted various order integrated correlation functions of protons and antiprotons from the
measured cumulants. In this section, we also discuss the results from the HRG model and transport model calculations. In Sec. IV, we present the summary. Detailed discussions on the efficiency correction, and the estimation of the statistical uncertainties are presented in Appendices A and B, respectively. 

\section{Experimental Data Analysis} 

\subsection{Data set and event selection}
The data presented in the paper were obtained using the Time Projection
Chamber (TPC)~\cite{Ackermann:2002ad} and the Time-of-Flight detectors (TOF)~\cite{Llope:2012zz} of the Solenoidal Tracker at RHIC (STAR)~\cite{Ackermann:2002ad}. 
The event-by-event proton ($N_{p}$) and antiproton ($N_{\bar{p}}$) multiplicities
are measured for Au+Au minimum-bias events at $\sqrt{s_{\mathrm
{NN}}}$ = 7.7, 11.5, 14.5, 19.6, 27, 39, 54.4, 62.4 and 200 GeV for collisions occurring within a certain $Z$-position ($V_{z}$)
range of the collision vertex (given in Table~\ref{table_events})  from the TPC center along the beam
line. These data sets were taken with a minimum-bias trigger, which was defined using a coincidence of hits in the
zero degree calorimeters (ZDCs)~\cite{Adler:2000bd}, vertex position detectors (VPDs)~\cite{Llope:2003ti}, and/or beam-beam
counters (BBCs)~\cite{Bieser:2002ah}. The range of $|V_{z}|$ is chosen to optimize the
event statistics and uniformity of the response of the detectors used in the analysis. 
\begin{table*}
\caption{Total number of events for Au+Au collisions analysed for various collision 
  energies ($\sqrt{s_{\rm NN}}$) obtained after all of the event selection criteria are applied. The $Z$-vertex 
  ($V_{z}$) range, the chemical freeze-out temperature 
  ($T_{\mathrm {ch}}$) and  baryon chemical potential ($\mu_{\mathrm 
    B}$) for 0-5\% Au+Au collisions~\cite{Adamczyk:2017iwn} are also given. 
\label{table_events}}
\begin{center}
\begin{tabular}{cccccc}
\hline 
$\sqrt{s_{NN}}$ (GeV) & No. of events ($\times10^{6}$) & $|V_{z}|$ (cm) &
                                                                   $T_{\mathrm 
                                                                   {ch}}$   (MeV) & $\mu_{\mathrm B}$ (MeV) \\ 
\hline 
200     &    238      &  30        & 164.3   & 28\\ 
\hline 
62.4    &    47      &  30        & 160.3   &  70\\ 
\hline 
54.4    &    550  &  30       &    160.0        &   83     \\
\hline 
39       &    86   &  30      &  156.4  & 103 \\
\hline 
27       &    30    &  30    &  155.0     & 144 \\
\hline 
19.6    &    15   & 30     &  153.9 & 188\\
\hline 
14.5    &    20        & 30    &  151.6 & 264 \\
\hline 
11.5    &    6.6        & 30    &  149.4 & 287 \\
\hline 
7.7      &   3          & 40   & 144.3 &  398\\
\hline 
\end{tabular}
\end{center}
\end{table*}

\begin{figure*}[htp]
\includegraphics[scale=0.5]{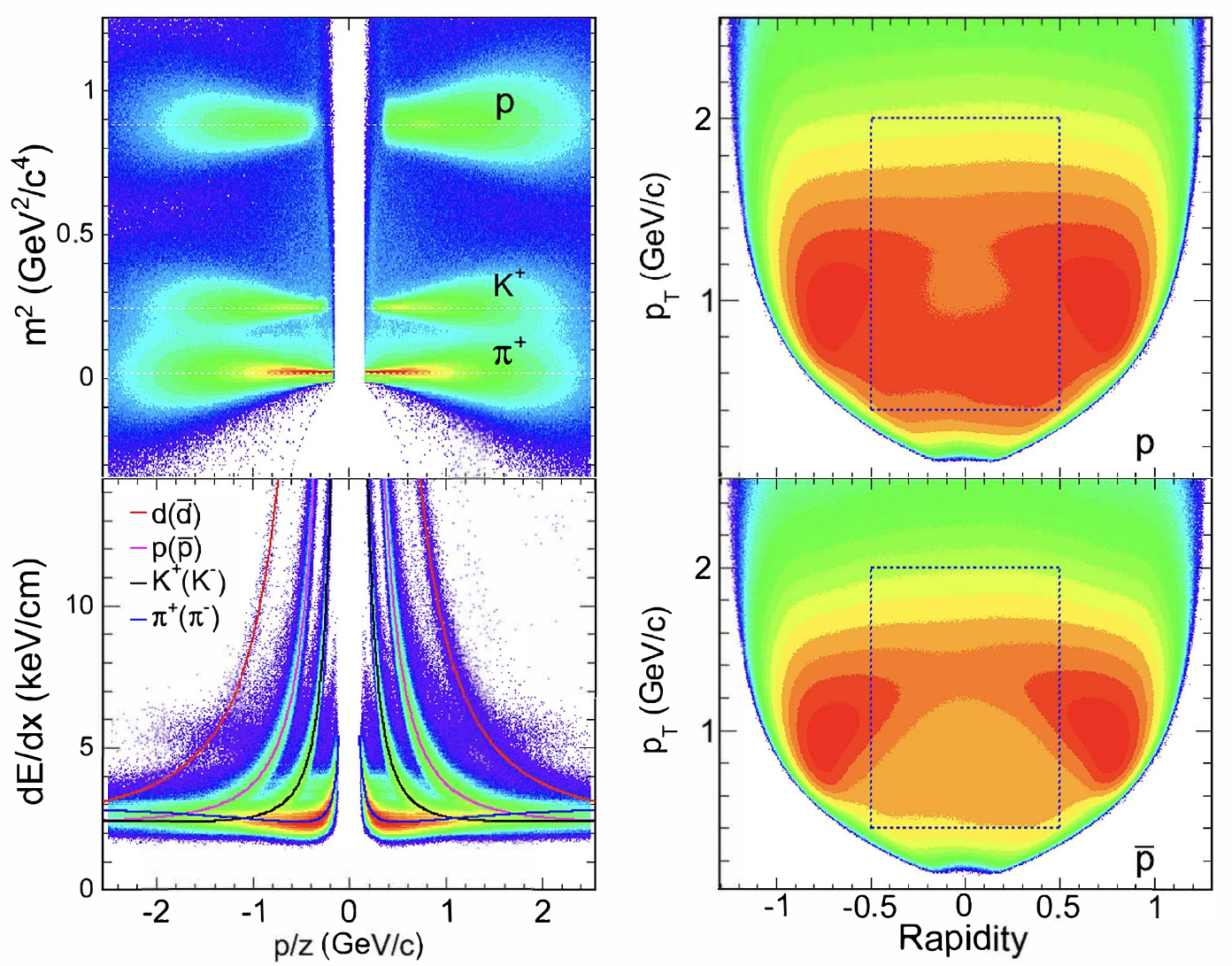}
\caption 
{(Color online) 
Top left panel: The mass squared ($m^{2}$) versus rigidity for charged 
tracks in Au+Au collisions at $\sqrt{s_{\mathrm    {NN}}}$  = 39 
GeV. The rigidity is defined as momentum/z, where z is the dimensionless ratio of particle charge to the electron charge magnitude.
Bottom left panel: The specific ionization energy loss ($dE/dx$) as a 
function of rigidity measured in the TPC for the same data set. Also 
shown as solid lines are the theoretical expectations for each particle species. Right
panels: Rapidity ($y$) versus transverse momentum ($p_{\rm T}$). The color reflects the relative yields of protons (top) and antiprotons (bottom) using the TPC PID for Au+Au collisions at 
$\sqrt{s_{\mathrm   {NN}}}$  = 39 GeV. The dashed boxes represent
the acceptance used in the current analysis. Two blobs at large rapidities are contaminated by particles other than (anti)protons. This contamination is rejected in later steps of the analysis. 
}
\label{pid}
\end{figure*}
\begin{table*}
\begin{center}
\caption{Proton and antiproton track selection criteria at all energies. The $\rm N_{Fit}$ and $\rm N_{HitPoss}$ represent
the number of hits used in track fitting and the maximum number of possible hits in the TPC. } 
\label{table_tcuts}
\vspace{.2cm}
\begin{center}
\begin{tabular}{ccccccccccc}
\hline
$|y|$   &  & $p_T$ (GeV/$c$)         &  & DCA (cm)      &  & $\rm N_{Fit}$ &  & $\rm N_{Fit}/N_{ HitPoss}$ &  & No. of $dE/dx$ points \\ \hline
$<$ 0.5 &  & 0.4-2.0 &  & $<$ 1 &  & $>$ 20          &  & $>$ 0.52            &  & $>$ 5             \\ \hline
\end{tabular}
\end{center}
\end{center}
\end{table*}

In order to reject background events which involve interactions with
the beam pipe, the transverse radius of the event vertex is required to be within 2 cm (1 cm for 14.5 GeV) of the center of STAR~\cite{Adamczyk:2017iwn}. 
We use two methods to determine the $V_z$: one from a fast scintillator-based vertex position detector, and the other from the most probable point of common origin of the tracks, which are reconstructed from the hits measured in the TPC. To remove pile-up events at energies above 27 GeV, we require the $V_z$ difference between the two methods to be within 3 cm. Further, a detailed study of the TPC tracks as a function of the TOF matched tracks with valid TOF information is carried out and outlier events are rejected.
To ensure the quality of the data, a run-by-run study of several variables -- such as the total number
of uncorrected charged particles measured in the TPC, average transverse
momentum ($\langle p_{\rm T} \rangle$), mean pseudorapidity
($\eta$) and azimuthal angle ($\phi$) in an event -- is carried
out. Outlier runs beyond $\pm$
3$\sigma$, where $\sigma$ corresponds to the standard deviation of run-by-run
distributions of a variable, are not included in the current analysis.  In addition, the distance of closest approach (DCA) of the charged-particle track from the primary vertex, and especially the signed transverse DCA (DCA$_{xy}$) are studied to remove bad events (The signed transverse DCA refers to the DCA with respect to the primary vertex in the transverse plane.  Its sign is the sign of the vector product of the DCA vector and the track momentum). These classes of bad events are primarily related to unstable beam conditions during the data taking and inaccurate space-charge calibration of the TPC.

Table~\ref{table_events} gives the total number of minimum-bias events analyzed for each $\sqrt{s_{\mathrm {NN}}}$  and the corresponding chemical freeze-out
temperature ($T_{\mathrm {ch}}$) and baryon chemical potential
($\mu_{\mathrm B}$) values for central 0-5\%
Au+Au collisions. The beam energy values in the BES program
are chosen so that the difference in $\mu_{B}$ values is not larger
than 100~MeV between adjacent collision energies.

\begin{figure*}[htp]
\includegraphics[scale=0.85]{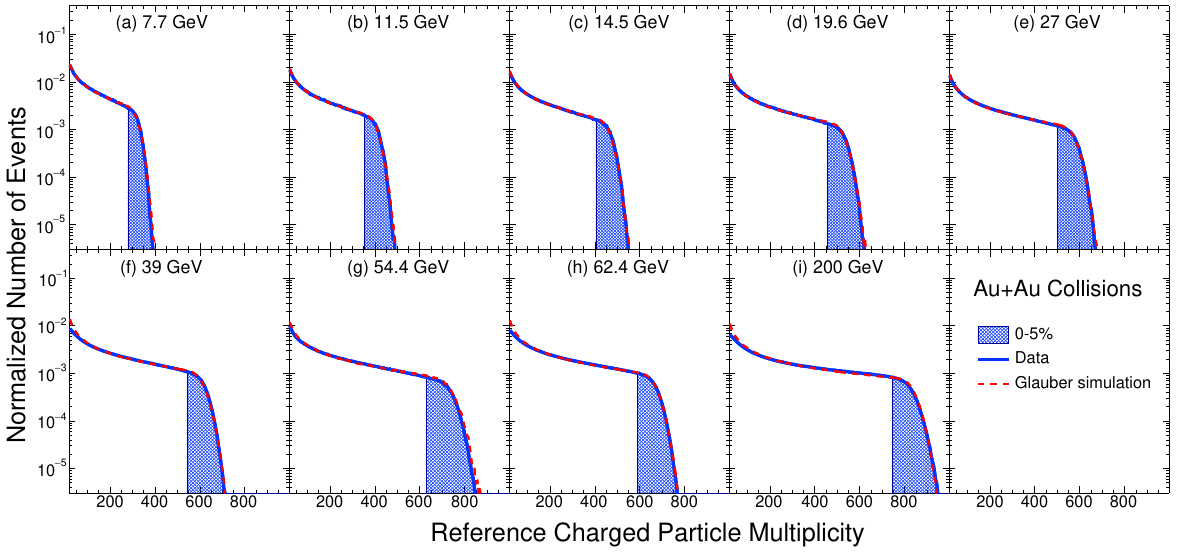}
\caption 
{(Color online)  The uncorrected reference charged particle multiplicity  ($N_{\rm 
    {ch}}$) distributions within pseudorapidity $|\eta| <$ 1 by excluding protons and antiprotons 
  in Au+Au collisions at $\sqrt{s_{\mathrm  {NN}}}$ = 7.7 - 200 
  GeV. These distributions are used for centrality determination. The 
  shaded region at each $\sqrt{s_{\mathrm  {NN}}}$ corresponds to 
  0-5\% central collisions. The dashed line corresponds to Monte Carlo
Glauber model simulations~\cite{Miller:2007ri}.}
\label{Fig:cen}
\end{figure*}
\subsection{Track selection, particle identification and acceptance}
The proton and antiproton track selection criteria for all the $\sqrt{s_{\mathrm    {NN}}}$  are presented in Table~\ref{table_tcuts}.
In order to suppress contamination by tracks from secondary vertices, a requirement of less than 1 cm is placed on
DCA between each track and the
event vertex. Tracks are required to have at least 20 points used in track fitting out of
a maximum of 45 possible hits in the TPC. To prevent multiple counting of split tracks, more than 52\% of the
maximum-possible fit points are required.  A condition is also placed on the number of points ($>$ 5) used to extract the energy loss ($dE/dx$)
values, which is used to identify the (anti)protons from the charged particles
detected in the TPC. The results presented here are within kinematics 
$|y|<$0.5 and 0.4 $< p_{\rm T} <$ 2.0 GeV/$c$.

Particle identification (PID) is carried out using the TPC and TOF by
measuring the $dE/dx$ and time of flight, respectively. 
Figure~\ref{pid} (left top panel) 
shows a typical plot of the square of the mass ($m^{2}$) associated with a track measured in the TPC as a function of rigidity (defined as momentum/z, where z is the dimensionless ratio of particle charge to the electron charge magnitude) for Au+Au collisions at  $\sqrt{s_{\mathrm
    {NN}}}$ = 39 GeV.  The $m^{2}$ is given by:
\begin{equation}
\label{msquare}
m^{2} = p^{2} \left( \frac {c^{2} t^{2}}{L^{2}} -1 \right ),
\end{equation}
 where $p$, $t$, $L$, and $c$ are the momentum, time-of-flight of the
 particle, path length, and speed of light, respectively. 
Protons and antiprotons can be identified by selecting charged tracks for which 
0.6 $<$ $m^{2}$ $<$ 1.2 $\mathrm{GeV}^{2}/c^{4}$. 

Figure~\ref{pid} (left bottom panel) shows the $dE/dx$ of 
measured charged particles plotted as a function of the rigidity. The measured values of $dE/dx$  are compared to the
expected theoretical values ~\cite{Bichsel:2006cs} (shown as solid
lines in Fig.~\ref{pid}) to select the
proton and antiproton tracks. A quantity called $N_{\sigma, p}$ for
charged tracks in the TPC is defined as:
\begin{equation}
\label{eqnsigma}
N_{\sigma, p}= (1/\sigma_{R}) \ln \left( \frac{\langle dE/dx \rangle}{\langle dE/dx \rangle_{p}^{\rm th}} \right),
\end{equation}
where $\langle dE/dx \rangle$ is the truncated mean value of the track energy loss measured in
the TPC, $\langle dE/dx \rangle_{p}^{\rm th}$ is the corresponding
theoretical value for a proton (or antiproton) in the STAR
TPC~\cite{Bichsel:2006cs} and $\sigma_{R}$ is the $dE/dx$ resolution which is momentum-dependent and of the order of 7.5\% for the momentum range of this analysis. Assuming that the $N_{\sigma, p}$ distribution in a given momentum
range is Gaussian, it should peak at zero for proton tracks and the
values represent the deviation from the theoretical values for proton
tracks in terms of standard deviations ($\sigma_{R}$). Momentum-dependent selection criteria are used for TPC tracks to select protons
or antiprotons. For 0.4 $<$
$p_{\rm T}$ $<$ 0.8 GeV/$c$ and momentum ($p$) less than 1 GeV/$c$, $|N_{\sigma, p}| < $ 2.0 is chosen and for 0.8 $<$
$p_{\rm T}$ $<$ 2.0 GeV/$c$ and momentum ($p$) less than 3 GeV/$c$, in
addition to $|N_{\sigma, p}| < $ 2.0, the track is required to have
0.6 $<$ $m^{2}$ $<$ 1.2 $\mathrm{GeV}^{2}/c^{4}$ from TOF.
The purity is estimated by referring to the 
$N_{\sigma, p}$ distributions from the TPC in various $p_{\mathrm T}$
ranges (within 0.4 to 0.8 GeV/$c$) to estimate the contamination from other hadrons within the PID 
selection criteria. For the higher $p_{\mathrm T}$ range, the $m^{2}$
distributions from the TOF are studied after applying the $N_{\sigma, p}$
criteria  and the contamination from other hadrons within the PID 
selection criteria is estimated. The purities of the proton and antiproton samples are better than 97\% for all 
the  $p_{\mathrm T}$ ranges and {\snn} studied.

Figure~\ref{pid} (right panels) shows the $p_{\rm T}$ versus $y$ for protons and antiprotons selected by the TPC with $|N_{\sigma, p}| < $ 2.0 in Au+Au collisions at $\sqrt{s_{\mathrm  {NN}}}$ = 39
GeV. The acceptance is uniform in $y$-$p_{\rm T}$ and is the same for other $\sqrt{s_{\mathrm  {NN}}}$
studied here. This is a major advantage of collider-based
experiments over fixed-target experiments. The boxes show the acceptance criteria used in this
analysis. The addition of the TOF extends the PID capabilities to higher $p_{\rm T}$, thereby allowing for the detection of $\sim$ 80\% of the total protons per unit rapidity (or antiprotons per unit rapidity) produced in the collisions at midrapidity. This is
a significant improvement compared to the previous analysis reported in Ref.~\cite{Adamczyk:2013dal}. The
uniform and large acceptance at midrapidity in $y$, $p_{\rm T}$ and 
$\phi$ allows STAR to measure and compare the cumulants in Au+Au collisions at $\sqrt{s_{\mathrm  {NN}}}$ = 7.7 to 200 GeV.

\subsection{Centrality selection}
Centrality selection plays a crucial role in the fluctuation
analysis. There are two effects related to the centrality selection
which need to be addressed. These are (a) the self-correlation~\cite{Luo:2013bmi,Chatterjee:2019fey} and (b) centrality
resolution/fluctuations effects~\cite{Luo:2013bmi,Chatterjee:2019fey,Zhou:2018fxx,Sugiura:2019toh,Chatterjee:2020nnn}.

One of the main self-correlation effects arises when particles used for the fluctuation analysis
are also used for the centrality definition. This can be significantly reduced by removing the
particles used in the fluctuation analysis from the centrality
definition. Hence, we exclude protons and antiprotons from charged particles for the centrality
selection. 

The centrality resolution effect arises due to the fact that the number of participant nucleons and particle multiplicities fluctuate even if the impact parameter is fixed. Through a model simulation it has been shown that the larger the $\eta$ acceptance used for centrality selection, the closer are the values of the cumulants to the actual
values~\cite{Luo:2013bmi}. This is because the centrality resolution is improved by increasing the number of particles for the centrality definition with wider acceptance.
Therefore, to suppress the effect of centrality resolution, one should use the maximum available acceptance of charged particles for centrality selection. In addition, it may be mentioned that
the choice of centrality definition also affects the way volume fluctuations
(discussed later) contribute to the measurements.

These are the driving considerations for the centrality selection for net-proton studies
presented in this paper and they are discussed below. The basic strategy is to maximize the
acceptance window for the centrality determination as allowed by the
detectors, and to not use protons and antiprotons for the centrality selection. In addition, the
centrality definition method given below is determined after
several optimization studies using data and models. These studies
were carried out by varying the acceptances in $\eta$ and charged particle types in order to understand
the effect of the choice of centrality determination method on the analysis~\cite{Chatterjee:2019fey}. 
The effect of self-correlation potentially arising due to the decay of heavier hadrons into protons and antiprotons
and other charged particles has been verified to be negligible from a study using standard heavy-ion collision event generators, HIJING~\cite{Gyulassy:1994ew} and UrQMD~\cite{Bass:1998ca,Chatterjee:2019fey}.

\begin{table}[]
\caption{The uncorrected number of charged particles other than protons and antiprotons
  ($N_{\rm{ch}}$) within the pseudorapidity $|\eta| <$ 1.0 used for the centrality 
    selection for various collision centralities expressed in \% centrality in Au+Au collisions at  $\sqrt{s_\mathrm{NN}}$ = 7.7 -- 200 GeV.  
\label{table_cen}}
\begin{tabular}{clccccccccc}
\hline
\multicolumn{2}{c}{\multirow{3}{*}{Centrality (\%)}} & \multicolumn{9}{c}{\multirow{2}{*}{{$N_{\rm{ch}}$ values at different {\snn} (GeV)}}}   \\
\multicolumn{2}{c}{}                                  & \multicolumn{9}{c}{}                                     \\ \cline{3-11} 
\multicolumn{2}{c}{}                                  & 200 & 62.4 & 54.4 & 39  & 27  & 19.6 & 14.5 & 11.5 & 7.7 \\ \hline
\multicolumn{2}{c}{0-5}                               & 725 & 571  & 621  & 522 & 490 & 448  & 393  & 343  & 270 \\ \hline
\multicolumn{2}{c}{5-10}                              & 618 & 482  & 516  & 439 & 412 & 376  & 330  & 287  & 225 \\ \hline
\multicolumn{2}{c}{10-20}                             & 440 & 338  & 354  & 308 & 289 & 263  & 231  & 199  & 155 \\ \hline
\multicolumn{2}{c}{20-30}                             & 301 & 230  & 237  & 209 & 196 & 178  & 157  & 134  & 105 \\ \hline
\multicolumn{2}{c}{30-40}                             & 196 & 149  & 151  & 136 & 127 & 116  & 103  & 87   & 68  \\ \hline
\multicolumn{2}{c}{40-50}                             & 120 & 91   & 90   & 83  & 78  & 71   & 63   & 53   & 41  \\ \hline
\multicolumn{2}{c}{50-60}                             & 67  & 51   & 50   & 47  & 44  & 40   & 36   & 30   & 23  \\ \hline
\multicolumn{2}{c}{60-70}                             & 34  & 26   & 24   & 24  & 22  & 20   & 19   & 15   & 11  \\ \hline
\multicolumn{2}{c}{70-80}                             & 16  & 12   & 10   & 11  & 10  & 9    & 13   & 7    & 5   \\ \hline
\end{tabular}
\end{table}

\begin{table*}[]
\caption{The average number of participant nucleons ($\langle N_{\rm
    {part}} \rangle$) for various collision centralities in Au+Au collisions 
    at  $\sqrt{s_\mathrm{NN}}$ = 7.7 -- 200 GeV from a Monte Carlo Glauber
    model. The numbers in parentheses are systematic uncertainties.
\label{table_npart}}
\begin{tabular}{clccccccccc}
\hline
\multicolumn{2}{c}{\multirow{3}{*}{Centrality (\%)}} & \multicolumn{9}{c}{\multirow{2}{*}{$\langle N_\mathrm{part} \rangle$ values at different {\snn} (GeV)} }   \\
\multicolumn{2}{c}{}                                  & \multicolumn{9}{c}{}                                     \\ \cline{3-11} 
\multicolumn{2}{c}{}                                  & 200         & 62.4         & 54.4       & 39        & 27       & 19.6 &       14.5 & 11.5 &      7.7      \\ \hline
\multicolumn{2}{c}{0-5}                               & 351 (2)   & 347 (3)    &346 (2)   & 342(2)    & 343 (2)  & 338 (2)    & 340(2)  & 338 (2)   & 337 (2)  \\ \hline
\multicolumn{2}{c}{5-10}                              & 299 (4)   & 294 (4)    &292 (6)   & 294 (6)   & 299 (6)  & 289 (6)    & 289 (6) & 291 (6)   & 290 (6)  \\ \hline
\multicolumn{2}{c}{10-20}                             & 234 (5)   & 230 (5)    &228 (8)   & 230 (9)   & 234 (9)  & 225 (9)    & 225 (8) & 226 (8)   & 226 (8)  \\ \hline
\multicolumn{2}{c}{20-30}                             & 168 (5)   & 164 (5)    &161 (10)  & 162 (10)  & 166 (11) & 158 (10)   & 159 (9)& 160 (9)   & 160 (10) \\ \hline
\multicolumn{2}{c}{30-40}                             & 117 (5)   & 114 (5)    &111 (11)  & 111 (11)  & 114 (11) & 108 (11)   & 109 (11) & 110 (11)  & 110 (11) \\ \hline
\multicolumn{2}{c}{40-50}                             & 78 (5)    & 76 (5)     &73 (10)   & 74 (10)   & 75 (10)  & 71 (10)    & 72 (10)  & 73 (10)   & 72 (10)  \\ \hline
\multicolumn{2}{c}{50-60}                             & 49 (5)    & 48 (5)     &45 (9)    & 46 (9)    & 47 (9)   & 44 (9)     & 45 (9) & 45 (9)    & 45 (9)   \\ \hline
\multicolumn{2}{c}{60-70}                             & 29 (4)    & 28 (4)     &26 (7)    & 26 (7)    & 27 (8)   & 26 (7)     & 26 (7) & 26 (7)    & 26 (7)   \\ \hline
\multicolumn{2}{c}{70-80}                             & 16 (3)    & 15 (2)     &13 (5)    & 14 (5)    & 14 (6)   & 14 (5)     & 14 (6) & 14 (6)    & 14 (4)   \\ \hline 
\end{tabular}
\end{table*}

\begin{figure*}[htp]
\includegraphics[scale=0.85]{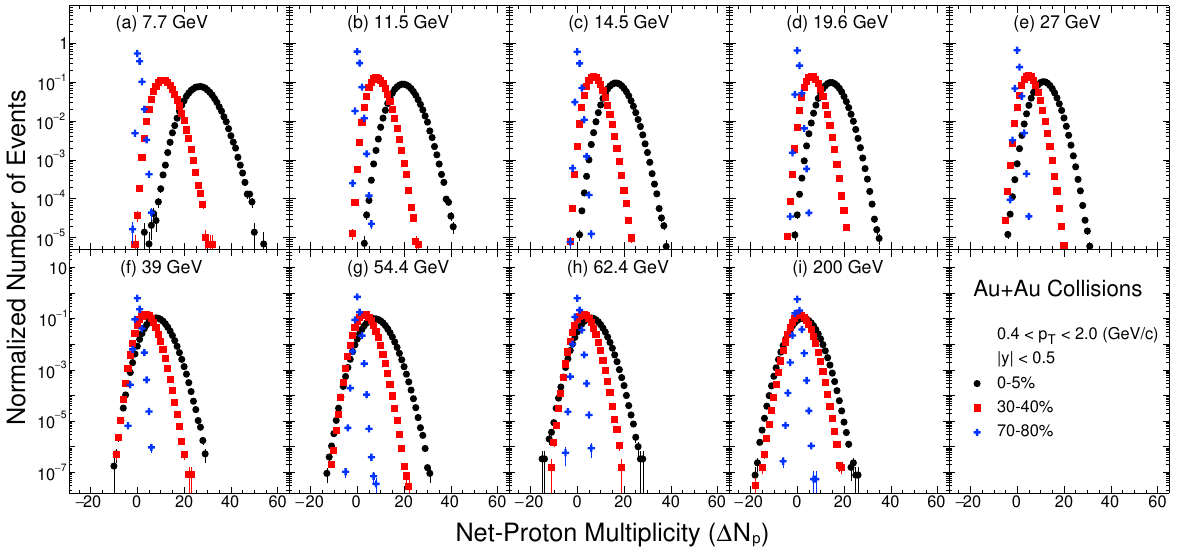}
\caption 
{(Color online) Net-proton multiplicity ($\Delta N_{p}$) distributions in Au+Au collisions at 
various $\sqrt{s_{\mathrm {NN}}}$ for 0-5\%, 30-40\% and 70-80\% collision centralities 
at midrapidity. The statistical errors are small and within the symbol size. 
The distributions are not corrected for either the finite-centrality-width
effect or for the reconstruction efficiencies of protons and antiprotons. 
}
\label{Fig:netp-dist}
\end{figure*}
In order to suppress the self-correlation, centrality
resolution and volume fluctuation effects with the available STAR detectors, a new
centrality measure is defined, and is different from other analyses reported by
STAR~\cite{Adamczyk:2017iwn}. 
The centrality is determined from the uncorrected charged particle
multiplicity within pseudorapidity $|\eta| <$ 1 ($N_\mathrm{ch}$) after excluding the protons and antiprotons. Strict particle identification criteria
are used to remove the proton and antiproton contributions. Charged
tracks with $N_{\sigma, p} < - 3$ are used and for those tracks
which have TOF information an additional criterion, $m^{2} <$
0.4~GeV$^{2}/c^{4}$, is applied. The resultant distribution of charged particles is corrected for luminosity and $V_{z}$
dependence at each $\sqrt{s_{\mathrm  {NN}}}$. The corrected charged particle distribution is then fit to a Monte Carlo Glauber Model~\cite{Abelev:2009bw,Miller:2007ri} to define the centrality
classes in the experiment (the percentage cross section and the associated cuts on the charged-particle multiplicity). In the fitting process,  a multiplicity-dependent efficiency has been applied~\cite{Abelev:2009bw}. 

Figure~\ref{Fig:cen} shows the reference charged particle multiplicity distributions after excluding protons and antiprotons used for centrality determination
for all of the {\snn} studied here. The lower boundaries of each centrality class based on $N_\mathrm{ch}$ are given in Table~\ref{table_cen}. Table~\ref{table_npart} gives the average number of participant
nucleons ($\langle N_{\rm {part}} \rangle$) for various collision
centralities for $\sqrt{s_{\mathrm  {NN}}}$ = 7.7 - 200 GeV obtained
from a Monte Carlo Glauber model simulation.

\subsection{Uncorrected net-proton multiplicity distributions}
Figure~\ref{Fig:netp-dist} shows the event-by-event net-proton
multiplicity ($\Delta N_{p}$) distributions from Au+Au collisions at
$\sqrt{s_{\mathrm {NN}}}$ = 7.7 -- 200 GeV for 0-5\%, 30-40\% and
70-80\% collision centralities. The $\Delta N_{p}$ distribution is
obtained by counting the number of protons and
antiprotons within the $y$-$p_{\rm T}$ acceptance on an event-by-event basis for a given
collision centrality and $\sqrt{s_{\mathrm {NN}}}$. The distributions
presented in Fig.~\ref{Fig:netp-dist} are not corrected
for the efficiency and acceptance effects. In general, the shape of the $\Delta N_{p}$ distributions is broader, more symmetric and closer to Gaussian, for central collisions than that for
peripheral collisions. The shape of the distributions also changes
with $\sqrt{s_{\mathrm {NN}}}$. Cumulants ($C_{n}$) up to the fourth order
are obtained from these distributions for each collision centrality
and $\sqrt{s_{\mathrm {NN}}}$.

\subsection{Definition of cumulants and integrated correlation functions} \label{sub:def}
In this subsection, we give the definition of the cumulants used in this paper. Let $N$ represent any entry in
the data sample, its deviation from its mean value ($\langle N \rangle$, referred to as the
first moment) is then given by $\delta N=N-\langle N \rangle$.
Any $r$th-order central moment is defined as:
\begin{equation}
\mu_r = \,\langle  (\delta N)^r \rangle. 
\end{equation}
The cumulants of a given data sample could be written in terms of moments as follows:
\begin{eqnarray}
 C_1 &=&\langle N\rangle, \nonumber \\ 
 C_2 &=&\langle (\delta N)^2 \rangle = \mu_2,  \nonumber \\ 
  C_3 &=& \langle (\delta N)^3 \rangle = \mu_3,\\
  C_4 &=& \langle  (\delta N)^4 \rangle - 3\langle  (\delta N)^2 \rangle^2 \nonumber \\  
    &=&\mu_4-3\mu_2^{2},  \nonumber \\ 
 C_n(n>3)  &=& \mu_n  - \sum\limits_{m = 2}^{n - 2}
{\left(
\begin{array}{l}
 n - 1 \\
 m - 1 \\
 \end{array} \right)C_m } \mu_{n - m}.  \nonumber
\end{eqnarray}
The relations between cumulants and various moments are given as:
\begin{eqnarray}
M=C_{1},~~\sigma^{2}=C_{2},~~S=\frac{C_{3}}{(C_{2})^{3/2}}, ~~\kappa=\frac{C_{4}}{(C_{2})^{2}}.
\end{eqnarray}
where $M$, $\sigma^{2}$, $S$ and $\kappa$ are mean, variance, skewness and kurtosis, respectively. 
The products $\kappa \sigma^{2}$ and $ S \sigma$
can be expressed in terms of the ratio of cumulants as:
\begin{eqnarray}
\sigma^{2}/M=\frac{C_{2}}{C_{1}},~~ 
S\sigma=\frac{C_{3}}{C_{2}},~~ 
\kappa \sigma^{2}=\frac{C_{4}}{C_{2}}.
\end{eqnarray}
With the above definition, we can calculate various order cumulants (moments)
and cumulant ratios (moment products) from the measured event-by-event
net-proton, proton and antiproton distributions for each centrality at a given $\sqrt{s_{\mathrm {NN}}}$.
For two independent variables $X$ and $Y$, the cumulants of the
probability distributions of their sum ($X+Y$), are just the addition of
cumulants of the individual distributions for $X$ and $Y$ $i.e.$
$C_{n,X+Y}=C_{n,X}+C_{n,Y}$ for the $n$th-order cumulant. For
a distribution of difference between $X$ and $Y$, the cumulants are 
$C_{n,X-Y}=C_{n,X}+(-1)^{n}C_{n,Y}$, where the even-order cumulants
are the addition of the individual cumulants, while the odd-order
cumulants are obtained by taking their difference. If the protons and antiprotons are distributed as independent Poissonian distributions,  the various order cumulants of net-proton, proton and antiproton 
distributions can be expressed as: 
\begin{eqnarray} \nonumber
C_{n,p}&=&C_{1,p},~C_{n,\bar{p}}=C_{1,\bar{p}}, \label{eq:Poi1}  \nonumber  \\  
C_{n,p-\bar{p}}&=&C_{1,p}+(-1)^{n}C_{1,\bar{p}} \nonumber
\end{eqnarray} 
where the net-proton multiplicity distributions obey the Skellam distribution and the Poisson baseline/expectation values of the net-proton, proton and antiproton cumulant ratios are: 
\begin{eqnarray} \nonumber 
&&(\sigma^{2}/M)_{p,\bar{p}}=(S\sigma)_{p,\bar{p}}=(\kappa \sigma^{2})_{p,\bar{p}}=1, \\ \nonumber 
&&(\sigma^{2}/M)_{p-\bar{p}}=\frac{1}{(S\sigma)_{p-\bar{p}}}=\frac{C_{1,p}+C_{1,\bar{p}}}{C_{1,p}-C_{1,\bar{p}}},\\ \nonumber 
&&(\kappa \sigma^{2})_{p-\bar{p}}=1 \nonumber 
\end{eqnarray}
where $C_{1,p}$ and $C_{1,\bar{p}}$ are the mean values of proton and antiproton, respectively. 

On the other hand, it is expected that close to the CP, the three- and four-particle
correlations are dominant relative to two-particle
correlations~\cite{Stephanov:2008qz}. The various orders 
integrated correlation functions of proton and antiproton ($\kappa_{n}$, also known as factorial cumulants) are related to the corresponding
proton and antiproton cumulants ($C_{n}$) through the following
relations~\cite{Ling:2015yau,Bzdak:2016sxg,Kitazawa:2017ljq}:
\begin{equation}  \label{eq:CKtoCF}
\begin{split}
\kappa_{1} &=C_{1}=\langle N \rangle,  \\ 
\kappa_{2} &=-C_{1}+C_{2},   \\ 
\kappa_{3} &=2C_{1}-3C_{2}+C_{3},  \\ 
\kappa_{4} &=-6C_{1}+11C_{2}-6C_{3}+C_{4},   \\ 
C_{2}&=\kappa_{2}+\kappa_{1} ,  \\ 
C_{3} &=\kappa_{3}+3\kappa_{2}+\kappa_{1} ,  \\
C_{4} &=\kappa_{4}+ 6\kappa_{3}+7\kappa_{2}+\kappa_{1},
\end{split}
\end{equation}
where $C_1$ and $\kappa_1$ represent the mean values for protons or antiprotons. 
For proton and antiproton cumulant ratios $C_2/C_1$, $C_3/C_2$ and $C_4/C_2$, they can be expressed in terms of corresponding normalized correlation functions $\kappa_n/\kappa_1$ ($n>1$) as: 
\begin{eqnarray}
\frac{C_2}{C_1}&=&\frac{\kappa_{2}}{\kappa_{1}}+1, \label{eq:C2toC1} \\
\frac{C_3}{C_2}&=&\frac{\kappa_{3}/\kappa_1-2}{\kappa_{2}/\kappa_1+1}+3, \label{eq:C3toC2} \\
\frac{C_4}{C_2}&=&\frac{\kappa_{4}/\kappa_1+6\kappa_{3}/\kappa_1-6}{\kappa_{2}/\kappa_1+1}+7, \label{eq:C4toC2}
\end{eqnarray}
The higher-order integrated correlation functions $\kappa_n$ ($n>1$) are equal to zero when the distributions are Poisson. Thus, $\kappa_n$ can be used to quantify the deviations from the Poisson distributions in terms of $n$-particle correlations.  For simplicity, from here on, we refer to the $\kappa_n$ as correlation functions instead of integrated correlation functions.  

In the following subsections,  we discuss corrections that are related to collision centrality bin width (Sec. II F) and detection efficiency (Sec. II G). This is followed by the estimation of statistical and systematic uncertainties in sections II H and II I, respectively.  
\subsection{Centrality bin width correction}
Data presented in this paper are classified into the following centrality bins: 0-5\%, 5-10\%, 10-20\%, 20-30\%, 30-40\%, 40-50\%, 50-60\%, 60-70\% and 70-80\%. The finite size of centrality bins implies that
the average number of protons and antiprotons varies even within a
centrality class. This variation has to be accounted for while
calculating the cumulants in a broad centrality class. In addition, it is known that calculating cumulants in such broad centrality bins leads to a strong enhancement of cumulants and cumulant ratios due to initial volume fluctuations~\cite{Luo:2013bmi,He:2018mri}.

\begin{figure*}[htp]
\includegraphics[scale=0.75]{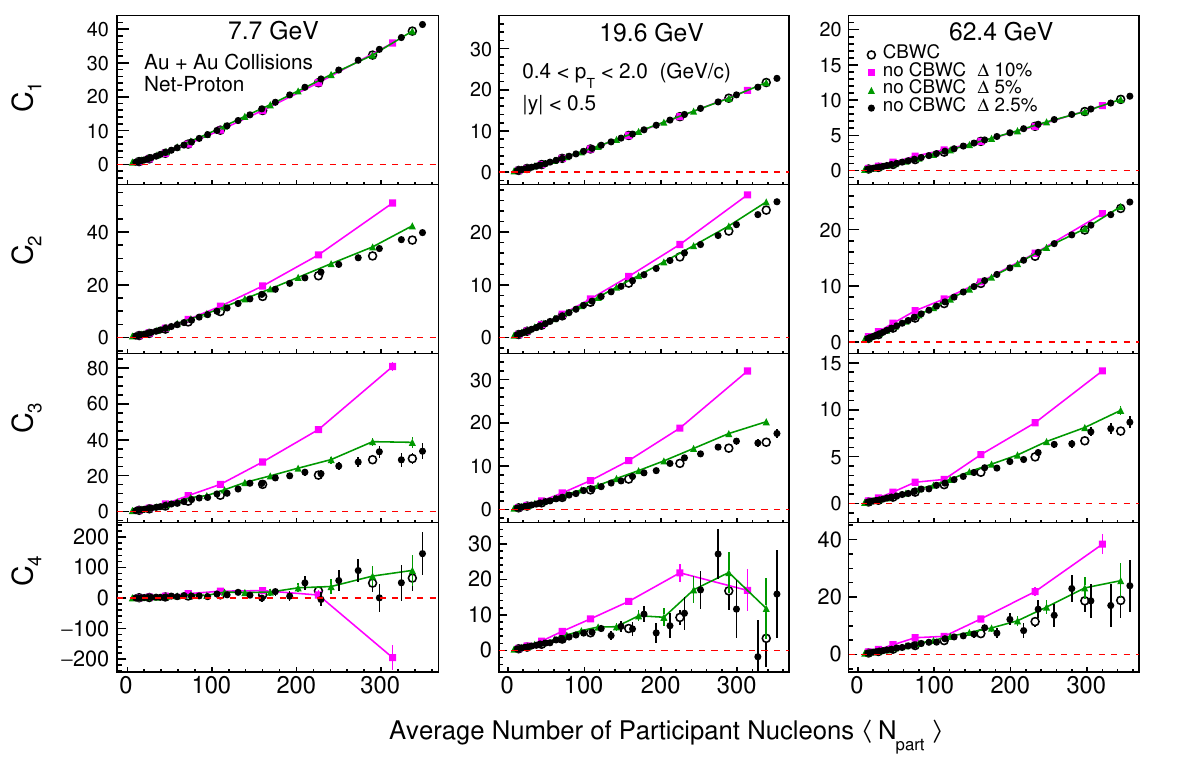}
\caption 
{(Color online) $C_{n}$ of net-proton distributions in Au+Au 
  collisions at $\sqrt{s_{\mathrm {NN}}}$ = 7.7, 19.6 and 62.4 GeV as 
  a function of $\langle N_{\mathrm {part}} \rangle$. The results are 
  shown for 10\%, 5\% and 2.5\% centrality bins without CBWC and for 
  nine centrality bins (0-5\%, 5-10\%, 10-20\%,..., 70-80\%) with CBWC. The bars are the statistical uncertainties.}
\label{Fig:cbwc}
\end{figure*}

A centrality bin width correction (CBWC) is the procedure used to 
take care of the measurements in a wide centrality bin and is based on 
weighting the cumulants measured at each multiplicity bin by the number of events in the bin~\cite{Luo:2013bmi,Chatterjee:2019fey,He:2018mri}.
This procedure is mathematically expressed in the equation below: 
\begin{eqnarray}
C_{n}  &=& \frac{{\sum\limits_r^{} {n_r C_{n} ^{r}  }}}{{\sum\limits_r^{} {n_r } }} = \sum\limits_r^{} {\omega _r C_{n}^{r}  },
\end{eqnarray}
where the $n_r$ is the number of events at the $r$th multiplicity bin for
the centrality determination, the $C_{n}^{r}$ represents the $n$th-order
cumulant of particle number distributions at $r$th multiplicity. The
corresponding weight for the $r$th multiplicity bin is $\omega_r={{ {n_r  } }}/{{\sum\limits_r^{} {n_r
} }}$.

\begin{figure}[htp]
\includegraphics[scale=0.5]{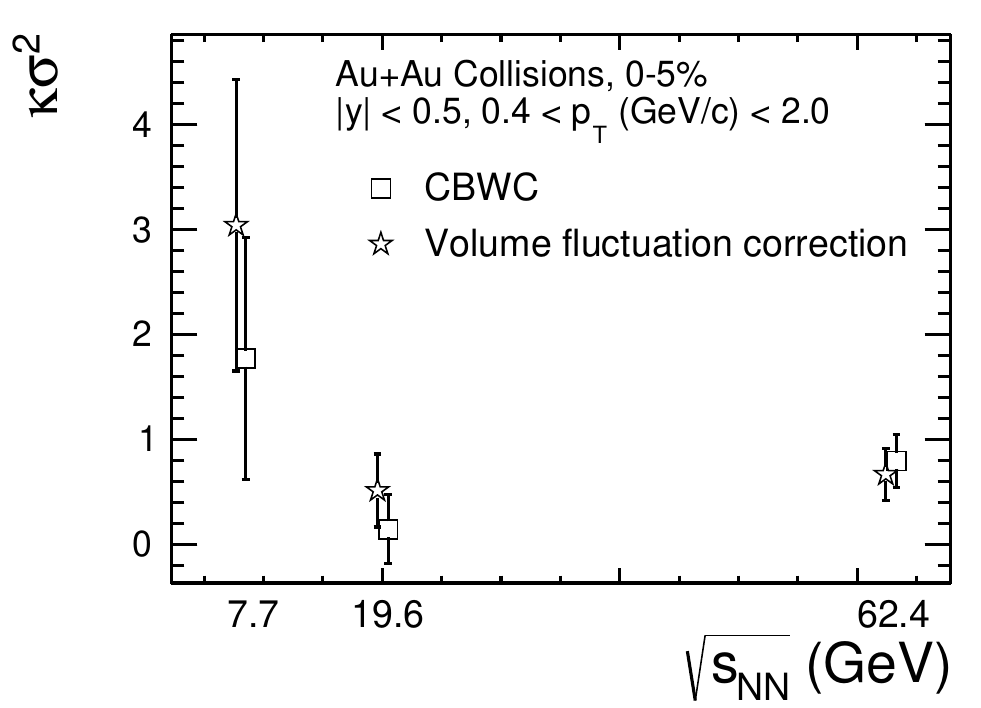}
\caption 
{(Color online) {\KV} as a function of collision energy for
Au+Au collisions for 0-5\% centrality. The data have
been corrected for volume fluctuation effects using CBWC, a
data driven approach, and a model-dependent volume fluctuation correction
method. The bars are the statistical uncertainties.}
\label{Fig:VFC}
\end{figure}
\begin{figure*}[htp]
\includegraphics[scale=0.5]{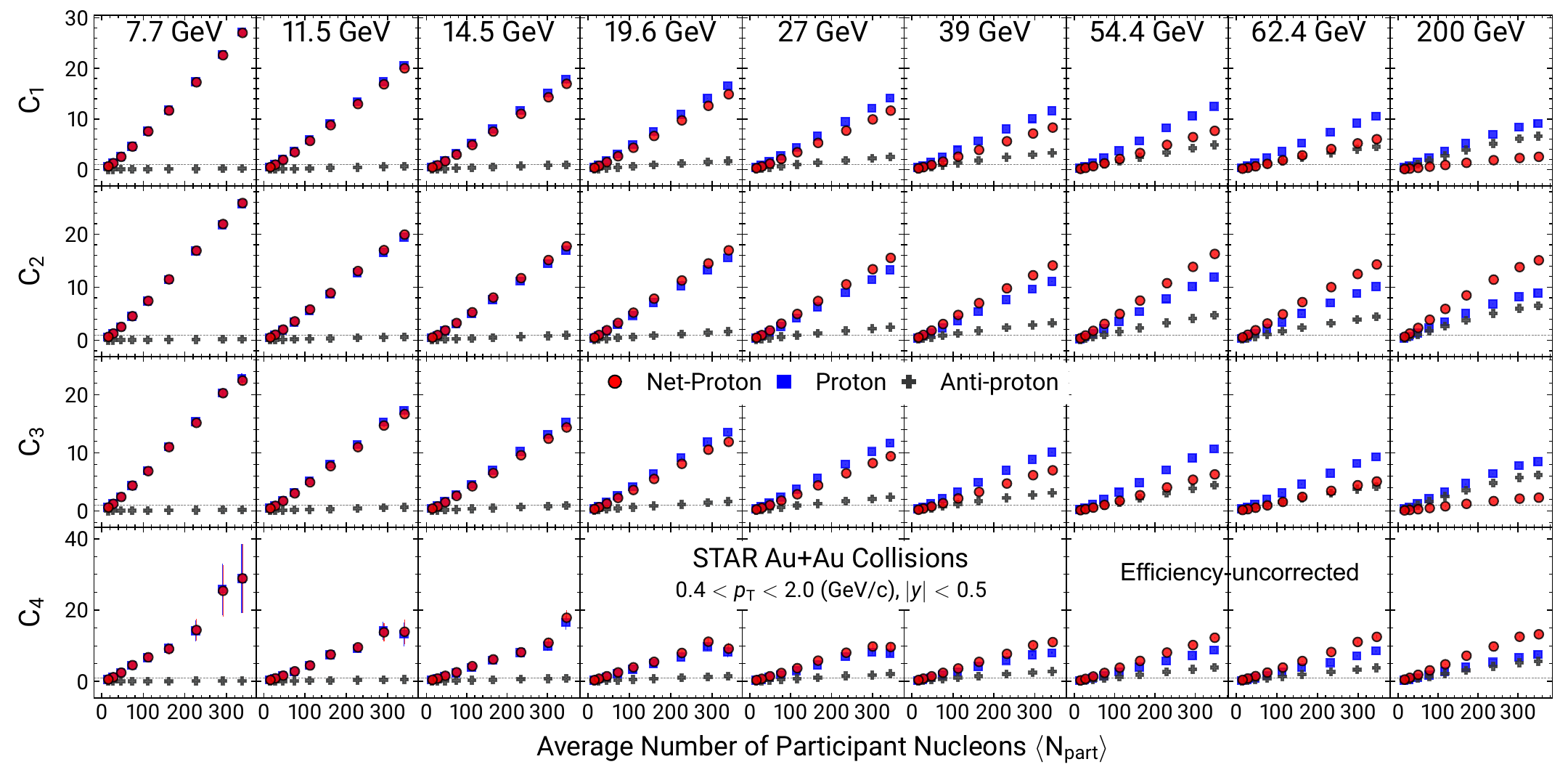}
\caption 
{(Color online) Efficiency-uncorrected $C_{n}$ of net-proton, proton, and antiproton multiplicity distributions in Au+Au 
  collisions at $\sqrt{s_{\mathrm {NN}}}$ = 7.7-- 200 GeV as 
  a function of $\langle N_{\mathrm {part}} \rangle$. The results are
  CBW-corrected. The bars are the statistical uncertainties.}
\label{Fig:uncorrC}
\end{figure*}

As an example, Fig.~\ref{Fig:cbwc} shows the $C_{n}$ up to the fourth order as a function
of $\langle N_{\rm {part}} \rangle$ for three different collision energies:
$\sqrt{s_{\mathrm {NN}}}$ = 7.7, 19.6 and 62.4 GeV. For each $C_{n}$
case, four different results are shown. One of them is the CBWC result for nine collision centrality bins, which correspond to 0-5\%, 5-10\%, 10-20\%, 20-30\%,...,70-80\%. 
For comparison,  cumulants are also calculated for the other three cases, which are 10\%, 5\% and 2.5\% centrality bin width without CBWC. 
The higher-order cumulant results with 10\% centrality bins are found to have significant deviations compared to those with 5\% and 2.5\% centrality bins without CBWC. 
This finding means that it is important to correct for the CBW effect, as one normally
expects that, irrespective of the centrality bin width, the cumulant
values should exhibit the same dependence on $\langle N_{\rm {part}} \rangle$. It is found that the results get closer to CBWC results with narrower centrality bins and the results with 2.5\% centrality bins almost overlap with CBWC results, which indicates that the CBWC can effectively suppress the effect of the volume fluctuations on cumulants (up to the fourth order) within a finite centrality bin width.

For comparison, a different approach, the volume fluctuation correction (VFC) method~\cite{Skokov:2012ds,Braun-Munzinger:2016yjz}, which assumes independent production of protons, has been also applied at {\sNN} = 7.7, 19.6 and 62.4 GeV for 0-5\% Au+Au central collisions. The correction factors are determined by the Glauber model~\cite{Braun-Munzinger:2016yjz}. Figure~\ref{Fig:VFC} shows the comparison between the results based on CBWC and VFC methods. As can be seen from the plot, for the 0-5\% central collisions, the results of CBWC and VFC are found to be consistent within statistical uncertainties. However, UrQMD model studies reported in Ref.~\cite{Sugiura:2019toh}, indicate that the VFC method (as discussed in Ref.~\cite{Skokov:2012ds}) does not work, as the independent particle production model assumed in the VFC is expected to be broken. Therefore, we follow the data-driven method, CBWC, in this paper. 

\subsection{Efficiency correction}
Figure~\ref{Fig:uncorrC} shows the efficiency-uncorrected $C_{n}$ for
proton, antiproton and net-proton multiplicity distributions in Au+Au collisions at
{\snn} = 7.7 -- 200 GeV as a function of $\langle N_{\rm {part}}
\rangle$.  This section discusses the method of
efficiency correction.  One such method is called the binomial-model-based method~\cite{Kitazawa:2017ljq,Bzdak:2012ab,Luo:2014rea,Nonaka:2017kko,Luo:2018ofd}
and another is the unfolding method~\cite{Garg:2012nf,Esumi:2020xdo}. The cumulants presented in the subsequent sections are corrected for
efficiency and acceptance effects related to proton and antiproton reconstruction, unless specified otherwise. 

\begin{figure*}[htp]
\includegraphics[scale=0.65]{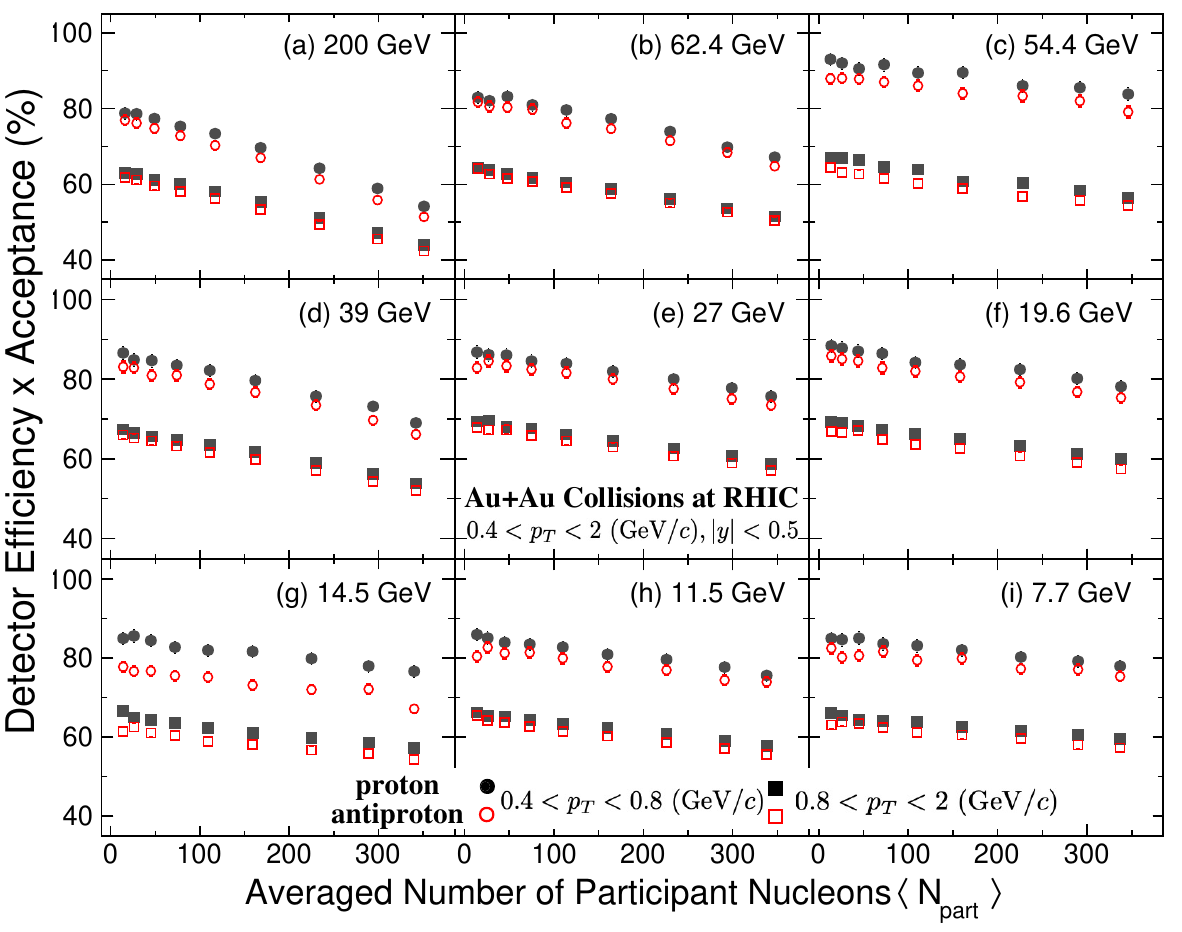}
\caption 
{(Color online)  Efficiencies of proton and antiproton 
  as a function of $\langle N_{\mathrm {part}} \rangle$ in Au+Au collisions for various 
  $\sqrt{s_{\mathrm {NN}}}$. For the lower 
$\mathrm p_{T}$ range ($0.4 < p_{\rm T} <  0.8$ GeV/$c$), only 
the TPC is used. For the higher $\mathrm p_{T}$ range ($0.8 < p_{\rm T} < $ 2.0 GeV/$c$),  both the TPC and TOF are used. 
}
\label{Fig:eff}
\end{figure*}

\subsubsection{Binomial model method}
The binomial-based method involves two steps.
First we obtain the efficiency of proton and antiproton reconstruction in the STAR detector and then correct the cumulants for efficiency and
acceptance effects using analytic expressions. The former uses the embedding process and the
latter invokes binomial model assumptions for the detector response function for the efficiencies. One can find more details in Appendix~\ref{appendix-1}. 

The detector acceptance and the efficiency of reconstructing proton and antiproton tracks are determined together by embedding Monte Carlo (MC)
tracks, simulated using the GEANT~\cite{Fine:2000qx} model of the STAR detector response, 
into real events at the raw data level. One important requirement is the matching of the distributions of reconstructed embedded tracks and real data tracks 
for quantities reflecting track quality and those used for track selection~\cite{Adamczyk:2017iwn}.
The ratio of the distribution of reconstructed to embedded Monte
Carlo tracks as a function of $p_{T}$ gives the efficiency $\times$ acceptance correction 
factor ($\varepsilon_\mathrm{TPC}(p_{T})$) for the rapidity interval studied. 
We refer to this factor as simply efficiency. 

The current analysis makes use of both the TPC and the TOF detectors. While the TPC identifies
low $p_T$ ($0.4< p_{T} <0.8$ GeV/$c$) protons and antiprotons with high purity, the TOF gives better particle identification
than the TPC in the higher $p_{T}$ range ($0.8 < p_{T} < 2.0$ GeV/$c$). However, not all TPC tracks 
have valid TOF information due to the limited TOF acceptance and the mismatching of the TPC tracks to TOF hits. This extra efficiency is called the TOF-matching efficiency ($\varepsilon _\mathrm{TOF}(p_T)$). 
The TOF-matching efficiency is particle-species-dependent and can be obtained using a data-driven technique, which is defined as the ratio of the number of (anti)proton
tracks detected in the TOF to the total number of (anti)proton tracks in the TPC
within the same acceptance~\cite{Adamczyk:2017iwn}. Thus, the final average (anti)proton efficiency within a certain $p_{T}$ range can be calculated as:
\begin{equation} \label{eq:eff}
\langle \varepsilon  \rangle = \frac{{\int\limits_{{p_{{T_1}}}}^{{p_{{T_2}}}} {\varepsilon ({p_T})f({p_T})d{p_T}} }}{{\int\limits_{{p_{{T_1}}}}^{{p_{{T_2}}}} {f({p_T})d{p_T}} }},
\end{equation}
where the $p_{T}$-dependent efficiency, $\varepsilon ({p_T})$, is defined as $\varepsilon ({p_T}) = {\varepsilon _\mathrm{TPC}}({p_T})$ for $0.4< p_{T} <0.8$ GeV/$c$ and $\varepsilon ({p_T}) = {\varepsilon _\mathrm{TPC}}({p_T})\times{\varepsilon _\mathrm{TOF}}({p_T})$ for $0.8 < p_{T} <2.0$ GeV/$c$. The function $f({p_T})$ is the efficiency-corrected $p_T$ spectrum for (anti)protons~\cite{Adamczyk:2017iwn}.

Figure~\ref{Fig:eff} shows the average efficiency ($\langle \varepsilon  \rangle$) for
protons and antiprotons at midrapidity ($|y| < $ 0.5) as a function of
collision centrality ($\langle N_{\mathrm {part}} \rangle$). For $0.4 < p_{\rm T} < $ 0.8 GeV/$c$ the efficiency is only from the TPC and for $0.8 < p_{\rm T}  < $ 2.0 GeV/$c$ it is the product of efficiencies from the TPC and TOF. In Fig.~\ref{Fig:eff},   only statistical uncertainties are presented and a $\pm$ 5\% systematic uncertainty associated with determining the efficiency is considered in the analysis.

\begin{figure*}[htp]
\includegraphics[scale=0.8]{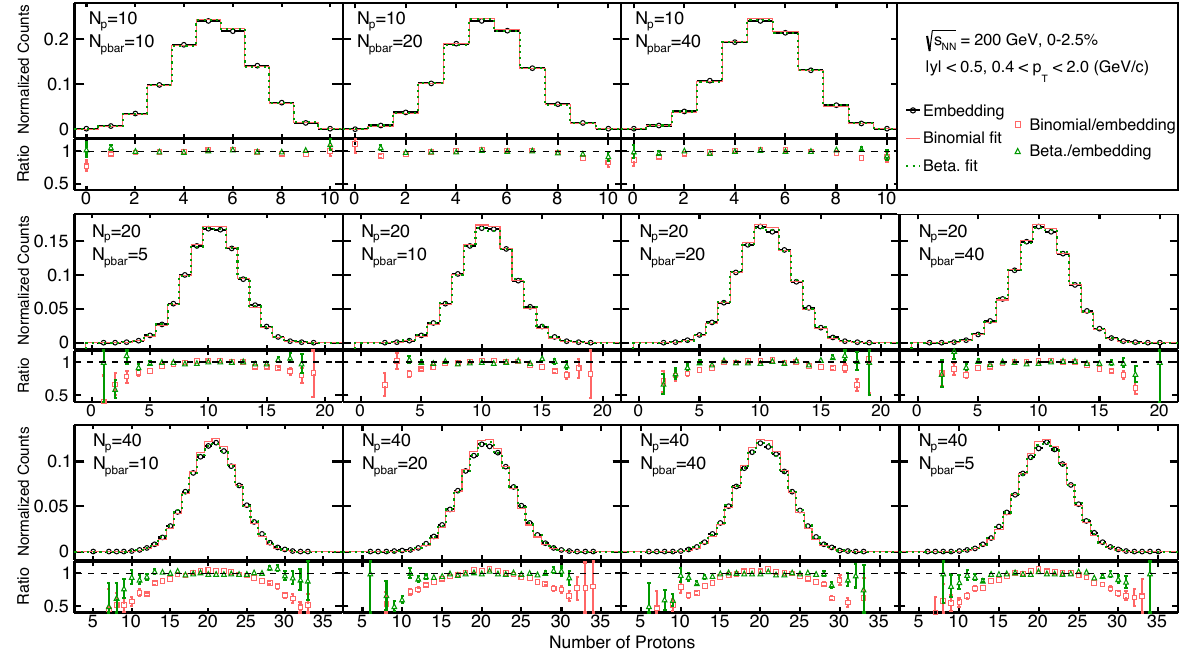}
\caption 
{(Color online)  Distributions of reconstructed protons (black
  circles) from embedding simulations in 200 GeV top 2.5\%-central Au+Au
collisions. Red lines are fits to the binomial distribution, 
		and green dotted lines represent the fit with the beta-binomial distributions 
		using the $\alpha$ that gives the minimum $\chi^{2}/{\rm ndf}$. 
		Each panel presents results for a different combination of the number of embedded protons and antiprotons as labeled in the legend. 
		The ratio of the fits to the embedding data is shown for each panel at the bottom.  
}
\label{fig:EmbeddingFitFinal}
\end{figure*}
\subsubsection{Unfolding method}
In this section we discuss the effect of efficiency correction on the
$C_{n}$ measurement if the assumption of binomial detector efficiency response breaks down due to some of the reasons given
in Refs.~\cite{Bzdak:2016qdc,Nonaka:2018mgw}. The technique is based
on unfolding of the detector response~\cite{Garg:2012nf,Esumi:2020xdo}. The response function is obtained by MC simulations
carried out in the STAR detector environment~\cite{Fine:2000qx}. MC tracks are
simulated through GEANT and embedded in the real data, and track
reconstruction is performed as is done in the real
experiment. Many effects can lead to non-binomial detector response in heavy-ion experiments. 
One of those effects could be track merging due to the extreme environment of high particle multiplicity densities in the detector. Hence, we have
performed the embedding simulations using the real data for 0-5\% Au+Au collisions
at $\sqrt{s_{\rm NN}}=$~200~GeV. The numbers of embedded tracks of $N_{\rm p}$ and
$N_{\rm \bar{p}}$ are varied within $5\leq N_{\rm p(\bar{p})}\leq 40$. Since we are measuring the net-proton multiplicity distributions, 
protons and antiprotons are embedded simultaneously. We have shown in Ref.~\cite{Adamczyk:2017wsl} that, for the event statistics in the current analysis, the efficiencies for kaon reconstruction follow binomial
distributions.

\begin{figure}[htp]
\includegraphics[scale=0.45]{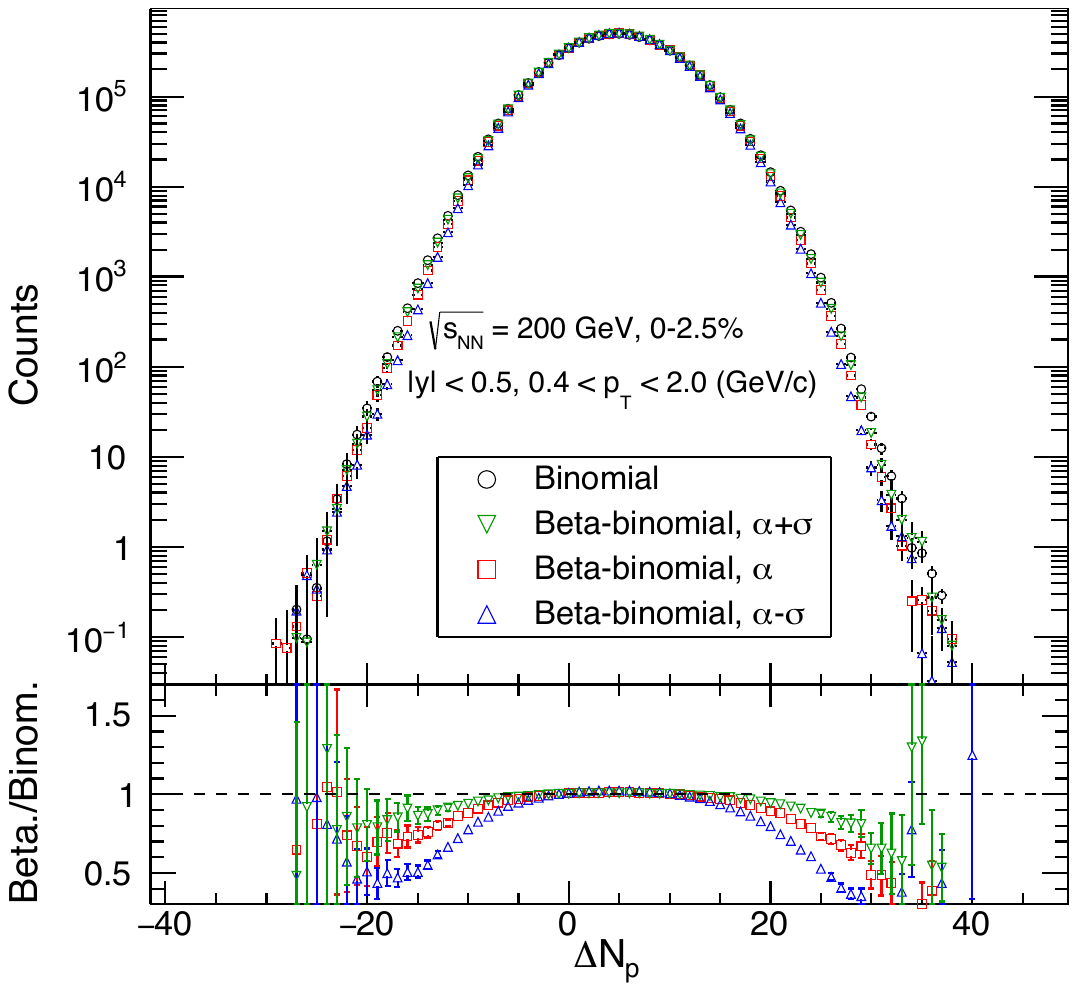}
\caption 
{(Color online)  Unfolded net-proton multiplicity distributions for $\sqrt{s_{\rm NN}}=$~200~GeV Au+Au collisions 
		 where the binomial distribution (black circle), 
		beta-binomial distributions with $\alpha+\sigma$ (green triangle), $\alpha$ (red square), 
		and $\alpha-\sigma$ (blue triangle) are utilized in response matrices. 
		Ratios of the beta-binomial unfolded distributions to that from binomial response matrices are 
		shown in the bottom panel. 
}
\label{fig:UnfHistoFinal}
\end{figure}

\begin{table*}[htbp] 
	\begin{center}
	\caption{ Net-proton cumulant ratios and their statistical errors for 0-5\% central Au+Au collisions at \sNN\ = 200 GeV,  (second column) from the conventional 
	efficiency correction with the binomial detector response, and 
	(third column) from unfolding with the beta-binomial detector response. 
	Systematic errors are also shown for the beta-binomial case. 
	The last column shows the difference between two results normalized by total uncertainty, which is equal to the statistical and systematic uncertainties summed in quadrature.}
	\label{table-un}
	\begin{tabular}{cccc}\hline 
		Cumulant ratio & Binomial $\pm$ statistical error &  Beta $\pm$ statistical error $\pm$ systematical error  & Significance \\ \hline 
		$C_{2}/C_{1}$  &  $1.3\pm \mathrm{neg.}$  &  $1.20\pm \mathrm{neg.}\pm0.03$ & $3.1$  \\ \hline 
		$C_{3}/C_{2}$  &  $0.13\pm0.01$  &  $0.13\pm0.01\pm\mathrm{neg.}$ & $4.8\times10^{-2}$  \\ \hline 
		$C_{4}/C_{2}$  &  $1.10\pm0.21$          &  $0.97\pm0.21\pm0.08$        & $4.2\times10^{-1}$  \\ \hline 
		$C_{5}/C_{1}$  &  $0.10\pm0.48$         &  $-0.14\pm0.44\pm0.11$              & $3.8\times10^{-1}$  \\ \hline 
		$C_{6}/C_{2}$  &  $-0.45\pm0.24$        &  $-0.14\pm0.20\pm0.07$       & $1.0$ \\ \hline 
	\end{tabular}
	\end{center}
\end{table*}

Figure~\ref{fig:EmbeddingFitFinal} shows the reconstructed protons
from the embedding data (black circles) of Au+Au collisions at \snn =
200 GeV and 0-2.5\% collision centrality.
Each panel represents a different number of embedded (anti)protons. These distributions are fitted by a binomial 
distribution (red solid line) at a fixed efficiency $\varepsilon$. 
The ratios of the fitted function to the embedding data are shown in the lower panels. 
The fitted $\chi^2/$ndf ranges from 5.2 to 17.8 and 
the tails of the distributions are not well described
by the binomial distribution for several combinations of embedded $N_{\rm p}$ and
$N_{\rm \bar{p}}$ tracks. We find that the embedding data is better
described by a beta-binomial distribution given by:
\begin{equation}
	\beta(n:N,a,b) = \int_{0}^{1}dpB(\varepsilon,a,b){\rm B}(n;N,\varepsilon),
\end{equation}
and with the beta distribution given as:
\begin{equation}
	\beta(\varepsilon;a,b) = \varepsilon^{a}(1-\varepsilon)^{b}/{\rm B}(a,b),
\end{equation}
where $B(a,b)$ is the beta function. 
The beta-binomial distribution is given by an urn model. 
Let us consider $N_{w}$ white balls and $N_{b}$ black balls in the urn. 
One draws a ball from the urn. If it is white (black), 
return two white (black) balls to the urn. 
This procedure is repeated with $N$ times, then the resulting distribution 
of $n$ white balls is given by the beta-binomial distributions as $\beta(n;N,N_{w},N_{b})$. 
This is actually equivalent to $\beta(n;N,\alpha,\varepsilon)$, 
where $N_{w}=\alpha N$ with $\varepsilon=N_{w}/(N_{w}+N_{b})$. 
A smaller $\alpha$ gives a broader distribution than the binomial, 
while the distribution becomes close to the binomial distribution with 
a larger value of $\alpha$. 

The beta-binomial distributions are numerically generated with various values of $\alpha$. 
These are compared to the embedding data to determine the best fit
parameter  value of $\alpha$.  The green lines in Fig.~\ref{fig:EmbeddingFitFinal} show the beta-binomial distribution 
for the value of $\alpha$ that gives the minimum $\chi^{2}/{\rm ndf}$. 
It is found that $\chi^{2}/{\rm ndf}\approx1$ for most $(N_{\rm p},N_{\bar{p}})$ combinations. 
With this additional parameter $\alpha$, it is found that the detector response 
is better described in the tails by a beta-binomial distribution
compared to a binomial distribution.

From the embedding simulations as discussed above, the $\varepsilon$ and $\alpha$ are parametrized as a function 
of $N_{\rm p}$ and $N_{\bar{\rm p}}$. Using the parametrization, a four-dimensional response matrix between generated and reconstructed protons and antiprotons is 
generated with 1 billion events. The limited statistics in the embedding simulations lead to uncertainties on the
$\alpha$ values. Therefore, two more
response matrices are generated using $\alpha-\sigma$ 
and $\alpha+\sigma$, where $\sigma$ is the statistical uncertainty on the $\alpha$ values determined by the embedding simulation. 
Furthermore, the standard response matrices are also generated 
with the binomial distribution as a reference using a multiplicity-dependent efficiency.
These response matrices are used to correct for the detector effects as a confirmation of 
this approach by comparing to the binomial correction method 
described in the previous section. 
The consistency of the unfolding method has been checked through a
detailed simulation and an analytic study.

Figure~\ref{fig:UnfHistoFinal} shows the unfolded net-proton distributions 
for 200 GeV Au+Au collisions at 0-2.5\% centrality. Results from four assumptions on the detector response are shown, 
one is the binomial detector response and the other three assume 
the beta-binomial distributions with different non-binomial $\alpha$ values. The ratios of the beta-binomial unfolded distributions to the binomial 
unfolded distributions are shown in the bottom panel. 
The unfolded distributions with beta-binomial response matrices are 
found to be narrower with a decreasing value of $\alpha$. Calculations are done for 0-2.5\% and 2.5-5.0\% centralities separately and averaged to determine the $C_n$ values for the 0-5\% centrality. The $C_n$ values and their ratios from data obtained using the binomial model method of efficiency correction and those using the binomial detector response matrix in the unfolding method are consistent. Table~\ref{table-un} summarizes the cumulant ratios and their errors. Results are also obtained from the unfolding method using the beta-binomial response function with non-binomial parameters in the range $\alpha \pm \sigma$. This range in values of $\alpha$ is used to generate the systematic uncertainties associated with the unfolding method. The deviations of those non-binomial efficiency-corrected results with respect to the conventional efficiency correction with binomial detector response is found to be 3.1 $\sigma$ for $C_2/C_1$ and less than 1.0 $\sigma$ for $C_4/C_2$ and for $C_3/C_2$. The $\sigma$ value is the statistical and systematic uncertainties added in quadrature.

These studies have been done for Au+Au collisions for the highest collision
energy of $\sqrt{s_{\rm NN}}=$~200~GeV and top-most 5\% centrality. 
This set of data provides the largest charged-particle-density
environment for the detectors, where we expect the maximum
non-binomial detector effects.  Even in this situation, the
differences in the two methods of efficiency correction are at a level of
less than one $\sigma$. Thus, we conclude that the non-binomial detector effects on higher-order 
cumulant ratios presented in this work are within the uncertainties
quoted for all of the BES-I energies.

\begin{figure}[htp]
\includegraphics[scale=0.4]{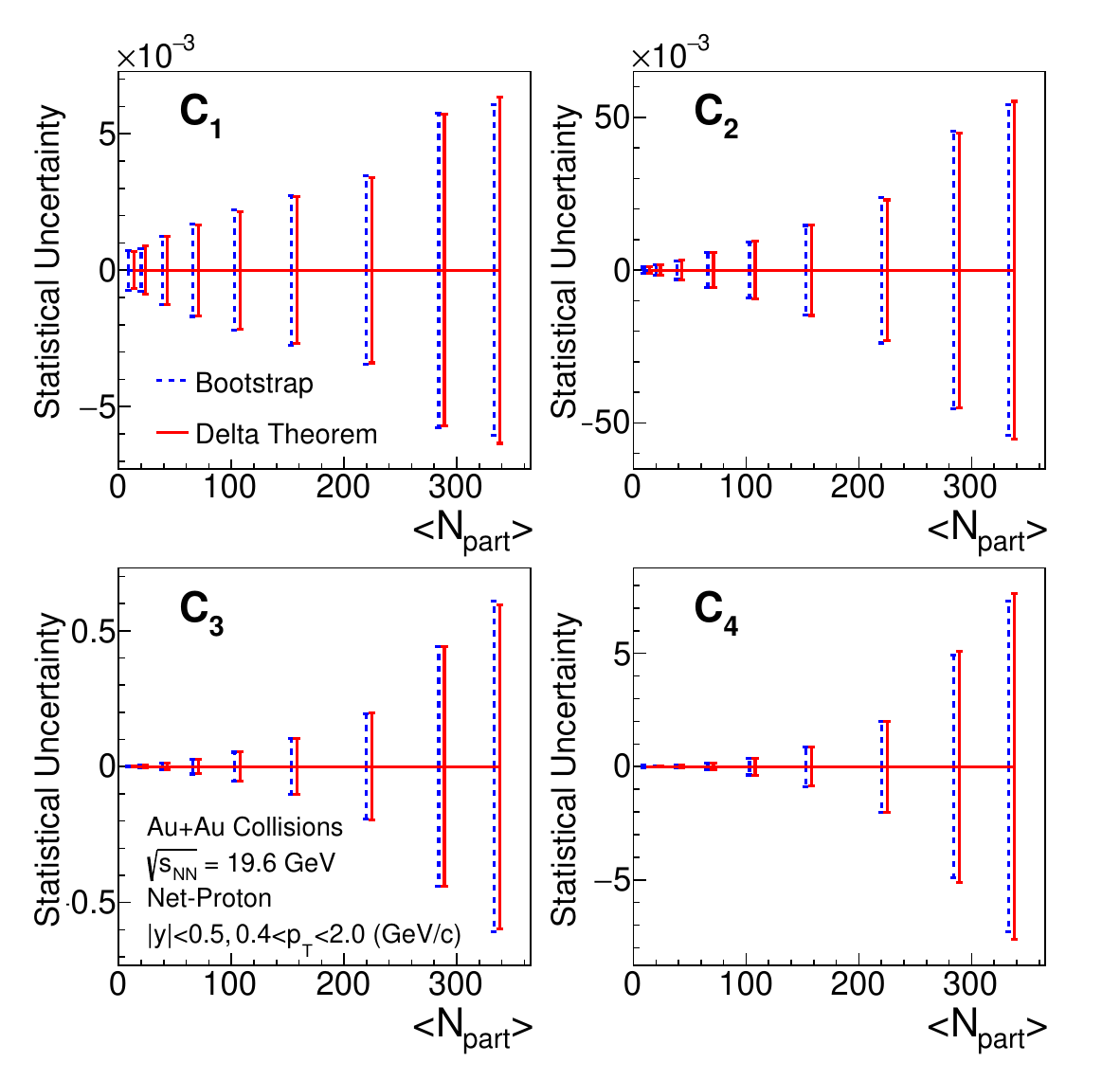}
\caption 
{(Color online) Comparison of the statistical uncertainties on $C_{n}$
of net-proton distributions in Au+Au collisions at \snn\ = 19.6 GeV
from the delta theorem and bootstrap methods. The results are presented as
 a function of $\langle N_{\mathrm {part}} \rangle$.}
\label{Fig:staterr}
\end{figure}

\begin{figure*}[htp]
\includegraphics[scale=0.8]{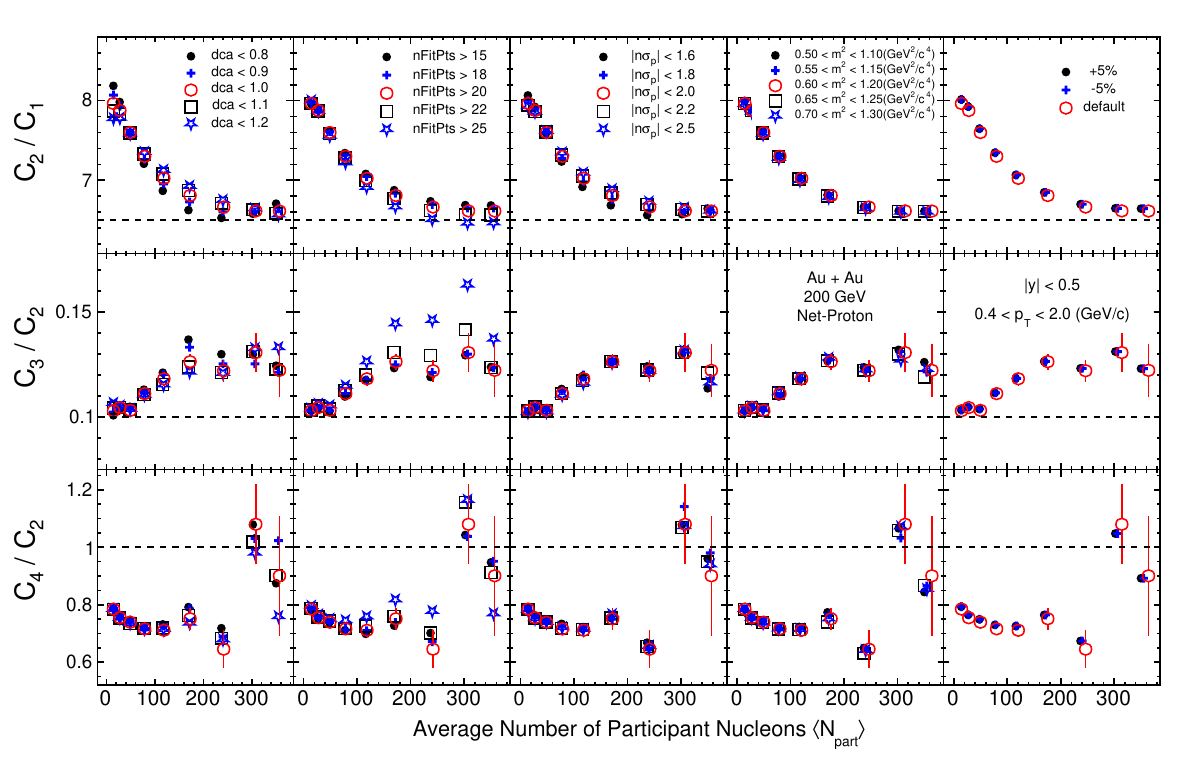}
\caption 
{(Color online)  Ratios of cumulants ($C_{n}$) as a function of $\langle N_{\rm
    {part}}\rangle$, for net-protons
  distributions in
  Au+Au collisions at $\sqrt{s_{\mathrm  {NN}}}$ = 200~GeV obtained by varying the analysis criteria in terms of track selection criteria, particle identification criteria and efficiency. Since variations with respect to default selection criteria are used to obtain the systematic uncertainties on the measurements, the errors are shown only for the default case.}
\label{Fig:syserr}
\end{figure*}
\subsection{Statistical uncertainty}
The higher-order cumulants are sensitive to the shape of the distribution, and estimating their statistical uncertainty is crucial due to the limited available statistics.  It has been shown that, among the various methods of
obtaining statistical uncertainty on cumulants, the delta theorem
method~\cite{Luo:2011tp} and the bootstrap
method~\cite{Luo:2013bmi,Luo:2014rea,Pandav:2018bdx,bootstrap,bootstrap1} are the most reliable ones. Below we briefly discuss the two methods and show that the
uncertainty values obtained up to the fourth-order cumulant from both methods are consistent.

\begin{table*}
\caption{Total systematic uncertainty as well as the absolute uncertainties from 
  individual sources, such as DCA and NhitsFit, for net-proton $C_{n}$ in 0-5\% central Au+Au collisions at 
  $\sqrt{s_{\rm {NN}}} = $ 7.7 - 200~GeV. The total systematic uncertainties are obtained by adding the uncertainties from 
  individual sources in quadrature.} 
\begin{center}
\label{table_syserr}
\begin{tabular}{cccccccc}
\hline 
$\sqrt{s_\mathrm{NN}}$ (GeV) & Cumulant &Total syst.& DCA & NhitsFit & $N_{\sigma, p}$ & $m^{2}$ & Efficiency\\
\hline 
 &$C_{1}$       &2.42    & 0.85&     0.78&     0.99&    0.028&      1.88\\
 &$C_{2}$       &2.03     &0.72&     0.60&     0.82&   0.032&      1.61\\
7.7
&$C_{3}$       &1.65      &0.60&      0.97&     0.54&      0.31&     1.02\\
&$C_{4}$       &16.20     &5.56&     12.54&      6.40&       2.68&      5.11\\
\hline 
&$C_{1}$     & 2.82&     1.76&     1.03&     1.13&    0.033&      1.59\\
&$C_{2}$      &2.34&     1.44&     0.73&      0.99&    0.020&    1.37\\
11.5
&$C_{3}$     &1.36&     0.64&     0.20&     0.85&     0.035&     0.82\\
&$C_{4}$      &7.37&     2.28&      4.10&    4.94&      2.60&     1.06\\
\hline 
&$C_{1}$               &    1.72&     0.77&     0.54&     0.76&    0.03&      1.22\\
&$C_{2}$               &     1.60&     0.69&     0.49&     0.74&    0.021&      1.13\\
14.5
&$C_{3}$               &    1.16&     0.52&     0.44&      0.51&    0.047&     0.78\\
&$C_{4}$               &     8.06&      2.89&      3.10&      5.41&     0.71&      4.15\\
\hline 
&$C_{1}$               & 1.46&     0.60&     0.62&     0.56&    0.045&      1.03\\
&$C_{2}$               &   1.46&     0.62&     0.62&     0.57&    0.041&      1.02\\
19.6
&$C_{3}$               &   0.68&     0.36&     0.26&    0.23&     0.13&     0.44\\
&$C_{4}$               &   3.65&      0.86&      1.99&     2.58&     0.59&      0.89\\
\hline 
&$C_{1}$               &     1.20&     0.51&     0.53&     0.47&     0.025&     0.83\\
&$C_{2}$               &    1.44&     0.67&     0.63&     0.57&    0.027&     0.96\\
27
&$C_{3}$               &    0.62&     0.33&     0.27&     0.23&    0.035&     0.39\\
&$C_{4}$               &   3.10&      1.58&      1.36&      1.80&     0.38&      1.36\\
\hline 
&$C_{1}$               &  0.94&     0.39&     0.45&     0.35&    0.026&     0.64\\
&$C_{2}$               &     1.48&     0.67&     0.67&     0.59&    0.033&     0.97\\
39 
&$C_{3}$               &   0.51&     0.29&     0.21&     0.17&    0.04&     0.313\\
&$C_{4}$               &   3.35&     1.00&      2.76&      1.43&     0.20&     0.65\\
\hline 
&$C_{1}$               & 0.81&	0.43&	0.33&	0.20&	0.034&	0.56\\
&$C_{2}$               &1.57&	 0.88&	0.65&	0.39&	0.064&	1.06\\
54.4
&$C_{3}$               &  0.42&	0.27&	        0.15&	0.078&	0.025&	0.27\\
&$C_{4}$               &2.95&	         1.18&	1.41&	        1.93&	         1.24&	0.21\\
\hline 
&$C_{1}$               &   1.04&       0.45&     0.49&     0.35&    0.044&     0.71\\
&$C_{2}$               &    2.15&     1.05&     1.087&     0.79&      0.11&      1.31\\
62.4
 &$C_{3}$               &  0.58&    0.14&      0.22&     0.30&       0.081&     0.41\\
&$C_{4}$               &    3.99&      2.40&      2.30&       1.38  &     1.21&      1.23\\
\hline 
&$C_{1}$               &    0.39&     0.19&      0.24&     0.11&   0.01&     0.22\\
&$C_{2}$               &    2.42&     1.11&      1.53&     0.77&    0.087&      1.31\\
200
&$C_{3}$               &    0.39&    0.24&     0.18&     0.19&    0.074&     0.14\\
&$C_{4}$               &     4.89&      2.69&      3.07&      1.80&      1.41&      1.42\\
\hline 
\end{tabular}%
\end{center}
\end{table*}

The delta theorem method gives a concise form of standard error propagation method.  This method of statistical
uncertainty estimation uses the central limit theorem (CLT). The variance of the statistic $\phi$ can be calculated as: 
\begin{equation} \label{eq:error}
V(\phi ) = \sum\limits_{i,j = 1}^m {\left( {\frac{{\partial \phi }}{{\partial {X_i}}}} \right)} \left( {\frac{{\partial \phi }}{{\partial {X_j}}}} \right){\rm Cov}({X_i},{X_j}),
\end{equation}
where the ${\rm Cov}(X_i,X_j)$ is the covariance between random variables $X_i$ and $X_j$. 
Thus, we need to know the covariance between $X_i$ and $X_j$ to calculate the statistical errors.

If particle multiplicities follow a Gaussian distribution with width $\sigma$, the statistical uncertainty of the cumulants and cumulant ratios at different orders
can be estimated as:
\begin{eqnarray} \label{eq:error_norm}
\mathrm{error} (C_{m}) \propto \frac{\sigma^{m}}{\sqrt{N}~\varepsilon^{\alpha}},~
\mathrm{error} (C_{n}/C_{2}) \propto \frac{\sigma^{n-2}}{\sqrt{N}~\varepsilon^{\beta}},
\end{eqnarray}
where $m$ and $n$ are integer numbers with $m \ge 1$ and $n \ge 2$, and $\alpha$ and $\beta$ are real numbers with $\alpha>0$ and $\beta>0$. The $N$ and $\varepsilon$ denote the number of events and the particle-reconstruction efficiency, respectively. Thus, one can find that the statistical uncertainty strongly depends on the width ($\sigma$) of the distributions. For similar event statistics, due to the increasing width of the net-proton distributions from peripheral to central collisions, the statistical uncertainties are larger in central collisions than those from peripheral collisions. Furthermore, the reconstruction efficiency increases the statistical uncertainties on the cumulants compared to their corresponding uncorrected case. A more detailed discussion can be found in Appendix~\ref{appendix-2}.

The bootstrap method finds the statistical uncertainties on the cumulants in a
Monte Carlo way by forming bootstrap samples. It makes use of a random
selection of elements with replacement from the original sample to
construct bootstrap samples over which the sampling variance of a
given order cumulant is calculated~\cite{bootstrap,bootstrap1}.
Let $X$ be a random sample representing the experimental dataset. Let $\mu_{r}$ be the estimator of a statistic (such as mean or variance etc.), on which we intend to find the statistical
error.  Given a parent sample of size $n$, construct $B$ number of
independent bootstrap samples $X^{*}_{1}$, $X^{*}_{2}$, $X^{*}_{3}$,
..., $X^{*}_{B}$, each consisting of $n$ data points randomly drawn
with replacement from the parent sample. Then evaluate the estimator in each bootstrap sample:
\begin{equation}
\mu_{r}^{*}=\mu_{r}(X^{*}_{b})      \qquad b=1,2,3, ..., B.
\end{equation}
Then obtain the sampling variance of the estimator as:
\begin{equation}
\mathrm{Var}(\mu_{r}) = \frac{1}{B-1}\sum_{b=1}^{B}\Big(\mu_{r}^{*} -\bar{\mu}_{r}\Big)^{2},
\end{equation}
where $\bar{\mu}_{r} = \frac{1}{B}\sum_{b=1}^{B}(\mu_{r}^{*})$.
The value of $B$ is optimized and, in general, the larger the value of $B$
the better the estimate of the error.

Figure~\ref{Fig:staterr} shows the statistical uncertainties on various orders of $C_{\rm
  n}$ obtained using the delta theorem and bootstrap methods for Au+Au
collisions at $\sqrt{s_{\mathrm  {NN}}}$ = 19.6~GeV. The results are
shown as a function of $\langle N_{\mathrm {part}} \rangle$ for each
$C_{n}$. The value of $B$ is 200.  Good agreement of the statistical uncertainties is seen
from both methods. The delta theorem method is used for obtaining
the statistical uncertainties on the results discussed below.

\begin{figure*}[htp]
\includegraphics[scale=0.85]{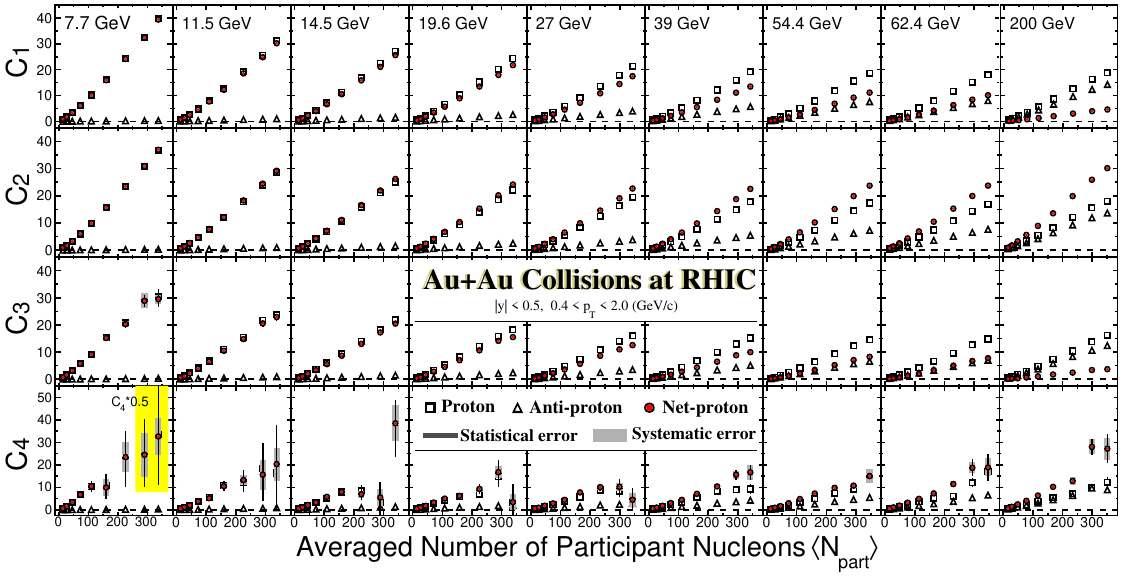}
\caption {(Color online) Collision centrality dependence of proton (open squares), antiproton (open triangles) and net-proton (filled circles) cumulants from (7.7 – 200 GeV) Au+Au collisions at RHIC. The data are from $|y| < 0.5$ and $0.4 <p_T<2.0$ GeV/$c$. Statistical and systematic uncertainties are shown as the narrow black and wide grey bands, respectively. Note that the net-proton and proton $C_4$ from 0-5\% and 5-10\% central Au+Au collisions at 7.7 GeV have been scaled down by a factor of 2, indicated in the yellow box.}
\label{Fig:Cum-cen}
\end{figure*}

\subsection{Systematic uncertainty}
Systematic uncertainties are estimated by varying the following requirements 
for $p(\bar{p})$ tracks: DCA, track quality (as reflected by the number of
fit points used in track reconstruction), $dE/dx$, and $m^{2}$ for $p(\bar{p})$ 
identification~\cite{Adamczyk:2013dal}. A $\pm$ 5\% systematic uncertainty
associated with determining the efficiency is also considered~\cite{Adamczyk:2017iwn}. 
All of the different sources of systematic uncertainty
are added in quadrature to obtain the final systematic uncertainties on
the $C_{n}$ and its ratios. Figure~\ref{Fig:syserr}
shows the variations of the cumulants ratios with the
changes in the above selection criteria for the net-proton distributions in
Au+Au collisions at $\sqrt{s_{\mathrm  {NN}}}$ = 200 GeV. 

Table~\ref{table_syserr} gives the systematic uncertainties on
the $C_{n}$ of the net-proton distribution for 0-5\%
central Au+Au collisions at $\sqrt{s_{\mathrm  {NN}}}$ = 7.7 - 200 GeV. 
The statistical and systematic uncertainties are presented separately in the figures.

\begin{figure*}[htp]
\hspace{-0.5cm}
\includegraphics[scale=0.5]{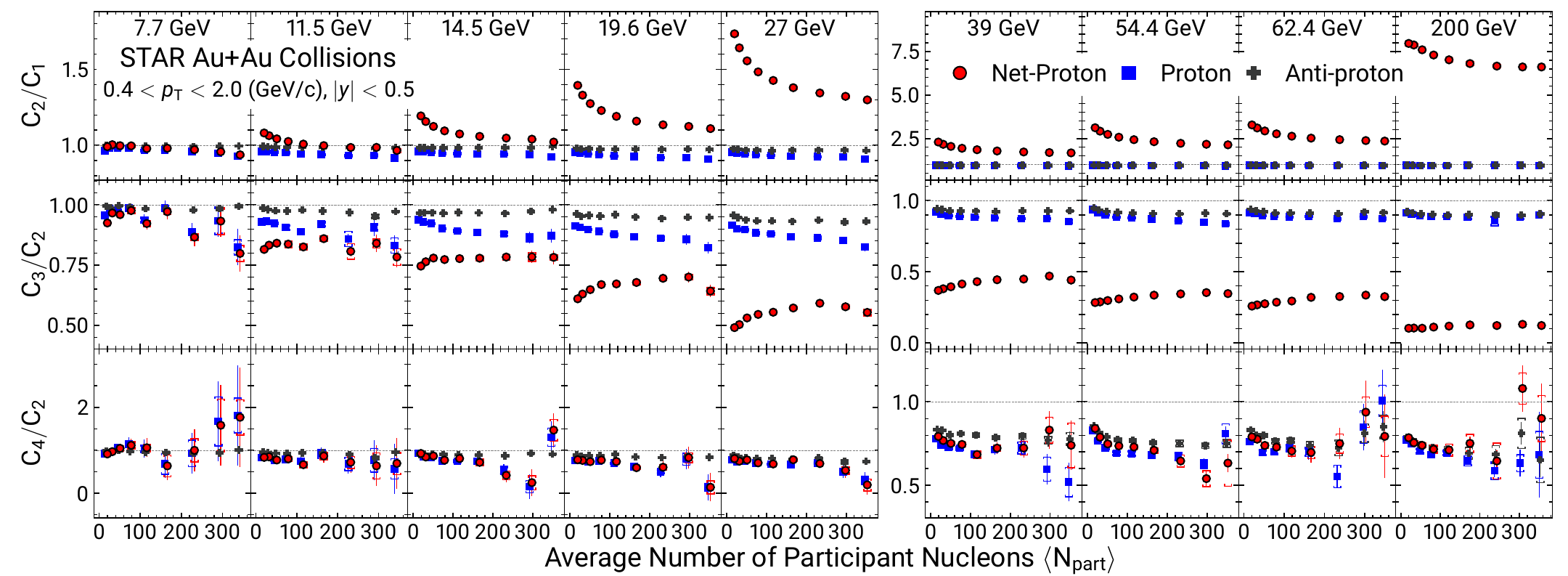}
\caption 
{(Color online) Collision centrality dependence of the cumulant
  ratios of proton, antiproton and net-proton multiplicity
  distributions for Au+Au collisions at \snn\ = 7.7, 11.5, 14.5, 19.6,
  27, 39, 54.4, 62.4 and 200 GeV. The bars and caps represent the statistical and systematic uncertainties, respectively.}
\label{Cum-ratio-cen}
\end{figure*}

\section{Results} 
In this section we present the efficiency-corrected
cumulants and cumulant ratios of net-proton, proton 
and antiproton multiplicity distributions in Au+Au collisions at {\snn} = 7.7, 11.5, 14.5, 19.6, 27,
39, 54.4, 62.4 and 200 GeV. The cumulant ratios are related to 
the ratios of baryon number susceptibilities ($\chi_{\mathrm B}$) 
computed in QCD-motivated models as: $\sigma^2$/$M$ =
$\chi^{B}_{\mathrm 2}/\chi^{B}_{\mathrm 1}$, ${\it{S}}$$\sigma$ = $\chi^{B}_{\mathrm 3}/\chi^{B}_{\mathrm 2}$, and 
$\kappa$$\sigma^2$ = $\chi^{B}_{\mathrm 4}/\chi^{B}_{\mathrm 2}$~\cite{Ejiri:2005wq,Cheng:2008zh,Stokic:2008jh,Gupta:2011wh,Gavai:2010zn}. 
Normalized correlation functions ($\kappa_{n}/\kappa_1$, $n>1$) for the proton and antiproton extracted from the measured $C_n$ are also presented.  The statistical uncertainties on $\kappa_n$ are obtained from the uncertainties on $C_n$ using the standard error propagation method. These results will be also compared to corresponding results from a hadron resonance gas (HRG)~\cite{Garg:2013ata} and hadronic-transport-based 
UrQMD model calculations~\cite{Xu:2016qjd,He:2017zpg}. 

In the following subsections, the dependence of the cumulants and correlation functions on collision energy, centrality, rapidity, and transverse momentum are presented. The corresponding physics implications are discussed. 

\subsection{Centrality dependence}
In this subsection, we show the $\langle N_{\mathrm {part}} \rangle$ (representing collision centrality) dependence of the cumulants, cumulant ratios and normalized correlation functions in Au+Au collisions 
at \snn\ = 7.7 -- 200 GeV. To understand the evolution of the centrality dependence of the cumulants and cumulant ratios, we invoke the central limit theorem and consider the distribution at any given centrality $i$ to be a superposition of several independent source distributions~\cite{Aggarwal:2010wy}.
Assuming the average number of sources for a given centrality is proportional to the corresponding $\langle
N_{\mathrm {part}}\rangle$, the $C_{n}$ should have a linear dependence
on  $\langle N_{\mathrm {part}}\rangle$ and the ratios $C_{2}/C_{1}$,
$C_{3}/C_{2}$ and $C_{4}/C_{2}$ should be constant as a function of
$\langle N_{\mathrm {part}}\rangle$.

\begin{figure*}[htp]
\hspace{-0.5cm}
\includegraphics[scale=0.5]{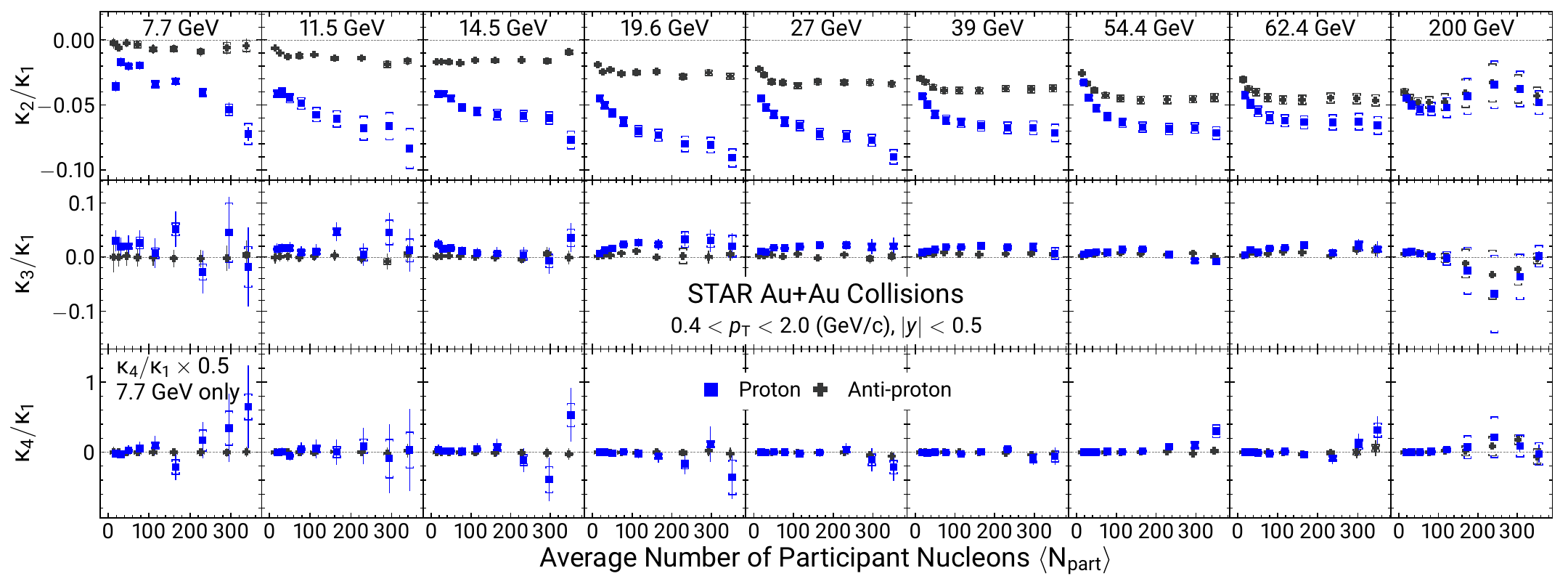}
\caption 
{(Color online) Collision centrality dependence of normalized correlation functions $\kappa_{n}/\kappa_1$ ($n=2, 3, 4$) for proton and antiproton multiplicity distributions in Au+Au collisions at \snn = 7.7, 11.5, 14.5, 19.6, 27, 39, 54.4, 62.4 and 200 GeV. The bars and caps represent the statistical and systematic uncertainties, respectively. For clarity, the $X$-axis values for protons are shifted and the values of proton and antiproton $\kappa_{4}/\kappa_1$ at \snn\ =  7.7 GeV are scaled down by a factor of 2.}
\label{Corr-func-cen}
\end{figure*}

Figure~\ref{Fig:Cum-cen} shows the $\langle N_{\mathrm{part}} \rangle$ dependence of $C_{n}$ for net-proton, proton and antiproton distributions in Au+Au collisions at \snn\ = 7.7 -- 200 GeV. Since the cumulants are extensive quantities, the $C_{n}$ for net-proton,
proton and antiproton increase with increasing $\langle N_{\mathrm{part}} \rangle$ for all of the \snn\ studied. The 
different mean values of the proton and antiproton distributions at each energy are determined by the interplay
between proton-antiproton pair production and baryon stopping effects. At the lower \snn, the effects of baryon
stopping at midrapidity are more important than at higher \snn, and therefore the net-proton $C_{n}$ has
dominant contributions from protons. The small mean values
for antiprotons at lower \snn\ are due to their low rate of production. At higher \snn, the pair production process
dominates the production of protons and antiprotons at midrapidity. The $\bar{p}/p$ ratio for 0-5\%
central Au+Au collisions at \snn\ = 200 GeV and 7.7 GeV are 0.769 and 0.007,
respectively~\cite{Abelev:2008ab,Adamczyk:2017iwn}. Large values of $C_{3}$ and $C_{4}$ also indicate
that the net-proton, proton and antiproton distributions are non-Gaussian. To facilitate plotting, the net-proton and proton $C_{4}$ from the 0-5\% and 5-10\% central Au+Au collisions at  \snn\ = 7.7 GeV are scaled down by a factor of 2.

Figure~\ref{Cum-ratio-cen} shows the  $\langle N_{\mathrm {part}}
\rangle$ dependence of cumulant ratios $C_{2}$/$C_{1}$,
$C_{3}$/$C_{2}$ and $C_{4}$/$C_{2}$ for net-proton, proton and antiproton distributions measured in Au+Au
collisions at \snn\ = 7.7 -- 200 GeV.  In terms of the moments of the
distributions, they correspond to $\sigma^{2}/M$ ($C_{2}$/$C_{1}$),
${\it{S}}$$\sigma$ ($C_{3}$/$C_{2}$) and $\kappa$$\sigma^2$
($C_{4}$/$C_{2}$).  The volume effects are canceled to the first order in these cumulant ratios. It is found that both of the proton and antiproton cumulant ratios  $C_{2}$/$C_{1}$ and $C_{3}$/$C_{2}$ show weak  variations with $\langle N_{\mathrm {part}} \rangle$. Based on the HRG model with the Boltzmann approximation, the orders of baryon number fluctuations can be analytically expressed as $C^{B}_{1}/C^{B}_{2}$ = $C^{B}_{3}/C^{B}_{2}$ = $\mathrm{tanh}(\mu_{B}/T)$ and $C^{B}_{4}/C^{B}_{2}$ = 1, where $\mu_{B}$ and $T$ are the baryon chemical potential and temperature of the system, respectively. The values of net-proton $C_{2}$/$C_{1}$ show a monotonic decrease with increasing $\langle N_{\mathrm {part}} \rangle$ while the values of $C_{3}$/$C_{2}$ show a slight increase with $\langle N_{\mathrm {part}} \rangle$. For a fixed centrality, both net-proton $C_{2}$/$C_{1}$ and $C_{3}$/$C_{2}$ show strong energy dependence, which can be understood as $C_{3}/C_{2} \propto \mathrm{tanh}(\mu_{B}/T)$ and $C_{2}/C_{1} \propto 1/\mathrm{tanh}(\mu_{B}/T)$. At high \snn, the net-proton $C_{3}$/$C_{2} \propto \mathrm{tanh}(\mu_{B}/T) \approx \mu_{B}/T \to 0$ and $C_{2}$/$C_{1} \propto 1/\mathrm{tanh}(\mu_{B}/T) \approx T/\mu_{B} > 1$. Since the $\mu_{B}/T \gg 1$ for the lower energies,  the values of net-proton $C_{2}/C_{1}$ and $C_{3}$/$C_{2}$ approach unity. Due to the connection between higher-order net-proton cumulant ratios and chemical freeze-out $\mu_B$ and $T$, those cumulant ratios have been extensively applied to probe the chemical freeze-out conditions and thermal nature of the medium created in heavy-ion collisions~\cite{Bazavov:2012vg,Borsanyi:2013hza,Gupta:2020pjd}. Finally, the net-proton and proton $C_{4}$/$C_{2}$ ratios have weak $\langle N_{\mathrm {part}} \rangle$ dependence for energies above \snn\ = 39 GeV. For energies below \snn\ = 39 GeV, the net-proton and proton $C_{4}$/$C_{2}$ generally show a decreasing trend with increasing $\langle N_{\mathrm {part}}\rangle$, except that, within current uncertainties, weak centrality dependences of $C_4/C_2$ are observed in Au+Au collisions at \snn\ = 7.7 and 11.5 GeV.

Figure~\ref{Corr-func-cen} shows the variation of normalized correlation functions $\kappa_{n}/\kappa_1$ ($n>1$) with $\langle N_{\rm part} \rangle$ for protons and antiprotons in Au+Au collisions at \snn\ = 7.7 
-- 200 GeV. As shown in Eqs.~(\ref{eq:C2toC1})--(\ref{eq:C4toC2}), the proton and antiproton cumulant ratios $C_2/C_1$, $C_3/C_2$ and $C_4/C_2$ can be expressed in terms of corresponding normalized correlation function $\kappa_n/\kappa_1$. Therefore, the results shown in Fig.~\ref{Corr-func-cen} provide important information on how different orders of multiparticle correlation functions of protons and antiprotons contribute to the cumulant ratios. The values of $\kappa_{1}$ are equal to mean $C_{1}$ values for protons and antiprotons, and linearly increase with $\langle N_{\mathrm{part}} \rangle$ as shown in Fig.~\ref{Fig:Cum-cen}. The normalized two-particle correlation functions, $\kappa_{2}/\kappa_1$, for protons and antiprotons are found to be negative for all $\langle N_{\mathrm{part}} \rangle$. The values of proton and antiproton $\kappa_{2}/\kappa_1$ become comparable at \snn\ = 200 GeV but exhibit larger discrepancies at lower energies. This can be understood as the interplay between baryon stopping and pair production of protons and antiprotons as a function of \snn. Within current uncertainties,  no statistically significant deviation from zero is observed in proton normalized correlation functions $\kappa_3/\kappa_1$ and $\kappa_4/\kappa_1$ as a function of collision centrality. As will be discussed later, however, one does observe non-monotonic energy dependence of proton $C_4/C_1$ in the 0-5\% central collisions; see Fig.~\ref{Corr-func-decom}. This is because, as defined in Eq.~(\ref{eq:CKtoCF}), the fourth-order cumulant $C_4$ contains contributions from second, third, and fourth-order correlation functions (factorial cumulants). In any case, high statistics data from the
second phase of the RHIC beam energy scan program (BES-II) are needed to understand the origin of the observed dependences on both collision energy and centrality.

\begin{figure*}[htp]
\hspace{-0.5cm}
\includegraphics[scale=0.5]{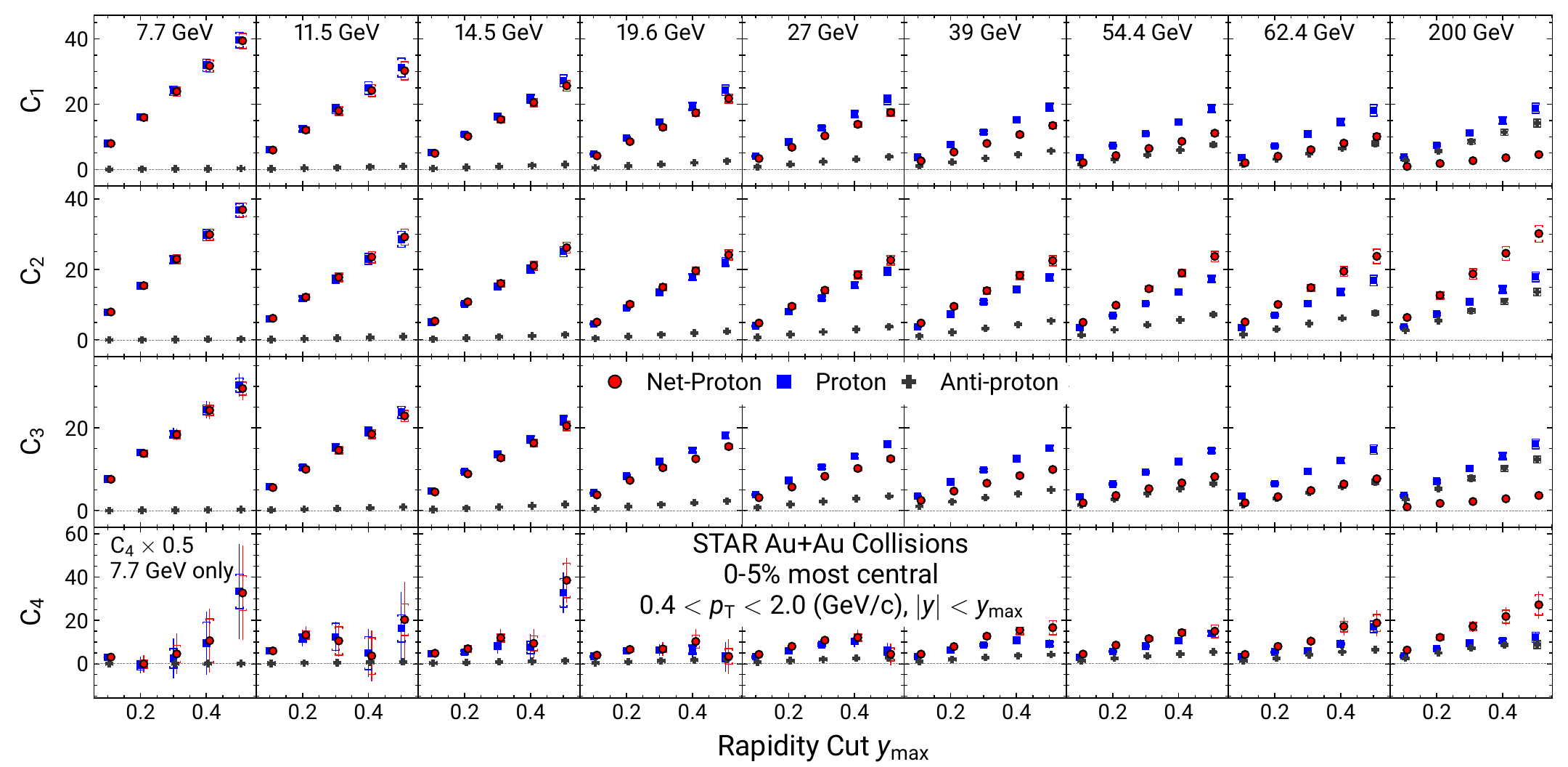}
\caption 
{(Color online) Rapidity acceptance dependence of cumulants of proton,
  antiproton and net-proton multiplicity distributions in 0-5\% central Au+Au
  collisions at \snn\ = 7.7, 11.5, 14.5, 19.6, 27, 39, 54.4, 62.4 and 200
  GeV. The bars and caps represent statistical and systematic uncertainties, respectively. For clarity,  the $X$-axis values for protons are shifted and the values of proton, antiproton and net-proton $C_4$ at \snn\ =  7.7 GeV are scaled down by a factor of 2.
}
\label{Fig:Cum-rap}
\end{figure*}

\begin{figure*}[htp]
\hspace{-0.5cm}
\includegraphics[scale=0.5]{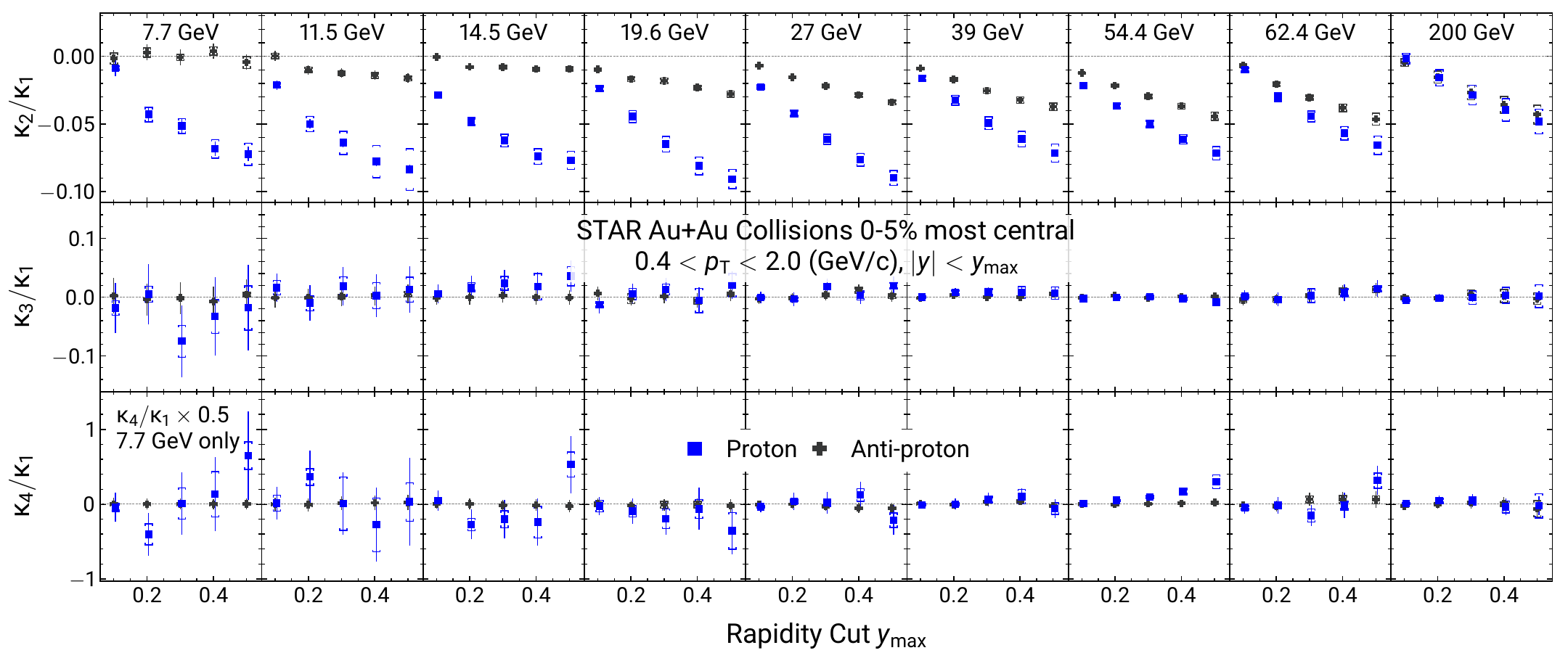}
\caption 
{(Color online) Rapidity acceptance dependence of normalized correlation 
  functions up to fourth order ($\kappa_{n}/\kappa_1$, $n$ = 2, 3, 4) for proton and antiproton multiplicity distributions in 
  0-5\% central Au+Au collisions at \snn\ = 7.7, 11.5, 14.5, 19.6, 27,
  39, 54.4, 62.4 and 200 GeV. The $X$-axis rapidity cut $y_{\mathrm{max}}$ is applied as $|y|<y_{\mathrm{max}}$. The bars and caps represent statistical and systematic uncertainties, respectively. For clarity,  the $X$-axis values for protons are shifted and the values of proton and antiproton $\kappa_{4}/\kappa_1$ at \snn\ =  7.7 GeV are scaled down by a factor of 2.}
\label{Corr-func-rap}
\end{figure*}

\begin{figure*}[htpb]
\hspace{-0.5cm}
\includegraphics[scale=0.5]{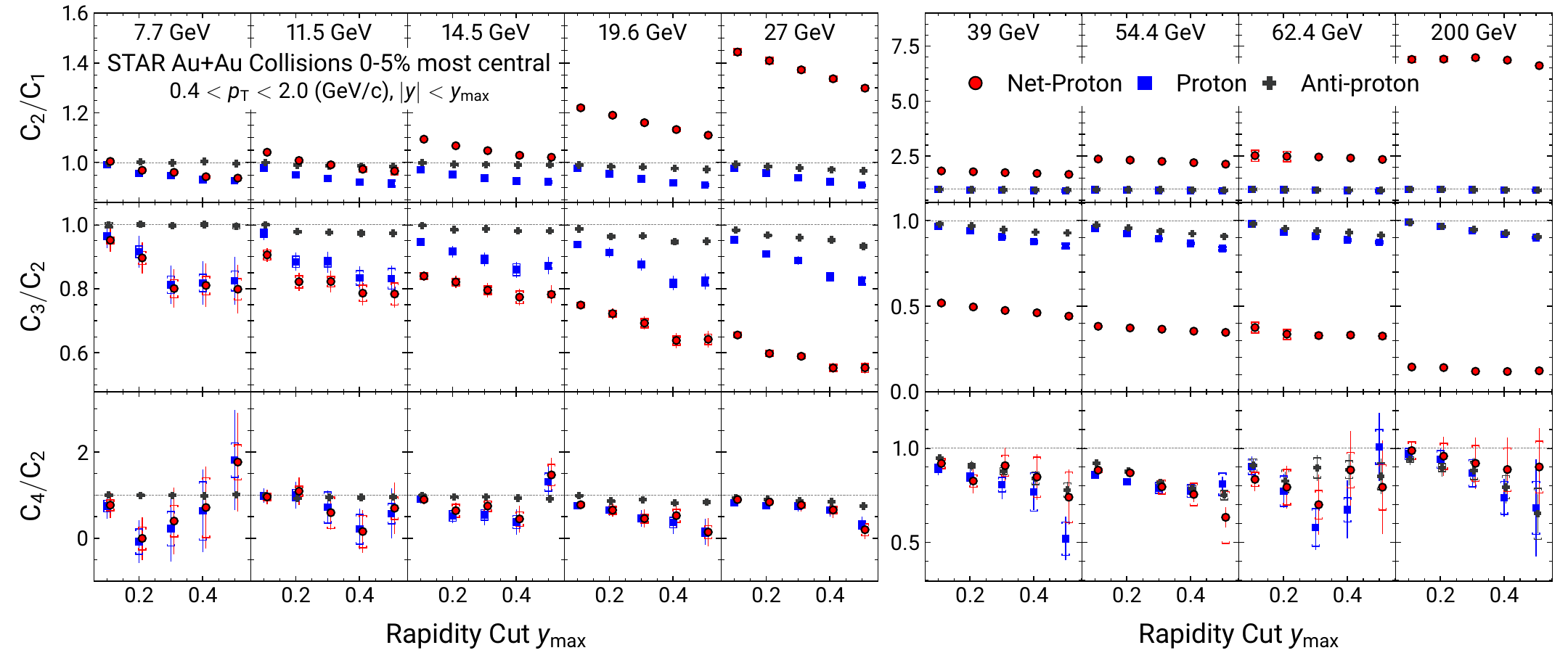}
\caption 
{(Color online) Rapidity-acceptance dependence of cumulant ratios of proton,
  antiproton and net-proton multiplicity distributions in 0-5\% central Au+Au
  collisions at \snn\ = 7.7, 11.5, 14.5, 19.6, 27, 39, 54.4, 62.4 and 200
  GeV. The bars and caps represent statistical and systematic uncertainties, respectively. For clarity,  the $X$-axis values for net-protons and protons are shifted.} 
\label{Fig:Cum-ratio-rap}
\end{figure*}

\subsection{Acceptance dependence}
In this subsection, we focus on discussing the acceptance dependence of the proton, antiproton and net-proton cumulants ($C_{n}$) and cumulant ratios in 0-5\% central Au+Au collisions at {\snn} = 7.7 -- 200 GeV.
It was pointed out in Refs.~\cite{Ling:2015yau,Bzdak:2016sxg,Bzdak:2017ltv,Brewer:2018abr} that when the rapidity acceptance ($\Delta y$) is much smaller than the typical correlation length ($\xi$) of the system ($\Delta y \ll \xi$), the cumulants ($C_n$) and correlation functions ($\kappa_n$) should scale with some power $n$ of the accepted mean particle multiplicities as $C_n,\kappa_n \propto (\Delta N)^n \propto (\Delta y)^n$. Meanwhile, in the regime where the rapidity acceptance becomes much larger than $\xi$ ($\Delta y \gg \xi$), the $C_n$ and $\kappa_n$ scale linearly with mean multiplicities or $\Delta y$. Thus, the rapidity acceptance dependence of the higher-order cumulants and correlation functions of proton, antiproton and net-proton distributions are important observables to search for a signature of the QCD critical point in heavy-ion collisions. On the other hand, that acceptance dependence of $C_n$ and $\kappa_n$ could be affected by the effects of non-equilibrium~\cite{Mukherjee:2016kyu,Wu:2018twy,Asakawa:2019kek}, smearing due to diffusion and hadronic re-scattering~\cite{Ohnishi:2016bdf,Sakaida:2017rtj,Nahrgang:2018afz,Asakawa:2019kek} in the dynamical expansion of the created fireball.

\subsubsection{Rapidity dependence}
Figure~\ref{Fig:Cum-rap} shows the rapidity ($-y_{max} < y < y_{max}$, $\Delta y= 2 y_{max}$) dependence of
the $C_{n}$ for proton, antiproton and net-proton distributions in
0-5\% central Au+Au collisions at \snn\ = 7.7 -- 200 GeV. The measurements are
made in the $p_{\mathrm T}$ range of 0.4  to 2.0 GeV/$c$. The rapidity
acceptance is cumulatively increased and the $C_{n}$ values for protons,
antiprotons, and net-protons increase
with increasing rapidity acceptance. For \snn\ $<$ 27 GeV, the proton
and net-proton $C_{n}$ have similar values, an inevitable consequence of the small production rate of antiproton at lower energies. 

Figure~\ref{Corr-func-rap} shows the variation of normalized correlation functions $\kappa_{n}/\kappa_1$ with 
rapidity acceptance for proton and antiproton in 0-5\%
central Au+Au collisions at \snn\ = 7.7 
-- 200 GeV.  The $\kappa_{2}/\kappa_{1}$ values for protons and antiprotons are negative
and monotonically increase in magnitude when enlarging the rapidity acceptance up to $y_{max}$=0.5 ($\Delta y$ = 1). For the antiproton, the values of $\kappa_{2}/\kappa_{1}$ show stronger deviations from zero at higher \snn.
As discussed around Fig.~\ref{Corr-func-cen}, the negative values of the two-particle correlation functions ($\kappa_2$) of protons and antiprotons are consistent with the expectation of the effect of baryon number conservation. Within current uncertainties, the rapidity acceptance dependences for the $\kappa_{3}/\kappa_{1}$ and $\kappa_{4}/\kappa_{1}$ of protons and antiprotons in Au+Au collisions at different \snn\ are not significant. The significances of the proton $\kappa_4/\kappa_1$ with $|y|<0.5$ deviating from zero are 1.04$\sigma$, 0.05$\sigma$, 1.27$\sigma$, 0.90$\sigma$, 0.95$\sigma$, 0.40$\sigma$, 2.91$\sigma$, 1.43$\sigma$, 0.11$\sigma$ for 0-5\% central Au+Au collisions at \snn\  = 7.7, 11.5, 14.5, 19.6, 27, 39, 54.4, 62.4 and 200 GeV, respectively, where the $\sigma$ is defined as the sum in quadrature of the statistical and systematic uncertainties.

\begin{figure*}[htp]
\hspace{-0.5cm}
\includegraphics[scale=0.5]{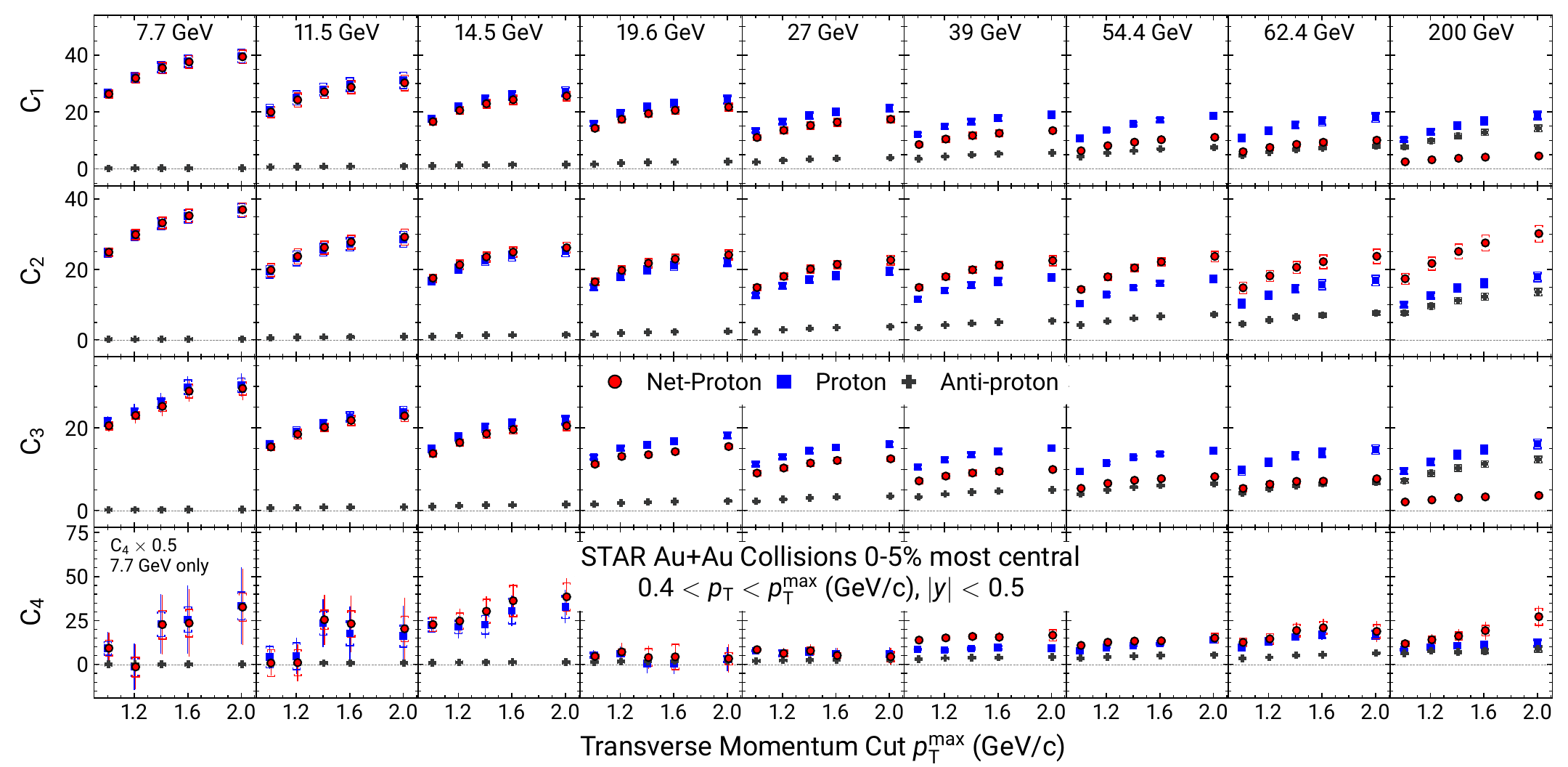}
\caption 
{(Color online) $\mathrm p_{T}$-acceptance dependence of cumulants of proton,
  antiproton and net-proton multiplicity distributions for 0-5\%
  central Au+Au collisions at \snn\ = 7.7, 11.5, 14.5, 19.6, 27, 39, 54.4, 62.4 and 200
  GeV. The bars and caps represent statistical and systematic uncertainties, respectively. For clarity,  the $X$-axis values for net-protons are shifted and the values of proton, antiproton and net-proton $C_4$ at \snn\ =  7.7 GeV are scaled down by a factor of 2. }
\label{Fig:Cum-pt}
\end{figure*}

\begin{figure*}[htp]
\hspace{-0.5cm}
\includegraphics[scale=0.5]{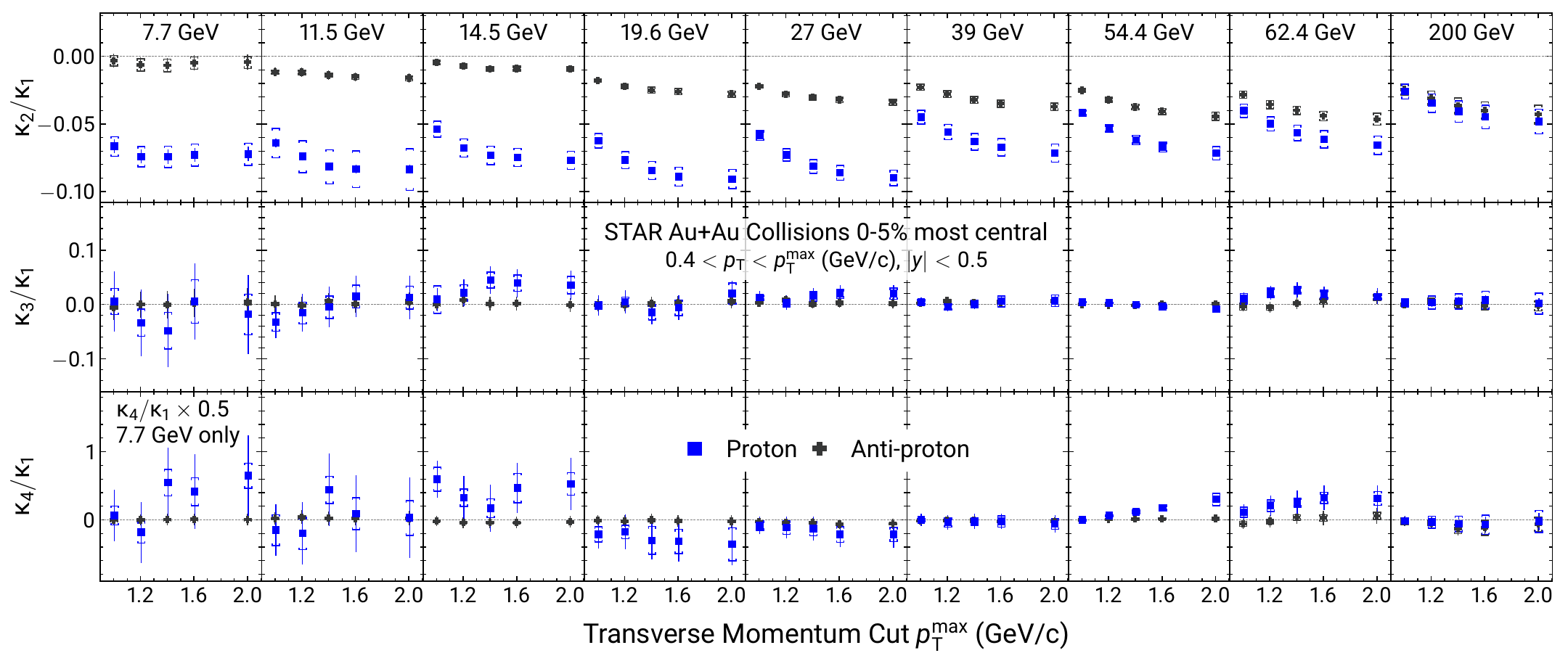}
\caption 
{(Color online) The $p_{\mathrm T}$-acceptance dependence of the normalized correlation 
  functions up to fourth order ($\kappa_{n}/\kappa_1$, $n$ = 2, 3, 4) for proton and antiproton multiplicity distributions in 
  0-5\% central Au+Au collisions at \snn\ = 7.7, 11.5, 14.5, 19.6, 27,
  39, 54.4, 62.4 and 200 GeV. The bars and caps represent statistical and systematic uncertainties, respectively. For clarity,  the $X$-axis values for protons are shifted and the values of proton and antiproton $\kappa_4/\kappa_1$ at \snn\ =  7.7 GeV are scaled down by a factor of 2.}
\label{Corr-func-pt}
\end{figure*}

\begin{figure*}[htpb]
\hspace{-0.5cm}
\includegraphics[scale=0.5]{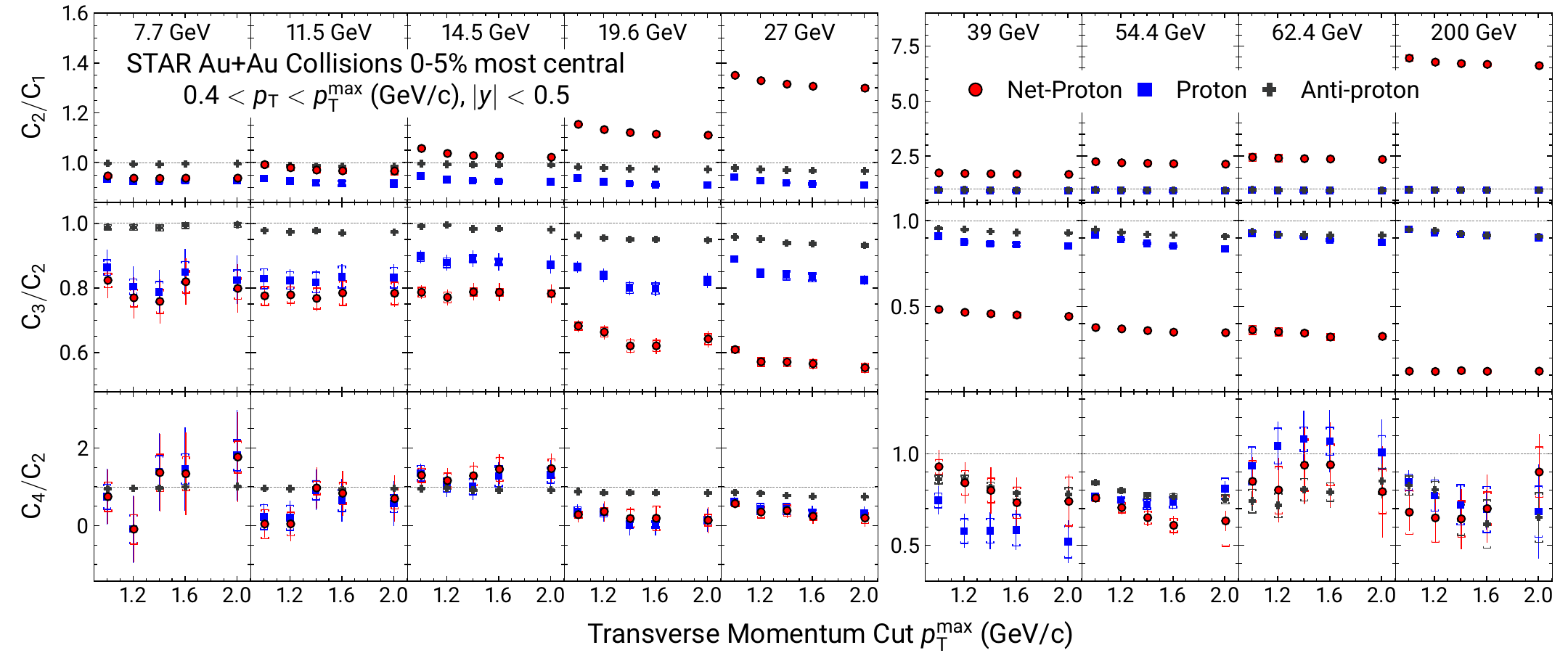}
\caption 
{(Color online) $\mathrm p_{T}$-acceptance dependence of cumulant ratios of
  proton, antiproton and net-proton multiplicity distributions for
  0-5\% central Au+Au collisions at \snn\ = 7.7, 11.5, 14.5, 19.6, 27,
  39, 54.4, 62.4 and 200 GeV. The bars and caps represent statistical and systematic uncertainties, respectively. For clarity,  the $X$-axis values for net protons are shifted.}
\label{Fig:Cum-pt-ratio}
\end{figure*}
Figure~\ref{Fig:Cum-ratio-rap} shows the rapidity acceptance
dependence of the cumulant ratios $C_{2}$/$C_{1}$, $C_{3}$/$C_{2}$,
and $C_{4}$/$C_{2}$ for protons, antiprotons, and net-protons in 0-5\% central Au+Au
collisions at \snn\ = 7.7 -- 200 GeV. Based on Eqs.~(\ref{eq:C2toC1}) to (\ref{eq:C4toC2}), the rapidity acceptance dependence of the cumulant ratios of proton and antiproton can be understood by the interplay between different orders of normalized correlation functions ($\kappa_n/\kappa_1$). The negative values of two-particle correlation functions ($\kappa_2$) for protons and antiprotons leads to a deviation of the corresponding $C_{2}$/$C_{1}$ and $C_{3}$/$C_{2}$ below unity.  Due to low production rate of antiproton at low energies, the values of $C_{2}$/$C_{1}$ and $C_{3}$/$C_{2}$ for the net-proton distributions approach the corresponding values for protons
when the beam energy decreases. The rapidity acceptance dependence of $C_{2}$/$C_{1}$, $C_{3}$/$C_{2}$ and $C_{4}$/$C_{2}$ values for protons and antiprotons are comparable at {\snn} = 200 GeV. However, among these ratios, protons and antiprotons start to deviate at lower beam energies. This is mainly due to baryon stopping and the larger fraction of transported protons compared with proton-antiproton pair production at midrapidity. The $C_{4}$/$C_{2}$ values for
proton, antiproton and net-proton distributions are consistent within uncertainties for {\snn} = 39, 54.4, 62.4 and 200 GeV. Significant deviations from unity are observed for 
proton and net-proton $C_{4}$/$C_{2}$ at {\snn} = 19.6 and 27 GeV, and 
the deviation decreases with decreasing $\Delta y$ acceptance, where the effects of baryon number conservation plays an important role. For energies below 19.6 GeV, the rapidity acceptance dependence of $C_4$/$C_2$ for protons, antiprotons and
net-protons is not significant within uncertainties.

\subsubsection{Transverse momentum dependence}
Figure~\ref{Fig:Cum-pt} shows the $p_{\mathrm T}$ acceptance dependence for the $C_{n}$ of proton, antiproton, and
net-proton distributions at midrapidity
($|y|<$ 0.5) for 0-5\% central Au+Au collisions at 
\snn\ = 7.7 -- 200 GeV. We fix the lower $p_{\mathrm T}$ cut at 0.4 GeV/$c$, and then the $p_{\mathrm T}$
acceptance is increased by varying the upper limit in steps between 1 and 2 GeV/$c$.
The average efficiency values used in the efficiency correction for various $p_T$ acceptances are calculated based on Eq.~(\ref{eq:eff}).
By extending the upper $p_{\mathrm T}$ coverage from 1 to 2 GeV/$c$, the mean numbers of protons increased about 50\% and 80\% at \snn\ = 7.7 and 200 GeV, respectively. 
It is found that the $C_{n}$ values for protons, antiprotons, and net protons increase with increasing $p_{\mathrm T}$ acceptance, except for a weak $p_{\mathrm T}$ acceptance dependence for $C_4$ observed at energies below 39 GeV.

Figure~\ref{Corr-func-pt} shows the variation of normalized correlation functions $\kappa_{n}/\kappa_1$ with 
$p_{\mathrm T}$  acceptance for proton and antiproton at midrapidity
($|y|<$ 0.5) in 0-5\% central Au+Au collisions at \snn\ = 7.7 -- 200
GeV. The $\kappa_{2}/\kappa_1$ values for protons and antiprotons are found to be negative and decrease with increasing $p_{\mathrm T}$ acceptance at higher \snn. The $\kappa_{2}/\kappa_1$ values for antiprotons approach zero when the beam energy is decreased, due to the small production rate of antiprotons at low energies. The negative values of $\kappa_{2}/\kappa_1$ for protons observed at low energies are mainly dominated by the baryon stopping. 

Figure~\ref{Fig:Cum-pt-ratio} shows the $p_{\mathrm T}$ acceptance dependence of $C_{2}$/$C_{1}$,
$C_{3}$/$C_{2}$ and $C_{4}$/$C_{2}$ for proton, antiproton and  
net-proton distributions in 0-5\% central Au+Au collisions at 
\snn\ = 7.7 -- 200 GeV. In general, most of the ratios show a weak dependence on $p_T$ acceptance for all 
of the \snn\ studied. The $C_{4}$/$C_{2}$ ratios of proton and net-proton distributions are
similar for all \snn\ below 27 GeV. 
The $C_{3}$/$C_{2}$  ratios for protons and antiprotons are similar at
higher beam energy. However, they differ from each other at the lower \snn.  
From the above differential measurements, it is found that the baryon number conservation strongly influences the cumulants and correlation functions in heavy-ion collisions, especially at low energies. It could be the main reason for the negative two-particle correlation functions for protons and antiprotons~\cite{He:2017zpg}.

\begin{figure*}[htp]
\includegraphics[scale=0.55]{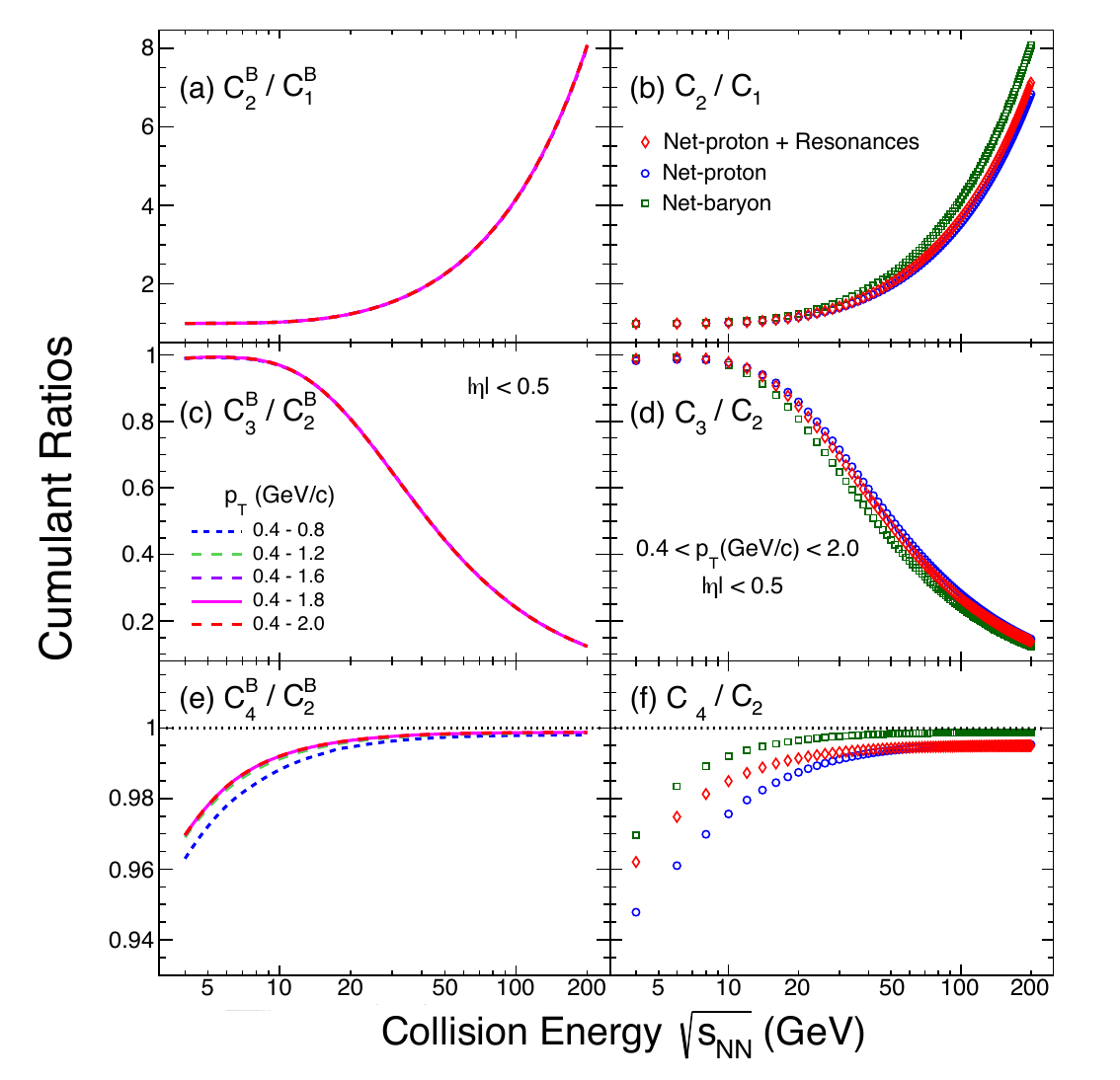}
\caption 
{(Color online) Left panel: Collision energy dependence of $C^{\rm B}_{\mathrm 
  2}$/$C^{\rm B}_{\mathrm 1}$,$C^{\rm B}_{\mathrm
  3}$/$C^{\rm B}_{\mathrm 2}$, and $C^{\rm B}_{\mathrm 4}$/$C^{\rm B}_{\mathrm 2}$
for various $p_{\mathrm T}$ acceptances from the hadron resonance gas model. Right
panel: The variation of net-proton and net-baryon $C_2/C_1$, $C_3/C_2$, and $C_4/C_2$ within
the experimental acceptance~\cite{Garg:2013ata}. Note: this simulation is done within a pseudorapidity window in order to 
make comparison between baryons of different mass. }
\label{Fig:hrg}
\end{figure*}

\subsection{Cumulants from models}
Although our results can be compared to several models~\cite{Li:2017via,Lin:2017xkd,Almasi:2017bhq,Yang:2016xga,Zhou:2017jfk,Zhao:2016djo,Xu:2016qjd,Vovchenko:2017ayq,Albright:2015uua,Fukushima:2014lfa,Netrakanti:2014mta,Morita:2014fda,Samanta:2019fef}, we have
chosen two models which do not have phase transition or critical point physics. They have contrasting physics processes to understand the following: (a) the effect of
measuring net-protons instead of net-baryons~\cite{Kitazawa:2012at,He:2016uei}, (b) the role of resonance
decay for net-proton measurements~\cite{Nahrgang:2014fza,Mishra:2016qyj,Bluhm:2016byc,Zhang:2019lqz}, (c) the effect of finite $p_{\mathrm
  T}$ acceptance for the measurements~\cite{Karsch:2015zna,He:2017zpg}, and (d) the effect of net-baryon
number conservation~\cite{Bzdak:2012an,He:2016uei,Braun-Munzinger:2019yxj}. Models without a critical point also provide an appropriate baseline for comparison to data.

\subsubsection{Hadron resonance gas model}
The hadron resonance gas model includes all the relevant degrees of freedom for the hadronic matter and also implicitly takes into account the interactions that are necessary for resonance 
formation~\cite{Garg:2013ata,Karsch:2010ck}. Hadrons and resonances of masses up to 3 GeV/$c^{2}$ are included. Considering a grand canonical ensemble
picture, the logarithm of the partition
 function ($Z$) in the HRG model  
is given as:
\begin{eqnarray}
\label{eq:eq1}
\ln Z(T, V, \mu) &=& \sum_{B}\ln Z_{i}(T, V, \mu_i) \nonumber \\
&+& \sum_{M}\ln Z_{i}(T, V, \mu_i)\ ,
\end{eqnarray}
where:
\begin{eqnarray}
&&\ln Z_{i}(T, V, \mu_{i})\\ 
&&=  \nonumber \pm\frac{Vg_{i}}{2\pi^2}\int 
d^3{p}\ln{\big\{1\pm\exp[(\mu_{i}-E)/T}]\big\},
\label{eq:eq2}
\end{eqnarray}
$T$ is the temperature, $V$ is the volume of the system, $\mu_{i}$ is the chemical potential, $E$ is
the energy, and $g_{i}$ is the degeneracy factor of the $i$th particle. The total chemical 
potential $\mu_{i}$ = $B_{i}\mu_{B}$ + $Q_{i}\mu_{Q}$ + $S_{i}\mu_{S}$, where $B_{i}$, $Q_{i}$ and 
$S_{i}$ are the baryon, electric charge  and strangeness number of the $i$th particle,
with corresponding chemical potentials $\mu_{B}$, $\mu_{Q}$ and $\mu_{S}$, respectively. The $+$ 
and $-$  signs in Eq.~(\ref{eq:eq2}) are for baryons ($B$) and mesons ($M$), respectively. The $n^{th}$-order generalized susceptibility for baryons can be expressed as \cite{Karsch:2010ck}:
 \begin{eqnarray}
 \label{eq:eq4}
 \chi_{x,\mathrm{baryon}}^{(n)}=\frac{x^n}{VT^3}
\int{d^{3}p}\sum_{k=0}^{\infty}{(-1)^k}
(k+1)^n \\ \nonumber \exp\bigg\{\frac{-(k+1)E } { T}\bigg\} {\exp\bigg\{ \frac{(k+1)\mu}{T}\bigg\}},
\,
 \end{eqnarray}
and for mesons:
 \begin{eqnarray}
 \label{eq:eq5}  
 \chi_{x,\mathrm{meson}}^{(n)}=\frac{x^n}{VT^3}
\int{d^{3}p}\sum_{k=0}^{\infty}(k+1)^n \\ \nonumber \exp\bigg\{
\frac {-(k+1)E } { T}\bigg\} {\exp\bigg\{ \frac{(k+1)\mu}{T}\bigg\}}.
\,
 \end{eqnarray}
The factor $x$ represents either $B$, $Q$ or $S$ of the $i$th particle, depending on whether
the computed $\chi_{x}$ represents baryon, electric charge or strangeness susceptibility.

\begin{figure*}[htp]
\includegraphics[scale=0.5]{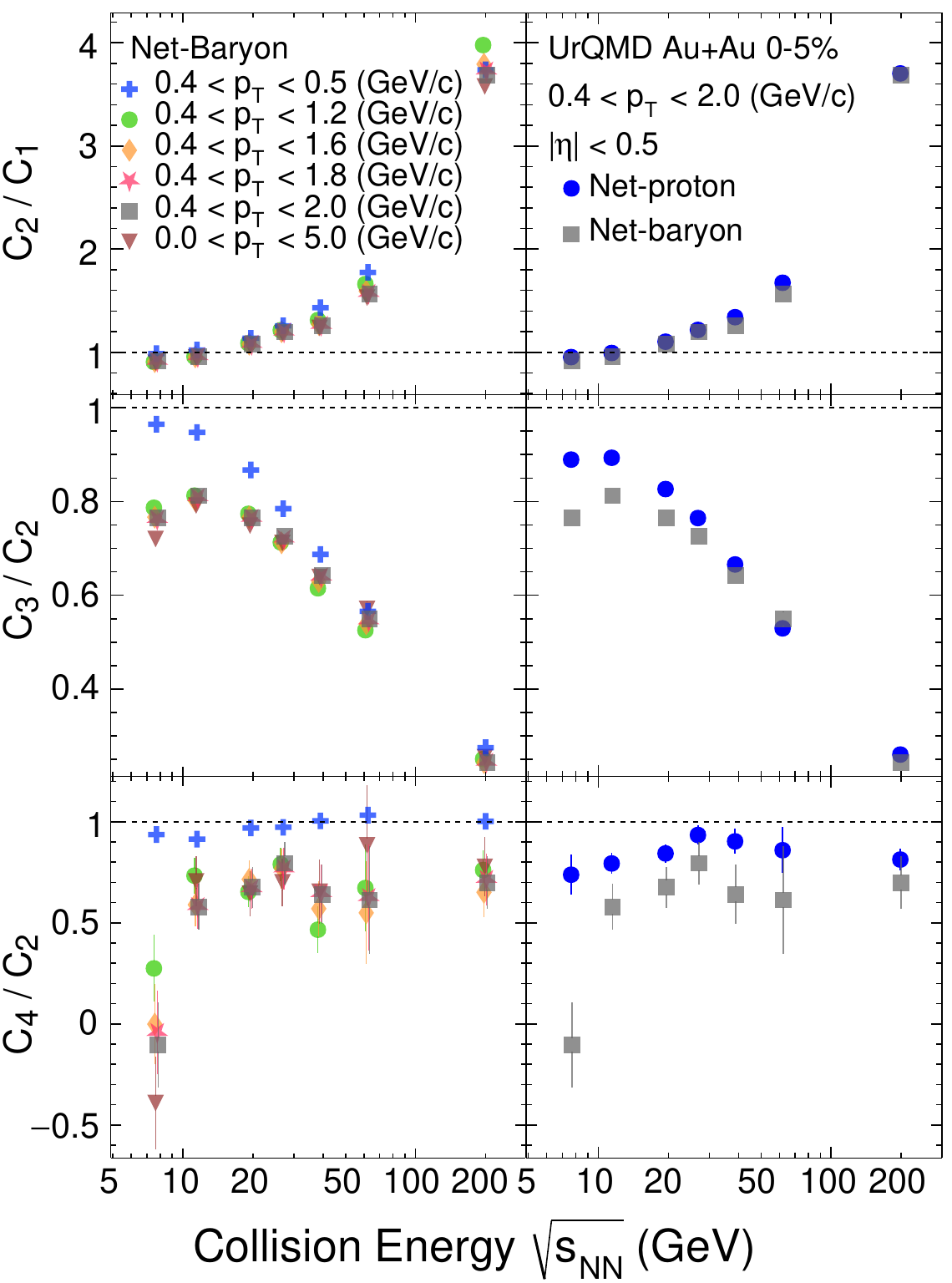}
\caption 
{(Color online) Left panel: UrQMD results on $p_{\mathrm T}$ acceptance dependence of 
  $C_{2}$/$C_{1}$, $C_{3}$/$C_{2}$, and $C_{4}$/$C_{2}$ ratios as a 
  function of \snn\ for net baryons. Right panel: Same ratios within 
  the experimental acceptance for net protons and net baryons. Note: similar to Fig~\ref{Fig:hrg}, this simulation is done within a pseudorapidity window in order to 
  make comparison between baryons of different mass. }
\label{Fig:URQMD}
\end{figure*}

For a particle of mass $m$ with $p_T$, $\eta$ and $\phi$, the 
volume element ($d^{3}p$) and energy ($E$) can be written as 
$d^{3}p=p_T m_T {\cosh}(\eta)$$d{p_{T}}$$d\eta$$d\phi$
and $E$ = $m_T\cosh\eta$, where $m_{T}$=$\sqrt{p_{T}^{2} + m^{2}}$. 
The experimental acceptance can be incorporated by considering the appropriate integration ranges 
in $\eta$, $p_{T}$, $\phi$ and charge states by considering the values of $|x|$.
The total generalized susceptibilities will then be the sum of the contributions from baryons and mesons as in
$\chi^{(n)}_x = \sum \chi^{(n)}_{x,\mathrm{baryon}} + \sum \chi^{(n)}_{x,\mathrm{meson}}$.

Figure~\ref{Fig:hrg} shows the  variation of $C^{\rm B}_{2}$/$C^{\rm B}_{1}$,
$C^{\rm B}_{3}$/$C^{\rm B}_{2}$ and $C^{\rm B}_{4}$/$C^{\rm B}_{2}$ as functions of \snn\ from a hadron resonance gas model~\cite{Garg:2013ata}. The results are shown for different $p_{\mathrm T}$
acceptances. The differences due to acceptance are very small, and the maximum effect is at
the level of 5\% for \snn\ = 7.7 GeV for
$C^{\rm B}_{\mathrm 4}$/$C^{\rm B}_{\mathrm 2}$. The HRG results
also show that the net-proton results with resonance decays are smaller compared to
net baryons and larger than net protons without the decay effect. Here also the effect is at
the level of 5\% for the lowest \snn\ and smaller at higher energies in the
case of $C^{\rm B}_{\mathrm 4}$/$C^{\rm B}_{\mathrm 2}$.
The corresponding effect on $C^{\rm B}_{\mathrm 3}$/$C^{\rm B}_{\mathrm
  2}$  and  $C^{\rm B}_{\mathrm   2}$/$C^{\rm B}_{\mathrm 1}$  is larger at the higher energies and of
the order of 17\% for net protons without resonance decay and
net baryons, while the effect is 10\% for net-proton with resonance
decays and net-baryons.

\subsubsection{UrQMD Model}
The UrQMD (ultra relativistic quantum molecular dynamics)
model~\cite{Bass:1998ca,Bleicher:1999xi} is a microscopic transport model where the phase space
description of the reactions are considered. It treats the propagation
of all hadrons as classical trajectories in combination with
stochastic binary scattering, color string formation and resonance
decays. It incorporates baryon-baryon, meson-baryon and meson-meson
interactions. The collisional term includes more than 50 baryon
species and 45 meson species. The model preserves the conservation of
electric charge, baryon number, and strangeness number as expected for
QCD matter. It also models the phenomenon of baryon stopping, an
essential feature encountered in heavy-ion collisions at
lower beam energies. In this model, the space-time evolution of the fireball is studied in
terms of excitation and fragmentation of color strings and formation
and decay of hadronic resonances. Since the model does not include the physics of the quark-hadron phase
transition nor the QCD critical point, the comparison of the data to
the results obtained from the UrQMD model will shed light on the contributions from
the hadronic phase and its associated processes, baryon number
conservation and effect of measuring only net protons relative to
net baryons.

\begin{figure*}[htp]
\includegraphics[scale=0.9]{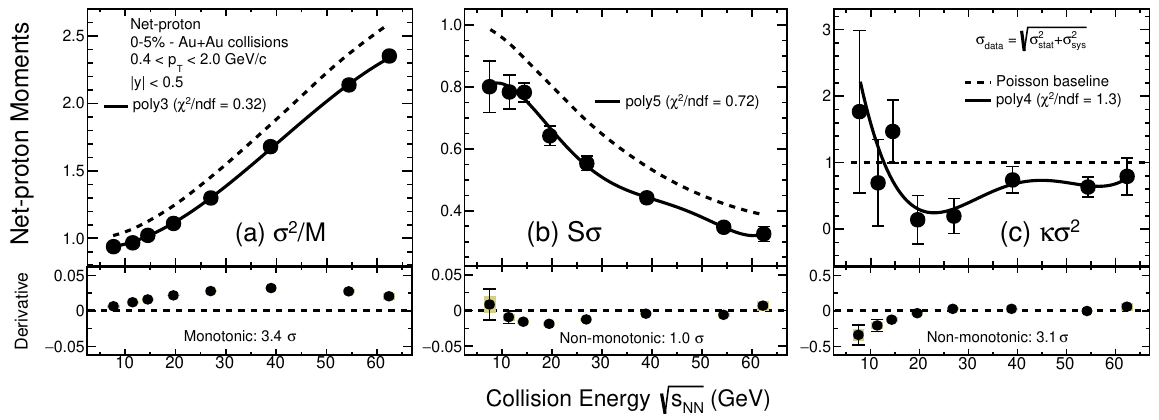}
\caption 
{(Color online) Upper panel: (a) $\sigma^{2}/M$,  (b) $S\sigma$ and (c) $\kappa\sigma^{2}$  of net-proton distributions for 0-5\% central Au+Au collisions from $\sqrt{s_\mathrm{NN}}$ = 7.7 - 62.4 GeV. The error bars on the data points are statistical and systematic uncertainties added in quadrature. The black solid lines are polynomial fit functions which well describe the cumulant ratios. The legends also specify the values of chi-squared per degree of freedom for the respective fits. The black dashed lines are the Poisson baselines. Lower panel: Derivative of the fitted polynomial as a function of collision energy.  The bar and the gold band on the derivatives represent the statistical and systematic uncertainties, respectively.}
\label{Fig:fitting_SDKV}
\end{figure*}
In Fig.~\ref{Fig:URQMD}, the panels on the left present the energy dependence of $C_{n}$ ratios of net-baryon distributions for various $p_{\mathrm T}$ acceptance. It is observed that the larger the $ p_T$ acceptance is, the smaller the cumulant ratios. 
Furthermore, with the same $ p_T$ acceptance, the values of net-baryon $C_{4}/C_2$ and $C_{2}/C_1$ ratios decrease with decreasing energies. Figure~\ref{Fig:URQMD} right panels show the comparison of the cumulant
ratios for net-baryon and net-proton distributions within the experimental
acceptance for various \snn. The differences between results from different acceptance are larger for UrQMD compared to the HRG model calculations with grand canonical ensemble. In UrQMD the difference between net baryons and
net protons is larger at the lower beam energies for a fixed
$p_{\mathrm T}$ and $y$ acceptance. The negative $C_4/C_2$ values of net-baryon distributions observed at low
energies could be mainly due to the effect of baryon number conservation. The effects of resonance weak decay and hadronic re-scattering on proton and net-proton number fluctuations 
in heavy-ion collisions have also been investigated in Ref.~\cite{Zhang:2019lqz} within the JAM (jet AA microscopic
transport) model. It is important to point out that in both the HRG model and UrQMD transport model calculations, a suppression in $C_4/C_2$ at low collision energy is observed, as is evident from the right plots of Fig.~\ref{Fig:hrg} and Fig.~\ref{Fig:URQMD}, respectively. In the case of the transport results, the suppression is attributed to the effect of baryon number conservation in strong interactions. However, the interpretation does not apply to the HRG calculation, since for the grand canonical ensemble (GCE), the event-by-event conservation is absent although, on average, the conservation law is preserved.  In addition to the law of conservation, quantum effects and the change of temperature and baryon chemical potential could play a role here. It is worth noting that the energy dependence of the suppression in $C_4/C_2$ depends on the details of modeling, especially on proton (baryon) rapidity distributions as they directly reflect the local baryon density. This effect is particularly important at lower energy region due to strong stopping in such collisions. Recently, Mohs, Ryu and Elfner reported rather different rapidity distributions for protons in Pb+Pb collisions around SPS energies, compared to those of UrQMD calculations. This is achieved by retuning parameters in string excitation and decay in the hadronic transport model SMASH~\cite{Mohs:2019iee}. In order to establish a non-critical baseline for the critical point search, more systematic theoretical studies of the higher-order cumulant as a function of collision energy with the reliable dynamical models are called for.

\subsubsection{Energy dependence}
Figure~\ref{Fig:fitting_SDKV} shows the collision-energy dependence of cumulant ratios (a) $\sigma^{2}/M$, (b) $S\sigma $ and  (c) $\kappa\sigma^{2}$ of net-proton distributions for 0-5\% central Au+Au collisions at $\sqrt{s_\mathrm{NN}}$  = 7.7 - 62.4 GeV. As shown in Fig.~\ref{Fig:fitting_SDKV}, a polynomial of order 4 (5) well describes the plotted collision-energy dependence of  $\kappa\sigma^{2}$ ($S\sigma$) of net-proton distributions for central Au+Au collisions with a $\chi^{2}$/ndf = 1.3(0.72).  The local derivative of the fitted polynomial function shown in the lower panel of Fig.~\ref{Fig:fitting_SDKV} changes sign, demonstrating the non-monotonic variation of the measurements with respect to collision energy. The statistical and
systematic uncertainties on derivatives are obtained by randomly varying the data points at each energy within their statistical and systematic uncertainties.

\begin{figure}[htp]
\hspace{-1cm}
\includegraphics[scale=0.85]{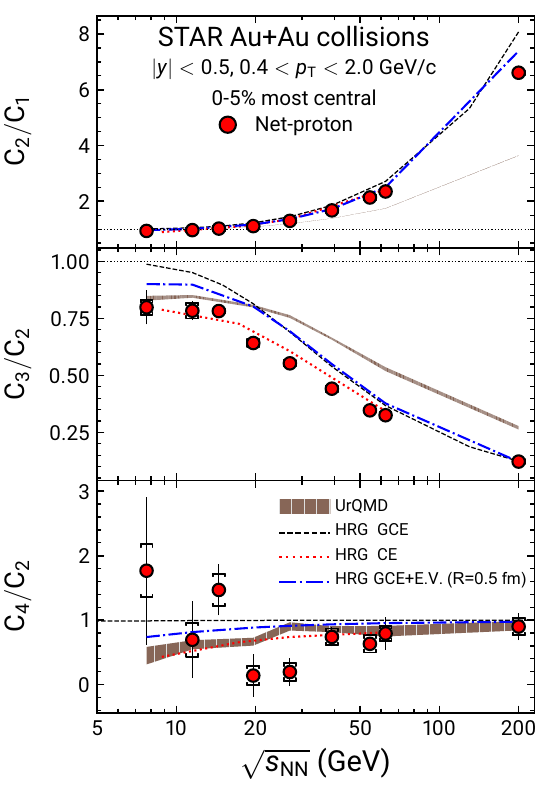}
\caption 
{(Color online) Collision energy dependence of $C_{2}$/$C_{1}$,
  $C_{3}$/$C_{2}$, and $C_{4}$/$C_{2}$ for net-proton multiplicity distributions in
  0-5\% central Au+Au collisions. The experimental net-proton measurements are compared to
  corresponding values from UrQMD and HRG models within the
  experimental acceptances. The bars and caps represent the statistical
and systematic uncertainties of the experimental data, respectively. The widths of
the bands reflect the statistical uncertainties for the model calculations. }
\label{Fig:data-model}
\end{figure}

\begin{table*} \label{tab:p_values}
\caption{The right-tail $p$ values of a chi-squared test between experimental data and various models (shown in Fig.~\ref{Fig:data-model}) for the energy dependence of the net-proton cumulant ratios in 0-5\% central Au+Au collisions at two ranges of collision energy: $\sqrt{s_{\rm {NN}}} = $ 7.7 -- 27 and 7.7 -- 62.4 GeV (the latter shown in the parentheses). Those $p$ values denote the probability of obtaining discrepancies at least as large as the results actually observed~\cite{Wasserstein2006}. The right-tail $p$ values are calculated via $p=\mathrm{Pr}(\chi^{2}_{n}>\chi^{2})$, where $\chi^{2}_{n}$ obeys the chi-square distribution with $n$ independent energy data points and the $\chi^{2}$ values are obtained in the chi-squared test.}
\vspace{0.2cm}
\begin{tabular}{ccccc}
\hline
Cumulant ratios & HRG GCE         & HRG CE                & HRG GCE+E.V. (R=0.5 fm) & UrQMD           \\ \hline
$C_2/C_1$       & \textless 0.001(\textless{}0.001)   & \textless 0.001(\textless{}0.001)           & \textless 0.001(\textless{}0.001)      & \textless 0.001(\textless{}0.001) \\ \hline
$C_3/C_2$       & \textless 0.001(\textless{}0.001)   & 0.0754 (\textless{}0.001) & \textless 0.001(\textless{}0.001)      & \textless 0.001(\textless{}0.001) \\ \hline
$C_4/C_2$       & 0.00553 (0.00174) & 0.0450 (0.128)            & 0.0145 (0.0107)      & 0.0221 (0.0577) \\ \hline
\end{tabular}
\end{table*}
The significance of the observed non-monotonic dependence of $\kappa\sigma^{2}$ ($S\sigma$) on collision energy, in the energy range $\sqrt{s_\mathrm{NN}}$  = 7.7 - 62.4 GeV, is obtained based on the fourth (fifth) order polynomial fitting procedure. This significance is evaluated by randomly varying the $\kappa\sigma^{2}$ and $S\sigma$ data points within their total Gaussian uncertainties (statistical and systematic uncertainties added in quadrature) at each corresponding energy.  This procedure is repeated $10^{6}$ times for $\kappa\sigma^{2}$ and for $S\sigma$. Out of $10^{6}$ trials, there are 1143 cases for $\kappa\sigma^{2}$ and 158640 cases for $S\sigma$ where the signs of the derivative at all $\sqrt{s_\mathrm{NN}}$ are found to be the same. Thus, the probability that at least one derivative at a given $\sqrt{s_\mathrm{NN}}$  has a different sign from the derivatives at remaining energies among the $10^{6}$ trials performed is 0.99886 (0.84136), which corresponds to a 3.1 $\sigma$ (1.0 $\sigma$) effect for $\kappa\sigma^{2}$ ($S\sigma$). Similarly, based on the third-order polynomial fitting procedure, the cumulant ratio  $\sigma^{2}/M$ on the other hand ($\chi^{2}$/ndf = 0.32), exhibits a monotonic dependence on collision energy with a significance of 3.4$\sigma$. Thus we find that the cumulant ratios as a function of collision energy change from a monotonic variation to a non-monotonic variation with \snn\ as we go to higher orders. This is consistent with the QCD-based model expectation that, the higher the order of the moments is, the more sensitive it is to physics processes such as a critical point~\cite{Stephanov:2008qz,Stephanov:2011pb}. A test of the non-monotonicity energy dependence with $\kappa\sigma^{2}$ is also carried out with the energy range $\sqrt{s_\mathrm{NN}}$ = 7.7 – 200 GeV and the resulting significance is 3.0 $\sigma$.

 \begin{figure*}[htpb]
\includegraphics[scale=0.5]{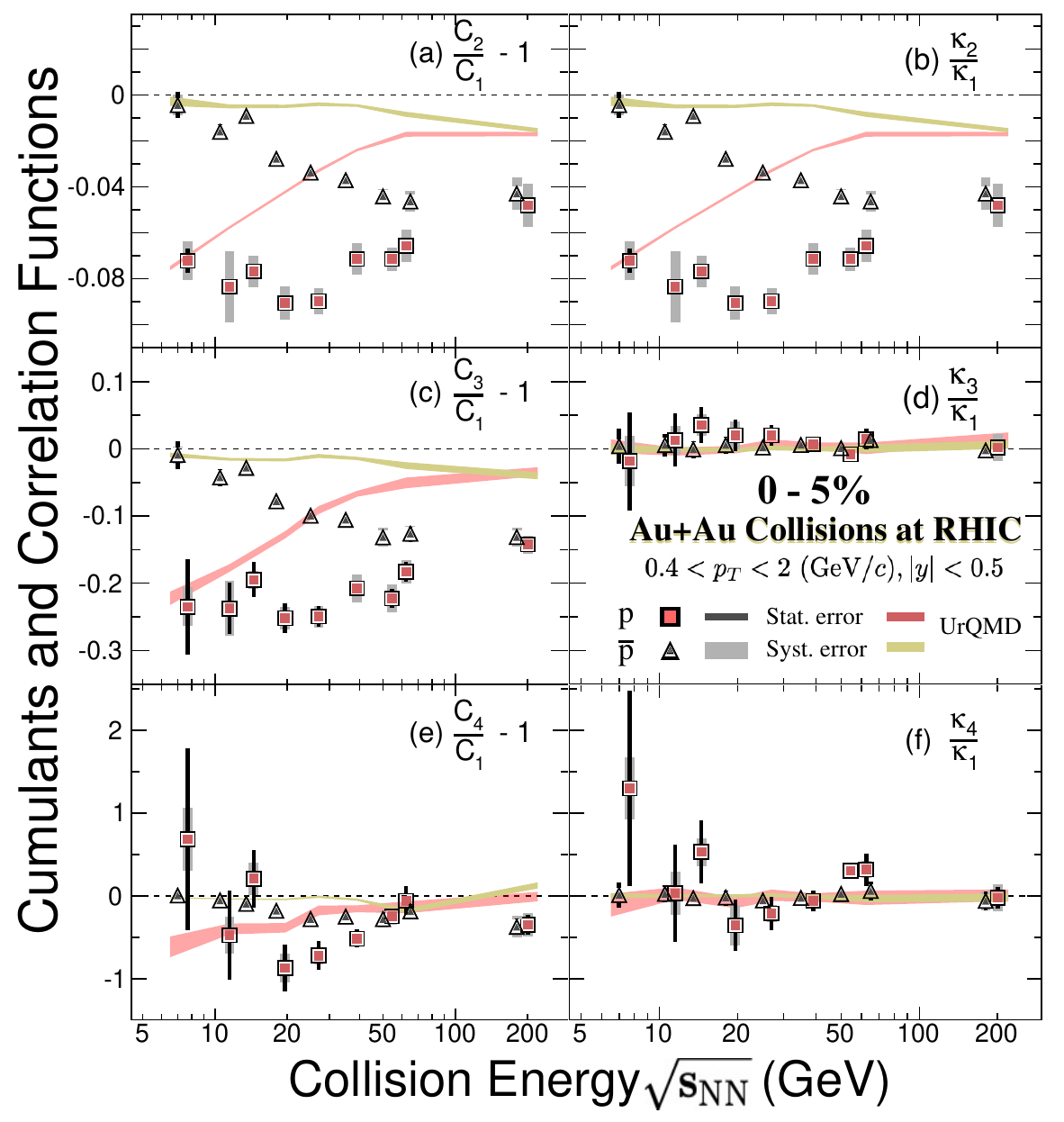}
\caption{(Color online) Collision energy dependence of the scaled (anti)proton cumulants and correlation functions in 0-5\% central
Au+Au collisions at \snn\ = 7.7, 11.5, 14.5, 19.6, 27,
  39, 54.4, 62.4, and 200 GeV. The error bars and bands represent the statistical
and systematic uncertainties, respectively. The results from UrQMD model calculation are also shown for comparison.}
\label{Corr-func-decom}
\end{figure*}

Figure~\ref{Fig:data-model} shows the collision-energy dependence of the cumulant ratios of net-proton multiplicity distributions for 0-5\% central Au+Au collisions.
The comparison has been made between experimental measurements and the
corresponding results from the HRG and UrQMD models. We observe that both models, which do not have phase transition effects, show monotonic variations of the
cumulant ratios with beam energy. However, the experimental
measurements of net-proton $C_{4}$/$C_{2}$ ratios show a
non-monotonic variation with \snn. On the other hand, the net-proton $C_{3}$/$C_{2}$ ($C_{2}$/$C_{1}$) in both
model and data show a smooth decrease (increase) trend with increasing \snn. Although both models show a smooth energy dependence, the third-order ratios in the middle panel are larger for UrQMD than that for (GCE) HRG at collision energies above 14.5 GeV. At lower energy, a suppression relative to the results of GCE  HRG is observed. On the other hand, the canonical ensemble (CE) HRG, presents a consistent suppression in all three panels. In this approach, the baryon number conservation is the main source of the suppression~\cite{Fu:2016baf,Braun-Munzinger:2020jbk}. It is interesting to point out that GCE models incorporating excluded volume effects (GCE E.V.) can also reproduce the suppression. The larger the repulsive volume, the stronger the suppression. Since the repulsive volume reflects the “baryon density”, the observed suppression GCE E.V. is due to the local density. For details, see Refs.~\cite{Fu:2013gga,Bhattacharyya:2013oya,Samanta:2019fef}. To quantify the level of agreement between the experimental measurements and the model calculations, the widely used $\chi^{2}$ test has been applied for two energy ranges (\snn\ = 7.7 -- 27 and 7.7 -- 62.4 GeV). The $\chi^{2}$ value is calculated as $\chi^{2}(R)=\sum_{\sqrt{s_\mathrm{NN}}}^{}\frac{\left | {R_{\rm data}-R_{\rm model}} \right |^{2}}{\mathrm{error}^{2}}$, where $R$ denotes the cumulant ratios ($C_2/C_1,~~C_3/C_2,~~C_4/C_2$) and the ‘error’ represents the statistical and systematic uncertainties of the data and the statistical uncertainties of the model added in quadrature. In addition, the obtained $\chi^{2}$ value can be converted to the corresponding right-tail $p$-value, which is the probability of obtaining discrepancies at least as large as the results actually observed~\cite{Wasserstein2006}. The resulting right tail $p$-values listed in Table~\ref{tab:p_values} are calculated via $p=\mathrm{Pr}(\chi^{2}_{n}>\chi^{2})$, where $\chi^{2}_{n}$ obeys the chi-square distribution with $n$ independent energy data points and the $\chi^{2}$ values are obtained in the chi-squared test. Usually, for the right tail $p$-value test, $p<0.05$ is the commonly used standard to reject the null hypothesis and claim a significant deviation between the data and model results. It is found that the $p$-values from the the $\chi^{2}$ test are smaller than 0.05 for all of the different variants of HRG and the UrQMD model at \snn\ = 7.7 -- 27 GeV, which means the deviations between data and model results are  
significant and cannot be explained by statistical fluctuations.  But, for the range \snn\  = 7.7 -- 62.4 GeV, the $p$-values of $C_4/C_2$ for the HRG CE and UrQMD model cases are 0.128 and 0.0577, respectively. Clearly as far as these tests are concerned, all of the above-mentioned models, showing monotonic energy dependences, do not fit the data in the most relevant energy region, \snn\ $\le$ 27 GeV. This result will be further tested with the high-precision data from RHIC BES-II program.

Based on Eq.~(\ref{eq:CKtoCF}), the cumulants can be expressed in terms of the sum of various-order multiparticle correlation functions.
In order to understand the contributions to the cumulants, one can present different orders of correlation functions separately. Figure~\ref{Corr-func-decom} shows the energy dependence of the cumulants and correlation functions normalized by the mean numbers of protons and antiprotons in 0-5\% central Au+Au collisions. By definition and as shown in Fig.~\ref{Corr-func-decom}, the values of $C_2/C_1-1$ are equal to $\kappa_2/\kappa_1$. It is observed that the normalized second and third-order cumulants minus unity ($C_2/C_1-1$, $C_3/C_1-1$) are negative and show an increasing (decreasing) energy dependence in magnitude for protons (antiprotons) with decreasing collision energies. From the right panels in Fig.~\ref{Corr-func-decom},  the third-order normalized correlation functions ($\kappa_3/\kappa_1$) of protons and antiprotons show flat energy dependence and are consistent with zero within uncertainties. Therefore, the energy dependence for $C_3/C_1$ is dominated by the negative two-particle normalized correlation functions ($\kappa_2/\kappa_1$), which is mainly due to the effects of baryon number conservation. The normalized four-particle correlation functions ($\kappa_4/\kappa_1$) of antiprotons show flat energy dependence and are consistent with zero within uncertainties. In panel (e) of Fig.~\ref{Corr-func-decom},  we observe a similar energy dependence trend for the normalized fourth-order cumulants ($C_4/C_1$) of protons as for the net-proton $C_4/C_2$ in 0-5\% central Au+Au collisions shown in Fig.~\ref{Fig:data-model}.  For \snn\ $\ge$ 19.6 GeV, the values of proton $C_4/C_1$ are dominated by the negative two-particle correlation function ($\kappa_2$) of protons (see panel (b) in Fig.~\ref{Corr-func-decom}). For \snn\ $<$ 19.6 GeV, the four-particle correlation function ($\kappa_4$) of protons plays a role in determining the energy dependence of proton $C_4/C_1$, which cannot be solely understood by the suppression effects due to negative values of $\kappa_2$ for protons.  As discussed in Refs.~\cite{Ling:2015yau,Bzdak:2016jxo}, the observed large values of the four-particle correlation function of protons ($\kappa_4$) could be attributed to the formation of proton cluster and related to the signature of a critical point or a first order phase transition. Therefore, it is necessary to perform precise measurements of the $\kappa_4/\kappa_1$ of protons below 19.6 GeV with high statistics data taken in the second phase of the beam energy scan at RHIC. In addition, we compare the experimental data in Fig.~\ref{Corr-func-decom} with UrQMD model calculations. The energy dependence of the second- and third-order normalized cumulants and correlation functions can be qualitatively described by the UrQMD model. However,  the non-monotonic energy dependence observed in the proton $C_4/C_1$ cannot be described by the UrQMD model. Furthermore, the three- and four-particle correlation functions ($\kappa_3$ and $\kappa_4$) for (anti)protons from UrQMD show flat energy dependence and are consistent with zero. This indicates that the higher-order (anti)proton correlation functions $\kappa_3$ and $\kappa_4$ are not sensitive to the effect of baryon number conservation within the current acceptance, and therefore can serve as good probes of critical fluctuations in heavy-ion collisions~\cite{He:2017zpg,Zhang:2019lqz}.  

\section{Summary and Outlook} 
In summary, we report a systematic study of the cumulants of the net-proton, proton, and antiproton multiplicity distributions from Au+Au collisions at $\sqrt{s_\mathrm{NN}}$ = 7.7 - 200 GeV. The data have been collected with the STAR experiment in the first phase of the RHIC beam energy scan acquired over the period of 2010 - 2017. The energy, centrality, and acceptance dependence of the correlation functions of protons and antiprotons are presented in this paper. Both cumulants and correlation functions up to fourth order at midrapidity ($|y|$$<$ 0.5) within 0.4 $<$ $p_{\mathrm T}$ $<$ 2.0 GeV/$c$ in Au+Au collisions are presented to search for the signatures of a critical point and/or a first-order phase transition over a broad region of baryon chemical potential. 

The protons and antiprotons are identified with greater than 97\%
purity using the TPC and TOF detectors of STAR. The centrality
selection is based on midrapidity pions and kaons only to avoid
self-correlation effects. The maximum-allowed rapidity acceptance around midrapidity
has been used for centrality determination to minimize the effect of
centrality resolution. The variation of the average number of protons and
antiprotons in a given centrality bin has been accounted for by applying a 
centrality bin-width correction, which also minimizes volume
fluctuation effects. The cumulants are corrected for the proton and
antiproton reconstruction efficiencies using a binomial
response function. Study of the unfolding technique for efficiency correction of cumulants has shown that, even in the 0-5\% central Au+Au collisions at \snn\ = 200 GeV, the case with the highest multiplicity, the results are consistent with the commonly-used binomial approach within current statistical uncertainties. The statistical errors on the
cumulants are based on the delta theorem method and are shown to be
consistent with those obtained by the bootstrap method. A detailed estimate
of the systematic uncertainties is also presented. Results on
cumulant ratios from different variants of the HRG and the UrQMD models are
presented to understand the effects of experimental acceptance, resonance decay, baryon number conservation, and net-proton versus net-baryon analysis. 
The cumulant ratios show a centrality and energy dependence, which are reproduced neither by purely hadronic-transport-based UrQMD model calculations nor by different variants of the hadron resonance gas model.
Specifically, the net-proton $C_{4}$/$C_{2}$ ratio for 0-5\%
central Au+Au collisions shows a non-monotonic variation with {\snn}, with a significance of 3.1$\sigma$. This is consistent with the expectations 
of critical fluctuations in a QCD-inspired model.  A $\chi^{2}$ test has been applied to quantify the level of agreement between experimental data and model calculations. The 
resulting $p$-values suggest that the models fail to explain the 0-5\% Au+Au collision data at \snn\ $\le$ 27 GeV. The $y$ and $p_{\mathrm T}$ acceptance dependence of the cumulants and their ratios
provide valuable data to understand the range of the correlations and their relation to the acceptance of the detector~\cite{Ling:2015yau,Brewer:2018abr}. Furthermore, the systematic analysis presented here can be used to constrain the freeze-out conditions in high-energy heavy-ion collisions using QCD-based approaches, and to understand the nature of thermalization in such collisions~\cite{Bazavov:2012vg,Borsanyi:2013hza,Gupta:2020pjd}. From the analysis of multiparticle correlation functions, one observes significant negative values for $\kappa_2$ of protons and antiprotons, which are mainly due to the effects of baryon number conservation in heavy-ion collisions.  The values of $\kappa_3$ of protons and antiprotons are consistent with zero for all of the collision energies studied. Further,  the energy dependence trend of proton $C_{4}$/$C_{1}$ below 19.6 GeV cannot be solely understood by the negative values of $\kappa_2$ for protons, and the four-particle correlation function of protons ($\kappa_4$) is found to play a role, which needs to be confirmed with the high statistics data taken in RHIC BES-II, which began data-taking in 2018. Upgrades to the STAR detector system have significantly improved the quality of the
measurements~\cite{bes2}.  Primarily the goal of BES-II is to make high-statistics
measurements, with extended kinematic range in rapidity and transverse
momentum for the measurements discussed in this paper. The extended
kinematic range in rapidity and transverse momentum are brought about
by upgrading the inner TPC (iTPC) to extend the measurement coverage to $|\eta| <$
1.5, the $p_{\mathrm T}$ acceptance down to 100 MeV/$c$ and improved
$dE/dx$ resolution. Particle identification capability will be extended to -1.6 $<
\eta <$ 1.0 with the addition of an endcap TOF (eTOF) detector. The collected event statistics to date, along with the goal for 2021, are listed in Table~\ref{table_bes2}.

\begin{table}[htp]
\caption{Total number of collected/expected events in BES Phase II for various collision 
  energies ($\sqrt{s_\mathrm{NN}}$)~\cite{bes2}. 
\label{table_bes2}}
\begin{center}
\begin{tabular}{cccc}
\hline 
$\sqrt{s_\mathrm{NN}}$ (GeV) & Year & No. of events ($\times 10^{6}$) & \\ 
\hline 
27     &    2018      &  500   & \\ 
19.6    &    2019      &  400    &  \\ 
17.3    &    2021      &  250    &  \\ 
14.5    &    2019  &  300      &         \\
11.5       &    2020   &  230    &  \\
9.2      &    2020    &  160     &   \\
7.7    &    2021   & 100    &  \\
\hline 
\end{tabular}
\end{center}
\end{table}
At the same time, STAR will take data in fixed-target mode to extend $\sqrt{s_\mathrm{NN}}$ to 3 GeV. With these upgrades, and with the benefits of extended kinematic coverage and the use of sensitive observables, the RHIC BES Phase-II program will allow measurements of unprecedented precision for exploring the QCD phase structure within $200 < \mu_{B} < 720$ MeV.

\section*{Acknowledgments}
We thank H. Elfner, S. Gupta, F. Karsch, M. Kitazawa, V. Koch, D. Mishra, J. M. Pawlowski, K. Rajagopal, K. Redlich,
and M. Stephanov for stimulating discussions related to this work.
We thank the RHIC Operations Group and RCF at BNL, the NERSC Center at LBNL, and the Open Science Grid consortium for providing resources and support.  This work was supported in part by the Office of Nuclear Physics within the U.S. DOE Office of Science, the U.S. National Science Foundation, the Ministry of Education and Science of the Russian Federation, National Natural Science Foundation of China, Chinese Academy of Science, the Ministry of Science and Technology of China and the Chinese Ministry of Education, the Higher Education Sprout Project by Ministry of Education at NCKU, the National Research Foundation of Korea, Czech Science Foundation and Ministry of Education, Youth and Sports of the Czech Republic, Hungarian National Research, Development and Innovation Office, New National Excellency Programme of the Hungarian Ministry of Human Capacities, Department of Atomic Energy and Department of Science and Technology of the Government of India, the National Science Centre of Poland, the Ministry  of Science, Education and Sports of the Republic of Croatia, RosAtom of Russia, German Bundesministerium fur Bildung, Wissenschaft, Forschung and Technologie (BMBF), Helmholtz Association, Ministry of Education, Culture, Sports, Science, and Technology (MEXT) and Japan Society for the Promotion of Science (JSPS).

\bibliography{ref}
\newpage
\appendix
\section{Efficiency Correction}  \label{appendix-1}
In order to correct the $C_{n}$ for efficiency effects,
one has to invoke a model assumption for the response of the
detector. The detector response is assumed to 
follow a binomial probability distribution function. The probability
distribution function of measured proton number $n_{p}$ and
antiproton number $n_{\bar{p}}$ can be expressed as~\cite{Bzdak:2012ab,Luo:2014rea}:
\begin{widetext}
\begin{equation}  \label{eq:conv} 
\begin{split}
 p({n_p},{n_{\bar p}}) &= \sum\limits_{{N_p} = n_p}^\infty  {\sum\limits_{{N_{\bar p}} = n_{\bar p}}^\infty  {P({N_p},{N_{\bar p}}) \times \frac{{{N_p}!}}{{{n_p}!\left( {{N_p} - {n_p}} \right)!}}{{({\varepsilon _p})}^{{n_p}}}{{(1 - {\varepsilon _p})}^{{N_p} - {n_p}}}} } \\
& \times  \frac{{{N_{\bar p}}!}}{{{n_{\bar p}}!\left( {{N_{\bar p}} - {n_{\bar p}}} \right)!}}{({\varepsilon _{\bar p}})^{{n_{\bar p}}}}{(1 - {\varepsilon _{\bar p}})^{{N_{\bar p}} - {n_{\bar p}}}} 
\end{split}
\end{equation}
\end{widetext}
where the $P({N_p},{N_{\bar p}})$ is the original joint probability
distribution of numbers of protons ($N_p$) and antiprotons ($N_{\bar
  p}$), and $\varepsilon _p$, $\varepsilon _{\bar p}$ are the
efficiency of reconstructing the protons and antiprotons, respectively. 
In order to arrive at an expression for efficiency-corrected
cumulants or moments, the bivariate factorial moments are first defined as:
\begin{widetext}
\begin{align}
& {F_{i,k}(N_p, N_{\bar p})} = \left \langle \frac{{{N_p}!}}{{\left( {{N_p} - i} \right)!}}\frac{{{N_{\bar p}}!}}{{\left( {{N_{\bar p}} - k} \right)!}}\right \rangle = \sum\limits_{{N_p} = i}^\infty  {\sum\limits_{{N_{\bar p}} = k}^\infty  {P({N_p},{N_{\bar p}})\frac{{{N_p}!}}{{\left( {{N_p} - i} \right)!}}\frac{{{N_{\bar p}}!}}{{\left( {{N_{\bar p}} - k} \right)!}}} }   \label{eq:fact1} \\ 
&  {f_{i,k}(n_p, n_{\bar p})} =\left \langle \frac{{{n_p}!}}{{\left( {{n_p} - i} \right)!}}\frac{{{n_{\bar p}}!}}{{\left( {{n_{\bar p}} - k} \right)!}}\right\rangle  = \sum\limits_{{n_p} = i}^\infty  {\sum\limits_{{n_{\bar p}} = k}^\infty  {p({n_p},{n_{\bar p}})\frac{{{n_p}!}}{{\left( {{n_p} - i} \right)!}}\frac{{{n_{\bar p}}!}}{{\left( {{n_{\bar p}} - k} \right)!}}} }  \label{eq:fact2}
\end{align}
\end{widetext}
The efficiency-corrected factorial moments are then given as:
\begin{equation}  \label{eq:relation} 
{F_{i,k}(N_p, N_{\bar p})} = \frac{{{f_{i,k}(n_p, n_{\bar p})}}}{{{{({\varepsilon _p})}^i}{{({\varepsilon _{\bar p}})}^k}}}.
\end{equation}
Then the $n$th order efficiency-corrected
moments of net-proton distributions are related to the efficiency-corrected factorial moments as:
\begin{widetext}
\begin{equation} \label{eq:mtof2}
\begin{array}{l}
{m_n}({N_p} - {N_{\bar p}}) = <{({N_p} - {N_{\bar p}})^n}> = \sum\limits_{i = 0}^n {{{( - 1)}^i}\left( {\begin{array}{*{20}{c}}
n\\
i 
\end{array}} \right)} <N_p^{n - i}N_{\bar p}^i>\\
 = \sum\limits_{i = 0}^n {{{( - 1)}^i}\left( {\begin{array}{*{20}{c}}
n\\
i 
\end{array}} \right)} \left[ {\sum\limits_{{r_1} = 0}^{n - i} {\sum\limits_{{r_2} = 0}^i {{s_2}(n - i,{r_1}){s_2}(i,{r_2}){F_{{r_1},{r_2}}}({N_p},{N_{\bar p}})} } } \right]\\
 = \sum\limits_{i = 0}^n {\sum\limits_{{r_1} = 0}^{n - i} {\sum\limits_{{r_2} = 0}^i {{{( - 1)}^i}\left( {\begin{array}{*{20}{c}}
n\\
i 
\end{array}} \right){s_2}(n - i,{r_1}){s_2}(i,{r_2}){F_{{r_1},{r_2}}}({N_p},{N_{\bar p}})} } } 
\end{array}
\end{equation}
\end{widetext}
The Stirling numbers of the first [$s_{1}(n,i)$] and second kind [$s_{2}(n,i)$], are defined as: 
 \begin{align}
&\frac{{N!}}{{(N - n)!}} = \sum\limits_{i = 0}^n {{s_1}(n,i)} {N^i}\\
&{N^n} = \sum\limits_{i = 0}^n {{s_2}(n,i)} \frac{{N!}}{{(N - i)!}}
\end{align}  
where $N$, $n$, and $i$ are non-negative integer numbers.
The efficiency-corrected cumulants of net-proton distributions can be
obtained from the efficiency-corrected moments by using the recursion relation:
\begin{equation} \label{eq:cumulants}
\begin{split}
&{C _r}({N_p} - {N_{\bar p}}) = {m_r}({N_p} - {N_{\bar p}}) \\
 &- \sum\limits_{s = 1}^{r - 1} {\left( \begin{array}{c}
r - 1\\
s - 1 
\end{array} \right)} {C _s}({N_p} - {N_{\bar p}}){m_{r - s}}({N_p} - {N_{\bar p}}) 
\end{split}
\end{equation}
where the $C_r$ denotes the $r$th-order cumulants of net-proton distributions.
 
If the protons and antiprotons have the same efficiency, $\varepsilon_p=\varepsilon_{\bar{p}}=\varepsilon$, the expressions for the first four efficiency-corrected cumulants can be explicitly written as:
\begin{widetext}
\begin{equation} \label{eq:effcorrPRL}
\begin{split}
C_1^{X-Y}&=\frac{\la x\ra -\la y\ra}{\varepsilon} \\
C_2^{X - Y} &= \frac{{C_2^{x - y} + (\varepsilon  - 1)( \la x \ra  +  \la y \ra )}}{{{\varepsilon ^2}}}\\
C_3^{X - Y} &= \frac{{C_3^{x - y} + 3(\varepsilon  - 1)(C_2^x - C_2^y) + (\varepsilon  - 1)(\varepsilon  - 2)( \la x \ra  -  \la y \ra )}}{{{\varepsilon ^3}}}\\
C_4^{X - Y} &= \frac{{C_4^{x - y} - 2(\varepsilon  - 1)C_3^{x + y} + 8(\varepsilon  - 1)(C_3^x + C_3^y) + (5 - \varepsilon )(\varepsilon  - 1)C_2^{x + y}}}{{{\varepsilon ^4}}}\\
 &+ \frac{{8(\varepsilon  - 1)(\varepsilon  - 2)(C_2^x + C_2^y) + ({\varepsilon ^2} - 6\varepsilon  + 6)(\varepsilon  - 1)( \la x \ra  +  \la y \ra )}}{{{\varepsilon ^4}}}
\end{split}
\end{equation}
\end{widetext}
where the $(X,Y)$ and $(x,y)$ are the numbers of $(p, \bar{p})$ produced and measured, respectively. 
The efficiency-corrected cumulants are sensitive to the efficiency and
depend on the lower order measured cumulants. 

In the current analysis, the proton and antiproton $p_{\mathrm T}$
range is from 0.4 to 2 GeV/$c$. This has been possible by using particle
identification information for the TPC in the $p_{\mathrm T}$ range 0.4 to
0.8 GeV/$c$ and the TPC+TOF in the momentum range 0.8 to 2 GeV/$c$. This
results in two different efficiencies for proton reconstruction and
two different values for antiprotons. Hence the above formulation
which holds for one single value of efficiency and
$\varepsilon=\varepsilon_p=\varepsilon_{\bar{p}}$ has to be modified
to take care of four different efficiency values, two each for the
proton and antiproton corresponding to different $p_{\mathrm T}$
ranges. Let $\varepsilon _{{p_1}}^{},\varepsilon _{{p_2}}^{}$
and $\varepsilon _{{{\bar p}_1}}^{},\varepsilon _{{{\bar p}_2}}^{}$
denote the efficiency for protons and antiprotons in the two sub-phase
spaces, and denote the corresponding numbers of protons and antiprotons in the
two sub-phase spaces by $N_{p_1}$, $N_{p_2}$ and $N_{\bar{p}_1}$,
$N_{\bar{p}_2}$, respectively. Using analogous formulations as above,
the bivariate factorial moments of protons and antiprotons
distributions are given as:
\begin{widetext}
\begin{equation} \label{eq:4Fto2F}
\begin{split}
{F_{{r_1},{r_2}}}({N_p},{N_{\bar p}}) &= {F_{{r_1},{r_2}}}({N_{{p_1}}} + {N_{{p_2}}},{N_{{{\bar p}_1}}} + {N_{{{\bar p}_2}}})= \sum\limits_{{i_1} = 0}^{{r_1}} {\sum\limits_{{i_2} = 0}^{{r_2}} {{s_1}({r_1},{i_1})} } {s_1}({r_2},{i_2})\la {({N_{{p_1}}} + {N_{{p_2}}})^{{i_1}}}{({N_{{{\bar p}_1}}} + {N_{{{\bar p}_2}}})^{{i_2}}} \ra\\
& = \sum\limits_{{i_1} = 0}^{{r_1}} {\sum\limits_{{i_2} = 0}^{{r_2}} {{s_1}({r_1},{i_1})} } {s_1}({r_2},{i_2})\la {\sum\limits_{s = 0}^{{i_1}} {\left( {\begin{array}{*{20}{c}}
{{i_1}}\\
s 
\end{array}} \right)N_{{p_1}}^{{i_1} - s}N_{{p_2}}^s\sum\limits_{t = 0}^{{i_2}} {\left( {\begin{array}{*{20}{c}}
{{i_2}}\\
t 
\end{array}} \right)N_{{{\bar p}_1}}^{{i_2} - t}N_{{{\bar p}_2}}^t} } } \ra \\
& = \sum\limits_{{i_1} = 0}^{{r_1}} {\sum\limits_{{i_2} = 0}^{{r_2}} {\sum\limits_{s = 0}^{{i_1}} {\sum\limits_{t = 0}^{{i_2}} {{s_1}({r_1},{i_1}){s_1}({r_2},{i_2})\left( {\begin{array}{*{20}{c}}
{{i_1}}\\
s 
\end{array}} \right)\left( {\begin{array}{*{20}{c}}
{{i_2}}\\
t 
\end{array}} \right)} } } } \la N_{{p_1}}^{{i_1} - s}N_{{p_2}}^sN_{{{\bar p}_1}}^{{i_2} - t}N_{{{\bar p}_2}}^t \ra \\
&= \sum\limits_{{i_1} = 0}^{{r_1}} {\sum\limits_{{i_2} = 0}^{{r_2}} {\sum\limits_{s = 0}^{{i_1}} {\sum\limits_{t = 0}^{{i_2}} {\sum\limits_{u = 0}^{{i_1} - s} {\sum\limits_{v = 0}^s {\sum\limits_{j = 0}^{{i_2} - t} {\sum\limits_{k = 0}^t {{s_1}({r_1},{i_1}){s_1}({r_2},{i_2})\left( {\begin{array}{*{20}{c}}
{{i_1}}\\
s 
\end{array}} \right)\left( {\begin{array}{*{20}{c}}
{{i_2}}\\
t 
\end{array}} \right)} } } } } } } } \\
 &\times {s_2}({i_1} - s,u){s_2}(s,v){s_2}({i_2} - t,j){s_2}(t,k) \times {F_{u,v,j,k}}(N_{{p_1}}^{},N_{{p_2}}^{},N_{{{\bar p}_1}}^{},N_{{{\bar p}_2}}^{}) 
\end{split}
\end{equation}
\end{widetext}
Similarly to Eq.~(\ref{eq:relation}) for the multivariate case, the
efficiency-corrected multivariate factorial moments of proton and
antiproton distributions in the current case are given as:
 \begin{equation} \label{eq:relation2}
{F_{u,v,j,k}}(N_{{p_1}}^{},N_{{p_2}}^{},N_{{{\bar p}_1}}^{},N_{{{\bar p}_2}}^{}) = \frac{{{f_{u,v,j,k}}(n_{{p_1}}^{},n_{{p_2}}^{},n_{{{\bar p}_1}}^{},n_{{{\bar p}_2}}^{})}}{{{{({\varepsilon _{{p_1}}})}^u}{{({\varepsilon _{{p_2}}})}^v}{{({\varepsilon _{{{\bar p}_1}}})}^j}{{({\varepsilon _{{{\bar p}_2}}})}^k}}}\end{equation}
where  ${{f_{u,v,j,k}}(N_{{p_1}}^{},N_{{p_2}}^{},N_{{{\bar p}_1}}^{},N_{{{\bar p}_2}}^{})}$ are the measured multivariate factorial moments of proton and antiproton distributions. 
By using  Eq. (\ref{eq:mtof2}), (\ref{eq:cumulants}),
(\ref{eq:4Fto2F}) and (\ref{eq:relation2}), one can obtain the
efficiency-corrected moments and cumulants of net-proton distributions for the case where 
the protons (antiprotons) have different efficiencies in two
sub-phase spaces. Through simulations as discussed in Refs.~\cite{Luo:2014rea,Nonaka:2016xje}, it has
been shown that this formulation works consistently. Another binomial-model-based efficiency correction method using track-by-track efficiency is discussed in Ref.~\cite{Luo:2018ofd}.

\section{Statistical Uncertainties Estimation} \label{appendix-2}
According to Eqs.~(\ref{eq:mtof2}), (\ref{eq:cumulants}) and~(\ref{eq:4Fto2F}), the efficiency-corrected moments are expressed in terms of the factorial moments, and thereby the factorial moments are the random variable $X_i$ in Eq.  (\ref{eq:error}).  The covariance of the multivariate moments can be written as:
\begin{equation} \label{eq:covm}{\rm Cov}({m_{r,s}},{m_{u,v}}) = \frac{1}{n}({m_{r + u,s + v}} - {m_{r,s}}{m_{u,v}}) \end{equation}
where $n$ is the number of events, $m_{r,s}=\langle X_1^{r}X_2^{s} \rangle$ and ${m_{u,v}}=\langle X_1^{u}X_2^{v}\rangle$ are the multivariate moments, and the $X_1$ and $X_2$ are random variables. In this paper, $X_1$ and $X_2$ represent proton and antiproton numbers, respectively. 
Based on Eq.~(\ref{eq:covm}), one can obtain the 
covariance for the multivariate factorial moments as:
\begin{widetext}
\begin{equation} \label{eq:covF}
\begin{split}
&{\rm Cov}({f_{r,s}},{f_{u,v}}) = {\rm Cov}\left(\sum\limits_{i = 0}^r {\sum\limits_{j = 0}^s {{s_1}(r,i){s_1}(s,j){m_{i,j}},} } \sum\limits_{k = 0}^u {\sum\limits_{h = 0}^v {{s_1}(u,k){s_1}(v,h){m_{k,h}}} } \right)\\
 &= \sum\limits_{i = 0}^r {\sum\limits_{j = 0}^s {\sum\limits_{k = 0}^u {\sum\limits_{h = 0}^v {{s_1}(r,i){s_1}(s,j){s_1}(u,k){s_1}(v,h)} }  \times {\rm Cov}({m_{i,j}},{m_{k,h}})} } \\
 &= \frac{1}{n}\sum\limits_{i = 0}^r {\sum\limits_{j = 0}^s {\sum\limits_{k = 0}^u {\sum\limits_{h = 0}^v {{s_1}(r,i){s_1}(s,j){s_1}(u,k){s_1}(v,h)} }  \times } } ({m_{i + k,j + h}} - {m_{i,j}}{m_{k,h}})\\
&= \frac{1}{n}({f_{(r,u),(s,v)}} - {f_{r,s}}{f_{u,v}})
\end{split}
\end{equation}
\end{widetext}
where the $f_{(r,u),(s,v)}$ is defined as:
\begin{equation}
\begin{array}{l} \label{eq:fact4}
{f_{(r,u),(s,v)}} = \left\langle {\frac{{{X_1}!}}{{({X_1} - r)!}}\frac{{{X_1}!}}{{({X_1} - u)!}}\frac{{{X_2}!}}{{({X_2} - s)!}}\frac{{{X_2}!}}{{({X_2} - v)!}}} \right\rangle \\
 = \sum\limits_{i = 0}^r {\sum\limits_{j = 0}^s {\sum\limits_{k = 0}^u {\sum\limits_{h = 0}^v {\sum\limits_{\alpha  = 0}^{i + k} {\sum\limits_{\beta  = 0}^{j + h} {{s_1}(r,i){s_1}(s,j){s_1}(u,k){s_1}(v,h)} } } } } } \\
 \times {s_2}(i + k,\alpha ){s_2}(j + h,\beta ){f_{\alpha ,\beta }}
\end{array}
\end{equation}
The definition of the bivariate factorial moments $f_{r,s}$, $f_{u,v}$, and $f_{\alpha,\beta}$ can be found in Eq. (\ref{eq:fact2}). The Equation~(\ref{eq:covF}) can be used in the standard error propagation formula, Eq.~(\ref{eq:error}), to obtain the statistical uncertainties of the efficiency-corrected cumulants. The detailed derivation of the analytical formulae for statistical uncertainties on cumulants and moments exists in the literature~\cite{Luo:2011tp,Luo:2014rea}. If we put $\varepsilon_p=\varepsilon_{\bar{p}}=1$, the statistical uncertainties on the cumulants and cumulant ratios up to the eighth-order expressed in terms of central moments  ($\mu_n$) are given below, where the uncertainties are the square roots of the variances.
\begin{widetext}
\begin{equation*}
\begin{aligned}
\mathrm{Var}(C_1)&=\mu_{2}/n \\
\mathrm{Var}(C_2)&=(- {\mu}_{2}^{2} + {\mu}_{4})/n \\
\mathrm{Var}(C_3)&=(9 {\mu}_{2}^{3} - 6 {\mu}_{2} {\mu}_{4} - {\mu}_{3}^{2} + {\mu}_{6})/n \\
\mathrm{Var}(C_4)&=(- 36 {\mu}_{2}^{4} + 48 {\mu}_{2}^{2} {\mu}_{4} + 64 {\mu}_{2} {\mu}_{3}^{2}  - 12 {\mu}_{2} {\mu}_{6} - 8 {\mu}_{3} {\mu}_{5} - {\mu}_{4}^{2} + {\mu}_{8})/n\\
\mathrm{Var}(C_5)&=({\mu}_{10} + 900 {\mu}_{2}^{5} - 900 {\mu}_{2}^{3} {\mu}_{4} - 1000 {\mu}_{2}^{2} {\mu}_{3}^{2} + 160 {\mu}_{2}^{2} {\mu}_{6} + 240 {\mu}_{2} {\mu}_{3} {\mu}_{5}  \\&+ 125 {\mu}_{2} {\mu}_{4}^{2} - 20 {\mu}_{2} {\mu}_{8} + 200 {\mu}_{3}^{2} {\mu}_{4} - 20 {\mu}_{3} {\mu}_{7} - 10 {\mu}_{4} {\mu}_{6} - {\mu}_{5}^{2})/n\\
\mathrm{Var}(C_6)&=(- 30 {\mu}_{10} {\mu}_{2} + {\mu}_{12} - 8100 {\mu}_{2}^{6} + 13500 {\mu}_{2}^{4} {\mu}_{4}  + 39600 {\mu}_{2}^{3} {\mu}_{3}^{2} - 2880 {\mu}_{2}^{3} {\mu}_{6} \\&- 9720 {\mu}_{2}^{2} {\mu}_{3} {\mu}_{5} - 3600 {\mu}_{2}^{2} {\mu}_{4}^{2} + 405 {\mu}_{2}^{2} {\mu}_{8} - 9600 {\mu}_{2} {\mu}_{3}^{2} {\mu}_{4} + 840 {\mu}_{2} {\mu}_{3} {\mu}_{7} - 400 {\mu}_{3}^{4} \\&+ 216 {\mu}_{2} {\mu}_{5}^{2} + 510 {\mu}_{2} {\mu}_{4} {\mu}_{6}  + 440 {\mu}_{3}^{2} {\mu}_{6} + 1020 {\mu}_{3} {\mu}_{4} {\mu}_{5} - 40 {\mu}_{3} {\mu}_{9} + 225 {\mu}_{4}^{3} \\&- 30 {\mu}_{4} {\mu}_{8} - 12 {\mu}_{5} {\mu}_{7} - {\mu}_{6}^{2})/n\\
\mathrm{Var}(C_7)&=(861 {\mu}_{10} {\mu}_{2}^{2} - 70 {\mu}_{10} {\mu}_{4} - 70 {\mu}_{11} {\mu}_{3} - 42 {\mu}_{12} {\mu}_{2} + {\mu}_{14} + 396900 {\mu}_{2}^{7} - 529200 {\mu}_{2}^{5} {\mu}_{4} \\& - 1102500 {\mu}_{2}^{4} {\mu}_{3}^{2} + 79380 {\mu}_{2}^{4} {\mu}_{6} + 299880 {\mu}_{2}^{3} {\mu}_{3} {\mu}_{5} + 176400 {\mu}_{2}^{3} {\mu}_{4}^{2} - 10080 {\mu}_{2}^{3} {\mu}_{8} + 558600 {\mu}_{2}^{2} {\mu}_{3}^{2} {\mu}_{4} \\& - 33600 {\mu}_{2}^{2} {\mu}_{3} {\mu}_{7} - 29400 {\mu}_{2}^{2} {\mu}_{4} {\mu}_{6} - 10584 {\mu}_{2}^{2} {\mu}_{5}^{2} + 137200 {\mu}_{2} {\mu}_{3}^{4} - 43120 {\mu}_{2} {\mu}_{3}^{2} {\mu}_{6} \\& - 76440 {\mu}_{2} {\mu}_{3} {\mu}_{4} {\mu}_{5} + 2310 {\mu}_{2} {\mu}_{3} {\mu}_{9} - 14700 {\mu}_{2} {\mu}_{4}^{3} + 1890 {\mu}_{2} {\mu}_{4} {\mu}_{8} \\& + 966 {\mu}_{2} {\mu}_{5} {\mu}_{7} + 343 {\mu}_{2} {\mu}_{6}^{2} - 15680 {\mu}_{3}^{3} {\mu}_{5} - 14700 {\mu}_{3}^{2} {\mu}_{4}^{2} + 1505 {\mu}_{3}^{2} {\mu}_{8} + 2590 {\mu}_{3} {\mu}_{4} {\mu}_{7} \\& + 2254 {\mu}_{3} {\mu}_{5} {\mu}_{6} + 1715 {\mu}_{4}^{2} {\mu}_{6} + 1911 {\mu}_{4} {\mu}_{5}^{2} - 42 {\mu}_{5} {\mu}_{9} - 14 {\mu}_{6} {\mu}_{8} - {\mu}_{7}^{2})/n \\
\mathrm{Var}(C_8)&=( - 28560 {\mu}_{10} {\mu}_{2}^{3} + 5600 {\mu}_{10} {\mu}_{2} {\mu}_{4} + 4256 {\mu}_{10} {\mu}_{3}^{2} - 56 {\mu}_{10} {\mu}_{6} + 5376 {\mu}_{11} {\mu}_{2} {\mu}_{3} - 112 {\mu}_{11} {\mu}_{5} \\& + 1624 {\mu}_{12} {\mu}_{2}^{2} - 140 {\mu}_{12} {\mu}_{4} - 112 {\mu}_{13} {\mu}_{3} - 56 {\mu}_{14} {\mu}_{2}  + {\mu}_{16} - 6350400 {\mu}_{2}^{8} + 12700800 {\mu}_{2}^{6} {\mu}_{4} \\& + 59270400 {\mu}_{2}^{5} {\mu}_{3}^{2} - 2399040 {\mu}_{2}^{5} {\mu}_{6}  - 15523200 {\mu}_{2}^{4} {\mu}_{3} {\mu}_{5} - 6174000 {\mu}_{2}^{4} {\mu}_{4}^{2} + 322560 {\mu}_{2}^{4} {\mu}_{8} \\& - 35280000 {\mu}_{2}^{3} {\mu}_{3}^{2} {\mu}_{4}  + 1626240 {\mu}_{2}^{3} {\mu}_{3} {\mu}_{7}  + 1340640 {\mu}_{2}^{3} {\mu}_{4} {\mu}_{6} + 677376 {\mu}_{2}^{3} {\mu}_{5}^{2} - 8467200 {\mu}_{2}^{2} {\mu}_{3}^{4} \\& + 2759680 {\mu}_{2}^{2} {\mu}_{3}^{2} {\mu}_{6}    + 5597760 {\mu}_{2}^{2} {\mu}_{3} {\mu}_{4} {\mu}_{5} - 119840 {\mu}_{2}^{2} {\mu}_{3} {\mu}_{9} + 882000 {\mu}_{2}^{2} {\mu}_{4}^{3} - 108360 {\mu}_{2}^{2} {\mu}_{4} {\mu}_{8}  \\& - 77952 {\mu}_{2}^{2} {\mu}_{5} {\mu}_{7}  - 26656 {\mu}_{2}^{2} {\mu}_{6}^{2} + 2007040 {\mu}_{2} {\mu}_{3}^{3} {\mu}_{5}  + 3684800 {\mu}_{2} {\mu}_{3}^{2} {\mu}_{4}^{2} - 160160 {\mu}_{2} {\mu}_{3}^{2} {\mu}_{8} \\&- 322560 {\mu}_{2} {\mu}_{3} {\mu}_{4} {\mu}_{7}  - 257152 {\mu}_{2} {\mu}_{3} {\mu}_{5} {\mu}_{6} - 172480 {\mu}_{2} {\mu}_{4}^{2} {\mu}_{6}  - 178752 {\mu}_{2} {\mu}_{4} {\mu}_{5}^{2} + 3808 {\mu}_{2} {\mu}_{5} {\mu}_{9} \\& + 1680 {\mu}_{2} {\mu}_{6} {\mu}_{8} + 512 {\mu}_{2} {\mu}_{7}^{2}  + 940800 {\mu}_{3}^{4} {\mu}_{4} - 71680 {\mu}_{3}^{3} {\mu}_{7} - 203840 {\mu}_{3}^{2} {\mu}_{4} {\mu}_{6} - 75264 {\mu}_{3}^{2} {\mu}_{5}^{2} \\& - 156800 {\mu}_{3} {\mu}_{4}^{2} {\mu}_{5}  + 8960 {\mu}_{3} {\mu}_{4} {\mu}_{9}  + 6496 {\mu}_{3} {\mu}_{5} {\mu}_{8} + 4480 {\mu}_{3} {\mu}_{6} {\mu}_{7} - 4900 {\mu}_{4}^{4} + 5040 {\mu}_{4}^{2} {\mu}_{8} \\& + 9856 {\mu}_{4} {\mu}_{5} {\mu}_{7}  + 4704 {\mu}_{4} {\mu}_{6}^{2} + 6272 {\mu}_{5}^{2} {\mu}_{6} - 16 {\mu}_{7} {\mu}_{9} - {\mu}_{8}^{2} )/n \\
\mathrm{Var}(\frac{C_2}{C_1})&=(- \frac{{\mu}_{2}^{2}}{\langle N\rangle^{2}}  + \frac{{\mu}_{4}}{\langle N\rangle^{2}} - \frac{2 {\mu}_{2} {\mu}_{3}}{\langle N\rangle^{3}} + \frac{{\mu}_{2}^{3}}{\langle N\rangle^{4}})/n\\
\mathrm{Var}(\frac{C_3}{C_2}) &= (9 {\mu}_{2} - \frac{6 {\mu}_{4}}{{\mu}_{2}} + \frac{6 {\mu}_{3}^{2}}{{\mu}_{2}^{2}} + \frac{{\mu}_{6}}{{\mu}_{2}^{2}} - \frac{2 {\mu}_{3} {\mu}_{5}}{{\mu}_{2}^{3}} + \frac{{\mu}_{3}^{2} {\mu}_{4}}{{\mu}_{2}^{4}})/n\\
\mathrm{Var}(\frac{C_4}{C_2})&=(- 9 {\mu}_{2}^{2} + 9 {\mu}_{4} + \frac{40 {\mu}_{3}^{2}}{{\mu}_{2}} - \frac{6 {\mu}_{6}}{{\mu}_{2}} - \frac{8 {\mu}_{3} {\mu}_{5}}{{\mu}_{2}^{2}} + \frac{6 {\mu}_{4}^{2}}{{\mu}_{2}^{2}} + \frac{{\mu}_{8}}{{\mu}_{2}^{2}} + \frac{8 {\mu}_{3}^{2} {\mu}_{4}}{{\mu}_{2}^{3}} - \frac{2 {\mu}_{4} {\mu}_{6}}{{\mu}_{2}^{3}} + \frac{{\mu}_{4}^{3}}{{\mu}_{2}^{4}})/n \\
\end{aligned}
\end{equation*}
\end{widetext}
\begin{widetext}
\begin{equation*}
\begin{aligned}
\mathrm{Var}(\frac{C_5}{C_1})&=(\frac{{\mu}_{10}}{\langle N\rangle^{2}} + \frac{900 {\mu}_{2}^{5}}{\langle N\rangle^{2}} - \frac{900 {\mu}_{2}^{3} {\mu}_{4}}{\langle N\rangle^{2}} - \frac{1000 {\mu}_{2}^{2} {\mu}_{3}^{2}}{\langle N\rangle^{2}} + \frac{160 {\mu}_{2}^{2} {\mu}_{6}}{\langle N\rangle^{2}} + \frac{240 {\mu}_{2} {\mu}_{3} {\mu}_{5}}{\langle N\rangle^{2}} + \frac{125 {\mu}_{2} {\mu}_{4}^{2}}{\langle N\rangle^{2}} \\& - \frac{20 {\mu}_{2} {\mu}_{8}}{\langle N\rangle^{2}} + \frac{200 {\mu}_{3}^{2} {\mu}_{4}}{\langle N\rangle^{2}} -  \frac{20 {\mu}_{3} {\mu}_{7}}{\langle N\rangle^{2}} - \frac{10 {\mu}_{4} {\mu}_{6}}{\langle N\rangle^{2}} - \frac{{\mu}_{5}^{2}}{\langle N\rangle^{2}} + \frac{600 {\mu}_{2}^{4} {\mu}_{3}}{\langle N\rangle^{3}}  - \frac{60 {\mu}_{2}^{3} {\mu}_{5}}{\langle N\rangle^{3}} - \frac{300 {\mu}_{2}^{2} {\mu}_{3} {\mu}_{4}}{\langle N\rangle^{3}}  \\&- \frac{200 {\mu}_{2} {\mu}_{3}^{3}}{\langle N\rangle^{3}} + \frac{20 {\mu}_{2} {\mu}_{3} {\mu}_{6}}{\langle N\rangle^{3}}  + \frac{30 {\mu}_{2} {\mu}_{4} {\mu}_{5}}{\langle N\rangle^{3}}  + \frac{20 {\mu}_{3}^{2} {\mu}_{5}}{\langle N\rangle^{3}} - \frac{2 {\mu}_{5} {\mu}_{6}}{\langle N\rangle^{3}} + \frac{100 {\mu}_{2}^{3} {\mu}_{3}^{2}}{\langle N\rangle^{4}} - \frac{20 {\mu}_{2}^{2} {\mu}_{3} {\mu}_{5}}{\langle N\rangle^{4}} + \frac{{\mu}_{2} {\mu}_{5}^{2}}{\langle N\rangle^{4}})/n\\
\mathrm{Var}(\frac{C_6}{C_2})&=(- \frac{30 {\mu}_{10}}{{\mu}_{2}} + \frac{{\mu}_{12}}{{\mu}_{2}^{2}} - 3600 {\mu}_{2}^{4} + 5400 {\mu}_{2}^{2} {\mu}_{4} + 30000 {\mu}_{2} {\mu}_{3}^{2} - 1800 {\mu}_{2} {\mu}_{6} - 8160 {\mu}_{3} {\mu}_{5} - 225 {\mu}_{4}^{2} \\&+ 345 {\mu}_{8} - \frac{3900 {\mu}_{3}^{2} {\mu}_{4}}{{\mu}_{2}} + \frac{840 {\mu}_{3} {\mu}_{7}}{{\mu}_{2}} - \frac{120 {\mu}_{4} {\mu}_{6}}{{\mu}_{2}} + \frac{216 {\mu}_{5}^{2}}{{\mu}_{2}} + \frac{2300 {\mu}_{3}^{4}}{{\mu}_{2}^{2}} - \frac{140 {\mu}_{3}^{2} {\mu}_{6}}{{\mu}_{2}^{2}} + \frac{240 {\mu}_{3} {\mu}_{4} {\mu}_{5}}{{\mu}_{2}^{2}} \\& - \frac{40 {\mu}_{3} {\mu}_{9}}{{\mu}_{2}^{2}} - \frac{12 {\mu}_{5} {\mu}_{7}}{{\mu}_{2}^{2}} + \frac{30 {\mu}_{6}^{2}}{{\mu}_{2}^{2}} - \frac{520 {\mu}_{3}^{3} {\mu}_{5}}{{\mu}_{2}^{3}}  + \frac{20 {\mu}_{3}^{2} {\mu}_{8}}{{\mu}_{2}^{3}} + \frac{52 {\mu}_{3} {\mu}_{5} {\mu}_{6}}{{\mu}_{2}^{3}} - \frac{2 {\mu}_{6} {\mu}_{8}}{{\mu}_{2}^{3}} + \frac{100 {\mu}_{3}^{4} {\mu}_{4}}{{\mu}_{2}^{4}} \\& - \frac{20 {\mu}_{3}^{2} {\mu}_{4} {\mu}_{6}}{{\mu}_{2}^{4}} + \frac{{\mu}_{4} {\mu}_{6}^{2}}{{\mu}_{2}^{4}})/n\\ 
\mathrm{Var}(\frac{C_7}{C_1})&=(\frac{861 {\mu}_{10} {\mu}_{2}^{2}}{\langle N\rangle^{2}} - \frac{70 {\mu}_{10} {\mu}_{4}}{\langle N\rangle^{2}} - \frac{70 {\mu}_{11} {\mu}_{3}}{\langle N\rangle^{2}} - \frac{42 {\mu}_{12} {\mu}_{2}}{\langle N\rangle^{2}} + \frac{{\mu}_{14}}{\langle N\rangle^{2}} + \frac{396900 {\mu}_{2}^{7}}{\langle N\rangle^{2}} - \frac{529200 {\mu}_{2}^{5} {\mu}_{4}}{\langle N\rangle^{2}} \\& - \frac{1102500 {\mu}_{2}^{4} {\mu}_{3}^{2}}{\langle N\rangle^{2}} + \frac{79380 {\mu}_{2}^{4} {\mu}_{6}}{\langle N\rangle^{2}} + \frac{299880 {\mu}_{2}^{3} {\mu}_{3} {\mu}_{5}}{\langle N\rangle^{2}} + \frac{176400 {\mu}_{2}^{3} {\mu}_{4}^{2}}{\langle N\rangle^{2}} - \frac{10080 {\mu}_{2}^{3} {\mu}_{8}}{\langle N\rangle^{2}} \\& + \frac{558600 {\mu}_{2}^{2} {\mu}_{3}^{2} {\mu}_{4}}{\langle N\rangle^{2}} - \frac{33600 {\mu}_{2}^{2} {\mu}_{3} {\mu}_{7}}{\langle N\rangle^{2}} - \frac{29400 {\mu}_{2}^{2} {\mu}_{4} {\mu}_{6}}{\langle N\rangle^{2}} - \frac{10584 {\mu}_{2}^{2} {\mu}_{5}^{2}}{\langle N\rangle^{2}} + \frac{137200 {\mu}_{2} {\mu}_{3}^{4}}{\langle N\rangle^{2}} \\& - \frac{43120 {\mu}_{2} {\mu}_{3}^{2} {\mu}_{6}}{\langle N\rangle^{2}} - \frac{76440 {\mu}_{2} {\mu}_{3} {\mu}_{4} {\mu}_{5}}{\langle N\rangle^{2}} + \frac{2310 {\mu}_{2} {\mu}_{3} {\mu}_{9}}{\langle N\rangle^{2}} - \frac{14700 {\mu}_{2} {\mu}_{4}^{3}}{\langle N\rangle^{2}} + \frac{1890 {\mu}_{2} {\mu}_{4} {\mu}_{8}}{\langle N\rangle^{2}} \\& + \frac{966 {\mu}_{2} {\mu}_{5} {\mu}_{7}}{\langle N\rangle^{2}} + \frac{343 {\mu}_{2} {\mu}_{6}^{2}}{\langle N\rangle^{2}} - \frac{15680 {\mu}_{3}^{3} {\mu}_{5}}{\langle N\rangle^{2}} - \frac{14700 {\mu}_{3}^{2} {\mu}_{4}^{2}}{\langle N\rangle^{2}} + \frac{1505 {\mu}_{3}^{2} {\mu}_{8}}{\langle N\rangle^{2}} + \frac{2590 {\mu}_{3} {\mu}_{4} {\mu}_{7}}{\langle N\rangle^{2}} \\& + \frac{2254 {\mu}_{3} {\mu}_{5} {\mu}_{6}}{\langle N\rangle^{2}} + \frac{1715 {\mu}_{4}^{2} {\mu}_{6}}{\langle N\rangle^{2}} + \frac{1911 {\mu}_{4} {\mu}_{5}^{2}}{\langle N\rangle^{2}} - \frac{42 {\mu}_{5} {\mu}_{9}}{\langle N\rangle^{2}} - \frac{14 {\mu}_{6} {\mu}_{8}}{\langle N\rangle^{2}} - \frac{{\mu}_{7}^{2}}{\langle N\rangle^{2}} + \frac{264600 {\mu}_{2}^{6} {\mu}_{3}}{\langle N\rangle^{3}} \\& - \frac{26460 {\mu}_{2}^{5} {\mu}_{5}}{\langle N\rangle^{3}} - \frac{220500 {\mu}_{2}^{4} {\mu}_{3} {\mu}_{4}}{\langle N\rangle^{3}} + \frac{1260 {\mu}_{2}^{4} {\mu}_{7}}{\langle N\rangle^{3}} - \frac{235200 {\mu}_{2}^{3} {\mu}_{3}^{3}}{\langle N\rangle^{3}} + \frac{11760 {\mu}_{2}^{3} {\mu}_{3} {\mu}_{6}}{\langle N\rangle^{3}} \\& + \frac{17640 {\mu}_{2}^{3} {\mu}_{4} {\mu}_{5}}{\langle N\rangle^{3}} + \frac{47040 {\mu}_{2}^{2} {\mu}_{3}^{2} {\mu}_{5}}{\langle N\rangle^{3}} + \frac{44100 {\mu}_{2}^{2} {\mu}_{3} {\mu}_{4}^{2}}{\langle N\rangle^{3}} - \frac{420 {\mu}_{2}^{2} {\mu}_{3} {\mu}_{8}}{\langle N\rangle^{3}} - \frac{840 {\mu}_{2}^{2} {\mu}_{4} {\mu}_{7}}{\langle N\rangle^{3}} \\& - \frac{1176 {\mu}_{2}^{2} {\mu}_{5} {\mu}_{6}}{\langle N\rangle^{3}} + \frac{39200 {\mu}_{2} {\mu}_{3}^{3} {\mu}_{4}}{\langle N\rangle^{3}} - \frac{1120 {\mu}_{2} {\mu}_{3}^{2} {\mu}_{7}}{\langle N\rangle^{3}} - \frac{1960 {\mu}_{2} {\mu}_{3} {\mu}_{4} {\mu}_{6}}{\langle N\rangle^{3}} - \frac{2352 {\mu}_{2} {\mu}_{3} {\mu}_{5}^{2}}{\langle N\rangle^{3}} \\& - \frac{1470 {\mu}_{2} {\mu}_{4}^{2} {\mu}_{5}}{\langle N\rangle^{3}} + \frac{42 {\mu}_{2} {\mu}_{5} {\mu}_{8}}{\langle N\rangle^{3}} + \frac{56 {\mu}_{2} {\mu}_{6} {\mu}_{7}}{\langle N\rangle^{3}} - \frac{3920 {\mu}_{3}^{2} {\mu}_{4} {\mu}_{5}}{\langle N\rangle^{3}} - \frac{2450 {\mu}_{3} {\mu}_{4}^{3}}{\langle N\rangle^{3}} + \frac{70 {\mu}_{3} {\mu}_{4} {\mu}_{8}}{\langle N\rangle^{3}} \\& + \frac{112 {\mu}_{3} {\mu}_{5} {\mu}_{7}}{\langle N\rangle^{3}} + \frac{70 {\mu}_{4}^{2} {\mu}_{7}}{\langle N\rangle^{3}} - \frac{2 {\mu}_{7} {\mu}_{8}}{\langle N\rangle^{3}} + \frac{44100 {\mu}_{2}^{5} {\mu}_{3}^{2}}{\langle N\rangle^{4}} - \frac{8820 {\mu}_{2}^{4} {\mu}_{3} {\mu}_{5}}{\langle N\rangle^{4}} - \frac{14700 {\mu}_{2}^{3} {\mu}_{3}^{2} {\mu}_{4}}{\langle N\rangle^{4}} \\& + \frac{420 {\mu}_{2}^{3} {\mu}_{3} {\mu}_{7}}{\langle N\rangle^{4}} + \frac{441 {\mu}_{2}^{3} {\mu}_{5}^{2}}{\langle N\rangle^{4}} + \frac{1470 {\mu}_{2}^{2} {\mu}_{3} {\mu}_{4} {\mu}_{5}}{\langle N\rangle^{4}} \\& - \frac{42 {\mu}_{2}^{2} {\mu}_{5} {\mu}_{7}}{\langle N\rangle^{4}} + \frac{1225 {\mu}_{2} {\mu}_{3}^{2} {\mu}_{4}^{2}}{\langle N\rangle^{4}} - \frac{70 {\mu}_{2} {\mu}_{3} {\mu}_{4} {\mu}_{7}}{\langle N\rangle^{4}} + \frac{{\mu}_{2} {\mu}_{7}^{2}}{\langle N\rangle^{4}} )/n \\
\end{aligned}
\end{equation*}
\end{widetext}
\begin{widetext}
\begin{equation*}
\begin{aligned}
\mathrm{Var}(\frac{C_8}{C_2})&=( - 27300 {\mu}_{10} {\mu}_{2} + \frac{4760 {\mu}_{10} {\mu}_{4}}{{\mu}_{2}} + \frac{3136 {\mu}_{10} {\mu}_{3}^{2}}{{\mu}_{2}^{2}} + \frac{112 {\mu}_{10} {\mu}_{3} {\mu}_{5}}{{\mu}_{2}^{3}} + \frac{70 {\mu}_{10} {\mu}_{4}^{2}}{{\mu}_{2}^{3}} - \frac{2 {\mu}_{10} {\mu}_{8}}{{\mu}_{2}^{3}} \\& + \frac{5376 {\mu}_{11} {\mu}_{3}}{{\mu}_{2}} - \frac{112 {\mu}_{11} {\mu}_{5}}{{\mu}_{2}^{2}} + 1624 {\mu}_{12} - \frac{140 {\mu}_{12} {\mu}_{4}}{{\mu}_{2}^{2}} - \frac{112 {\mu}_{13} {\mu}_{3}}{{\mu}_{2}^{2}} - \frac{56 {\mu}_{14}}{{\mu}_{2}} + \frac{{\mu}_{16}}{{\mu}_{2}^{2}} \\& - 3572100 {\mu}_{2}^{6} + 6747300 {\mu}_{2}^{4} {\mu}_{4} + 48686400 {\mu}_{2}^{3} {\mu}_{3}^{2} - 1693440 {\mu}_{2}^{3} {\mu}_{6} - 13335840 {\mu}_{2}^{2} {\mu}_{3} {\mu}_{5} \\& - 2425500 {\mu}_{2}^{2} {\mu}_{4}^{2} + 282240 {\mu}_{2}^{2} {\mu}_{8} - 25166400 {\mu}_{2} {\mu}_{3}^{2} {\mu}_{4} + 1545600 {\mu}_{2} {\mu}_{3} {\mu}_{7} + 664440 {\mu}_{2} {\mu}_{4} {\mu}_{6} \\& + 606816 {\mu}_{2} {\mu}_{5}^{2} - 1254400 {\mu}_{3}^{4} + 1881600 {\mu}_{3}^{2} {\mu}_{6} + 3974880 {\mu}_{3} {\mu}_{4} {\mu}_{5} - 119840 {\mu}_{3} {\mu}_{9} + 102900 {\mu}_{4}^{3} \\& - 78540 {\mu}_{4} {\mu}_{8} - 77952 {\mu}_{5} {\mu}_{7} - 784 {\mu}_{6}^{2} - \frac{439040 {\mu}_{3}^{3} {\mu}_{5}}{{\mu}_{2}} + \frac{1764000 {\mu}_{3}^{2} {\mu}_{4}^{2}}{{\mu}_{2}} - \frac{115360 {\mu}_{3}^{2} {\mu}_{8}}{{\mu}_{2}} \\& - \frac{268800 {\mu}_{3} {\mu}_{4} {\mu}_{7}}{{\mu}_{2}} - \frac{119168 {\mu}_{3} {\mu}_{5} {\mu}_{6}}{{\mu}_{2}} - \frac{31360 {\mu}_{4}^{2} {\mu}_{6}}{{\mu}_{2}} - \frac{131712 {\mu}_{4} {\mu}_{5}^{2}}{{\mu}_{2}} + \frac{3808 {\mu}_{5} {\mu}_{9}}{{\mu}_{2}} \\& - \frac{840 {\mu}_{6} {\mu}_{8}}{{\mu}_{2}} + \frac{512 {\mu}_{7}^{2}}{{\mu}_{2}} - \frac{62720 {\mu}_{3}^{2} {\mu}_{4} {\mu}_{6}}{{\mu}_{2}^{2}} + \frac{159936 {\mu}_{3}^{2} {\mu}_{5}^{2}}{{\mu}_{2}^{2}} + \frac{3920 {\mu}_{3} {\mu}_{4}^{2} {\mu}_{5}}{{\mu}_{2}^{2}} + \frac{8960 {\mu}_{3} {\mu}_{4} {\mu}_{9}}{{\mu}_{2}^{2}} \\& + \frac{224 {\mu}_{3} {\mu}_{5} {\mu}_{8}}{{\mu}_{2}^{2}} + \frac{896 {\mu}_{3} {\mu}_{6} {\mu}_{7}}{{\mu}_{2}^{2}} + \frac{28175 {\mu}_{4}^{4}}{{\mu}_{2}^{2}} + \frac{2100 {\mu}_{4}^{2} {\mu}_{8}}{{\mu}_{2}^{2}} + \frac{9856 {\mu}_{4} {\mu}_{5} {\mu}_{7}}{{\mu}_{2}^{2}} + \frac{3136 {\mu}_{5}^{2} {\mu}_{6}}{{\mu}_{2}^{2}} \\& - \frac{16 {\mu}_{7} {\mu}_{9}}{{\mu}_{2}^{2}} + \frac{56 {\mu}_{8}^{2}}{{\mu}_{2}^{2}} + \frac{62720 {\mu}_{3}^{3} {\mu}_{4} {\mu}_{5}}{{\mu}_{2}^{3}} + \frac{39200 {\mu}_{3}^{2} {\mu}_{4}^{3}}{{\mu}_{2}^{3}} - \frac{1120 {\mu}_{3}^{2} {\mu}_{4} {\mu}_{8}}{{\mu}_{2}^{3}} - \frac{7168 {\mu}_{3}^{2} {\mu}_{5} {\mu}_{7}}{{\mu}_{2}^{3}} \\& - \frac{4480 {\mu}_{3} {\mu}_{4}^{2} {\mu}_{7}}{{\mu}_{2}^{3}} - \frac{7840 {\mu}_{3} {\mu}_{4} {\mu}_{5} {\mu}_{6}}{{\mu}_{2}^{3}} - \frac{6272 {\mu}_{3} {\mu}_{5}^{3}}{{\mu}_{2}^{3}} + \frac{128 {\mu}_{3} {\mu}_{7} {\mu}_{8}}{{\mu}_{2}^{3}} - \frac{4900 {\mu}_{4}^{3} {\mu}_{6}}{{\mu}_{2}^{3}} - \frac{3920 {\mu}_{4}^{2} {\mu}_{5}^{2}}{{\mu}_{2}^{3}} \\& + \frac{140 {\mu}_{4} {\mu}_{6} {\mu}_{8}}{{\mu}_{2}^{3}} + \frac{112 {\mu}_{5}^{2} {\mu}_{8}}{{\mu}_{2}^{3}} + \frac{3136 {\mu}_{3}^{2} {\mu}_{4} {\mu}_{5}^{2}}{{\mu}_{2}^{4}} + \frac{3920 {\mu}_{3} {\mu}_{4}^{3} {\mu}_{5}}{{\mu}_{2}^{4}} \\& - \frac{112 {\mu}_{3} {\mu}_{4} {\mu}_{5} {\mu}_{8}}{{\mu}_{2}^{4}} + \frac{1225 {\mu}_{4}^{5}}{{\mu}_{2}^{4}} - \frac{70 {\mu}_{4}^{3} {\mu}_{8}}{{\mu}_{2}^{4}} + \frac{{\mu}_{4} {\mu}_{8}^{2}}{{\mu}_{2}^{4}})/n  \\
\end{aligned}
\end{equation*}
\end{widetext}
\end{document}